\documentclass[12pt]{iopart}
\usepackage{iopams}
\usepackage{graphicx}
\usepackage{color}
\usepackage{hyperref}
\usepackage{url}
\usepackage{subfiles}

\begin{document}

\title[Construction of optimized tight-binding models using \textit{ab initio} Hamiltonian]{Construction of optimized tight-binding models using \textit{ab initio} Hamiltonian: Application to monolayer $2H$-transition metal dichalcogenides}

\author{Sejoong Kim}
\address{University of Science and Technology (UST), Gajeong-ro 217, Daejeon 34113, Republic of Korea}
\address{Korea Institute for Advanced Study, Hoegiro 85, Seoul 02455, Republic of Korea}
\ead{sejoong@alum.mit.edu}
\vspace{10pt}

\begin{abstract}
We present optimized tight-binding models with atomic orbitals to improve \textit{ab initio} tight-binding models constructed by truncating full density functional theory (DFT) Hamiltonian based on localized orbitals. Retaining qualitative features of the original Hamiltonian, the optimization reduces quantitative deviations in overall band structures between the \textit{ab initio} tight-binding model and the full DFT Hamiltonian. The optimization procedure and related details are demonstrated by using semiconducting and metallic Janus transition metal dichalcogenides monolayers in the $2H$ configuration. Varying the truncation range from partial second neighbors to third ones, we show differences in electronic structures between the truncated tight-binding model and the original full Hamiltonian, and how much the optimization can remedy the quantitative loss induced by truncation. We further elaborate the optimization process so that local electronic properties such as valence and conduction band edges and Fermi surfaces are precisely reproduced by the optimized tight-binding model. We also extend our discussions to tight-binding models including spin-orbit interactions, so we provide the optimized tight-binding model replicating spin-related properties of the original Hamiltonian such as spin textures. The optimization process described here can be readily applied to construct the fine-tuned tight-binding model based on various DFT calculations. 
\end{abstract}

%
%
%
%
%

\section{\label{sec:Intro}Introduction}
Constructing Hamiltonians is an essential starting point to investigate physical systems. 
For condensed matter systems, electronic structures and related properties are derived from Hamiltonians. 
The tight-binding model, in which electrons are described in terms of hopping processes between neighboring localized orbitals, is one of modeling methods widely used to construct Hamiltonians for condensed matter systems. 
The tight-binding model constructed to investigate a given system can further be used to study extended systems derived from the original system, such as defected systems~\cite{PRB2014Pike}, nanostructures~\cite{PRB2004XiaoDong, PRB2005Schulz}, surfaces~\cite{PRB1976Pandey, PRB1994Davidson, JChemPhys1993}, strained systems~\cite{PRB2009Pereira, PhyRevB_98_075106_2018_Fang, PhyRevB_76_115202_2007_Jancu, PhyRevB_94_155416_2016_Pearce, PhyRevB_79_245201_2009_Niquet}, twisted systems~\cite{PRB_85_194558_2012_Moon, PRB_90_155406_2014_Moon, PRB_99_165430_2019_Moon}, etc. 

The tight-binding model construction begins with a relevant set of localized orbitals that mainly contributes to electronic structures around the Fermi energy~\cite{PhysRev_94_1498_1954_Slater_Koster, JPhysCondensMatter_2003_Papaconstantopoulos_Mehl, Handbook_TB_Papaconstantopoulos}. 
Together with the chosen set of localized orbitals, the number of non-zero hopping neighbors between localized orbitals determines the size of the tight-binding model~\cite{PhysRev_94_1498_1954_Slater_Koster, JPhysCondensMatter_2003_Papaconstantopoulos_Mehl, Handbook_TB_Papaconstantopoulos}. 
Once the overall tight-binding model is designed, the tight-binding model parameters can be determined by numerically fitting existing band structures, possibly obtained from experiments or computed by first-principle calculations based on the density functional theory (DFT)~\cite{PhysRev_136_B864_1964, PhysRev_140_A1133_1965}. 

From a general point of view, the fitting procedure is an optimization problem in which the tight-binding model parameters can be obtained by minimizing a cost function quantifying deviations between targeted electronic bands and ones computed by the tight-binding model~\cite{JPhysCondMatter2015Ridolfi}.
Depending on initial values of tight-binding parameters it is in principle possible to reach different local minima throughout the optimization procedure. As a result of the minimization, possibly obtained tight-binding parameters do not necessarily constitute the physically correct Hamiltonian. Even though the resulting tight-binding model well reproduces targeted band structures, the tight-binding model can fail to reproduce other important physical properties such as orbital-projected density of states, spin textures, etc. This is because the optimization process focuses only on minimizing deviations between targeted electronic bands and fitted ones. 
This issue on the fitting procedure might be resolved by adding to the cost function additional constraints on physical properties required to retain, such as orbital characters. 

Alternatively one can use DFT calculations to construct realistic tight-binding Hamiltonians. Not only do DFT calculations provide energy band structures targeted by the fitting procedure described above, but DFT calculations themselves also produce tight-binding Hamiltonian as a result of computations. DFT calculations based on atomic-like localized orbitals produce the tight-binding Hamiltonian by construction after the self-consistent calculations, and the Hamiltonian constructed by plane-wave based DFT calculations can be converted into the tight-binding one via the Wannier function transformation~\cite{PhysRev_52_191_1937_Wannier, RevModPhys_34_645_1962_Wannier, PhysRevB_56_12847_1997_Marzari, PhysRevB_65_035109_2002_Souza, RevModPhys_84_1419_2012_Marzari, PhysRevLett_94_026405_2005_Thygesen, PhysRevB_104_125140_Fontana, PhysRevB_61_10040_2000_Berghold, Multiscale_Model_Simul_17_167_Damle}. In this sense, the resulting DFT Hamiltonian written in terms of localized orbitals are called as the \textit{ab initio} tight-binding Hamiltonian~\cite{PhysRevB_89_035405_Jung, PhysRevB_92_205108_Fang}. The resulting \textit{ab initio} tight-binding Hamiltonian provides reliable electronic structures including not only band structures but also orbital characters.

Despite the advantages mentioned above, the \textit{ab initio} tight-binding Hamiltonian is very large, since localized orbitals of the Hamiltonian are generally extended over several unit cells. Since the coupling strength decreases as the inter-site distance increases, one can simplify the \textit{ab initio} tight-binding Hamiltonian by truncating the Hamiltonian within a finite range of couplings, i.e., first or second nearest-neighboring couplings as suggested in Ref.~\cite{PhysRevB_92_205108_Fang}. The truncated \textit{ab initio} tight-binding model, where significant contributions are included, qualitatively retains main properties of the original Hamiltonian. However, since hopping terms dropped out in the truncation are small but finite, the truncated model can lose a quantitative accuracy compared with the original system. In particular, for the case where it is required to precisely reproduce local features of band structures such as Fermi surfaces, spin textures around the Fermi energy, and band-edge properties, the quantitative loss induced by truncation cannot be simply ignored.

In this work we construct the optimized tight-binding model with atomic orbitals by using the truncated \textit{ab initio} tight-binding Hamiltonian. 
Recalling that the truncated \textit{ab initio} tight-binding Hamiltonian retains qualitatively reliable properties of the full DFT calculations as discussed in Ref.~\cite{PhysRevB_92_205108_Fang}, hopping parameters of the truncated \textit{ab initio} Hamiltonian can serve as a good initial set of tight-binding parameters for construction of optimized tight-binding model of atomic orbitals. 
The optimization process will reduce quantitative loss due to truncation, so the resulting tight-binding model with atomic orbitals can reproduce electronic structures better than the truncated \textit{ab initio} Hamiltonian. Thus one can construct the optimized tight-binding model whose size is smaller than the full \textit{ab initio} Hamiltonian, with accuracy of electronic structure properties secured.

To demonstrate the construction of the optimized tight-binding model based on the truncated $\textit{ab initio}$ Hamiltonian, we consider Janus transition metal dichalcogenide (TMDC) monolayers~\cite{NatNanoTech_12_744_2017_Lu, ACSNano_11_8192_2017_Zhang, JPhysCM_30_215301_2018_Shi}, where upper and lower chalcogen atom layers consist of different chalcogens, so the mirror symmetry with respect to the transition metal layer is naturally broken. 
Particularly, investigating semiconducting and metallic Janus $2H$-TMDC monolayers, we present the optimization procedure to fine-tune local electronic structures such as valence and conduction band edges around the semiconducting band gap position, and Fermi surfaces and spin textures of metallic systems. 

This paper is organized as follows. In Sec.~\ref{sec:DFT} we describe atomic configurations of monolayer Janus $2H$-TMDC and summarize details of DFT calculations, which produce the full \textit{ab initio} tight-binding Hamiltonian based on the Wannier functions. 
Section~\ref{subsec:TB} starts with the tight-binding model construction based on energy integrals and symmetry considerations, by following the Slater-Koster theory~\cite{PhysRev_94_1498_1954_Slater_Koster, PhysRevB_55_4168_1997_Stiles}. 
In Secs.~\ref{subsec:truncation} and \ref{subsec:optimization} we summarizes the truncation process of the full \textit{ab initio} tight-binding Hamiltonian, and the accompanying optimization procedure. 
In Sec.~\ref{subsec:tTB} we demonstrate band structures of truncated tight-binding models of monolayer Janus $2H$-TMDC by increasing the truncation range from the partial second nearest neighbors to the third ones. In Sec.~\ref{subsec:oTB} we discuss the optimized tight-binding models corresponding to truncated models. In Sec.~\ref{subsec:SOC} we extend the optimization of the truncated tight-binding models to the case where spin-orbit coupling is included. We summarize our conclusion in Sec.~\ref{sec:conclusions}. Detailed expressions of energy integrals and tight-binding model parameters are provided in the supplementary data (available online).

\section{\label{sec:DFT}DFT Calculations}
\begin{figure}[htp!]
\begin{center}
\includegraphics[width=1.0\columnwidth, clip=true]{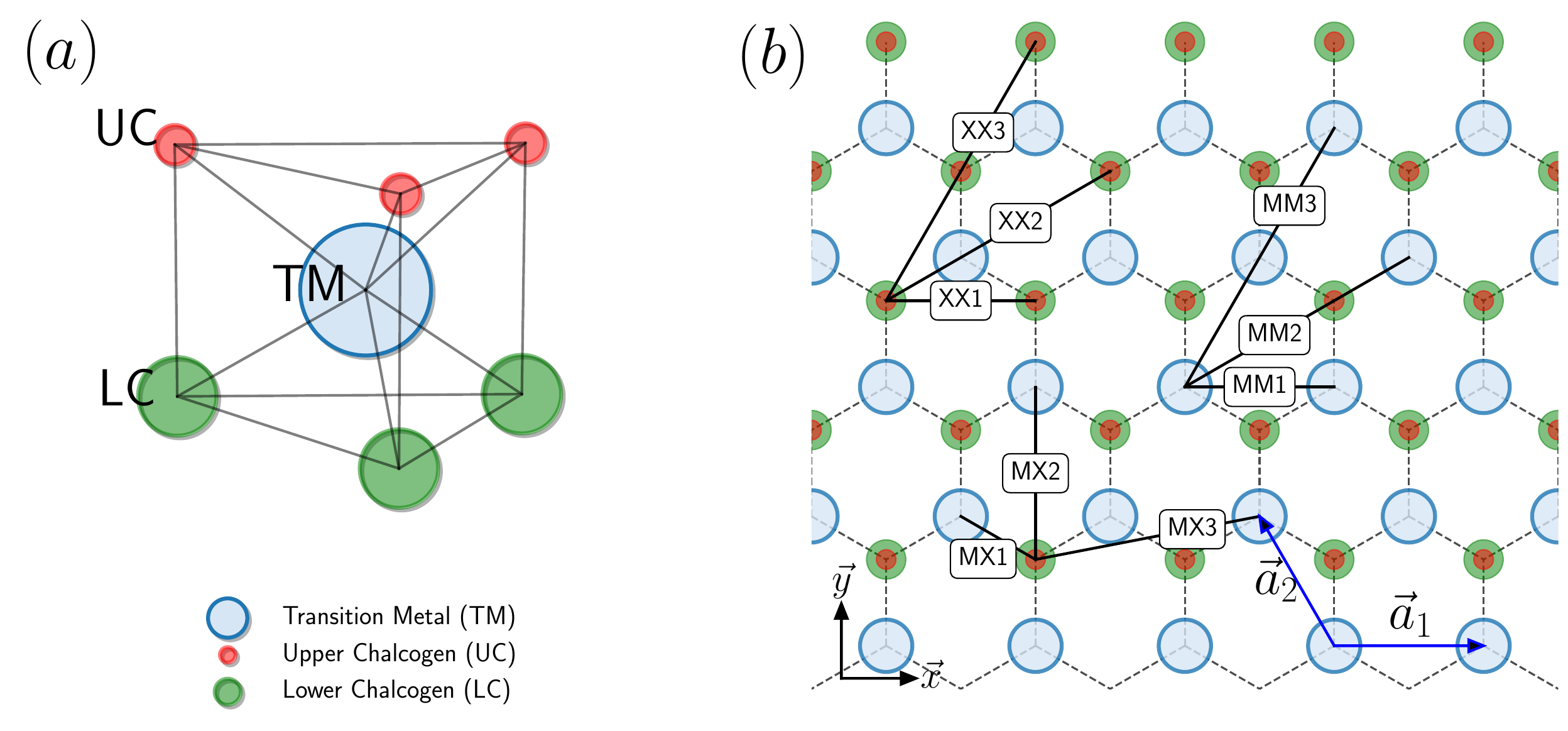} 
\end{center}
\caption{\label{CONFIG} (Color online) (a) Janus TMDC monolayer in $2H$ configuration. The largest blue circles, the middle-sized green ones, and the smallest red ones represent transition metal (TM), chalcogen on upper sub-layer (UC), and chalcogen on lower sub-layer (LC), respectively. (b) First, second, and third nearest neighbors between transition metals (MM), between chalcogen atoms (XX), and between transition metal and chalcogen atom (MX). Blue arrows denote two in-plane unit vectors $\vec{a}_{1}$ and $\vec{a}_{2}$.}
\end{figure}
In this work we consider the Janus TMDC monolayer in the $2H$ type, whose atomic configuration is depicted in figure~\ref{CONFIG}~\cite{NatNano2012Wang}. As shown in figure~\ref{CONFIG}(a), a transition metal atom is surrounded by three chalcogen atoms in the upper sublayer and the other three chalcogen atoms in the lower sublayer. Since upper and lower sublayers consist of different kinds of chalcogen atoms, the Janus TMDC monolayer has broken mirror-reflection symmetry with respect to the transition metal plane. As reference systems we choose MoSSe and NbSSe, which are semiconducting and metallic respectively. 

First we perform DFT calculations on Janus TMDC monolayers, MoSSe and NbSSe.
We use \textsc{Quantum  Espresso}~\cite{JPhys_CM_21_395502_2009,JPhys_CM_29_465901_2017} with the plane-wave (PW) basis, the PBE exchange-correlation functional~\cite{PhysRevLett_77_3865_1996} and norm-conserving pseudopotentials~\cite{PhysRevB_88_085117_2013, ComPhysComms_226_39_2018_Setten}. Here we choose fully relativistic pseudopotentials in order to take the effect of spin-orbit coupling into account~\cite{PhysRevB_21_2603_1980_Kleinman, PhysRevB_25_2103_1982_Bachelet}.    
We adopt $24\times24\times1$ $k$-point mesh, smearing temperature $0.005$ Ry with the cold smearing technique~\cite{PhysRevLett_82_3296_1999_Marzari}, and the kinetic energy cutoff $120$ Ry for self-consistent calculations. A vacuum size between periodic image layers is taken to be 30 $\textrm \AA$ in order to remove interactions between periodic images. 
Figures~\ref{Fig:MoSSe},~\ref{Fig:NbSSe},~\ref{Fig:MoSSe-SOC}, and~\ref{Fig:NbSSe-SOC} illustrate band structures computed by DFT calculations, which are used as reference band structures that tight-binding models are fitted to.  
DFT calculations also provide other electronic structure properties such as orbital-projected band energies, Fermi surfaces, and spin textures, which are compared with tight-binding models in Sec.~\ref{sec:discussion}. 
For the full \textit{ab initio} tight-binding Hamiltonian we use the maximally localized Wannier functions (MLWF)~\cite{PhysRevB_56_12847_1997_Marzari, PhysRevB_65_035109_2002_Souza, RevModPhys_84_1419_2012_Marzari} implemented in \textsc{Wannier90}~\cite{JPhysCM_32_165902_Pizzi} in order to construct five $d$-orbitals of the transition metal atom and six $p$-orbitals, each three of which belong to upper and lower chalcogen atoms, respectively.  

\section{\label{sec:Theory}Theory}
\subsection{\label{subsec:TB}Tight-Binding Model}
In the Slater-Koster scheme~\cite{PhysRev_94_1498_1954_Slater_Koster}, the tight-binding (TB) model reads 
\begin{equation}
\label{Eq:TB}\mathcal{H}_{m, n}(\mathbf{k}) = \sum_{\mathbf{R}}e^{i\mathbf{k}\cdot\mathbf{R}} E_{m, n}(\mathbf{R}), 
\end{equation}
where $E_{m, n}(\mathbf{R})$ is an energy integral, which is 
\begin{eqnarray}
\label{Eq:energy_integrals}E_{m, n}(\mathbf{R}) &=& \langle \phi_{m}(\mathbf{0}) | \mathcal{H} | \phi_{n}(\mathbf{R})\rangle,
\end{eqnarray}
where $|\phi_{m}(\mathbf{R})\rangle$ is the $m$th basis state, whose center is located at the lattice vector $\mathbf{R}$~\cite{PhysRevB_88_085433_Liu, PhysRevB_55_4168_1997_Stiles}. 

The energy integrals respect crystal symmetries of the system. Without the mirror-reflection symmetry with respect to the transition metal plane, monolayer Janus 2H-TMDC satisfies the $C_{3v}$ point-group symmetry, which includes symmetric operations $\{ \hat{E}, \hat{C}_{3}, \hat{C}_{3}^{2}, \hat{\sigma}_{v}, \hat{\sigma}_{v}^{\prime},\hat{\sigma}_{v}^{\prime\prime} \}$~\cite{Group_Theory_Dresselhaus}. 
Here $\hat{E}$ is the identity operation. 
$\hat{C}_{3}$ and $ \hat{C}_{3}^{2}$ are rotational operations by $2\pi/3$ and $4\pi/3$ around the $z$ axis, respectively. 
$\hat{\sigma}_{v}$, $\hat{\sigma}_{v}^{\prime}$, and $\hat{\sigma}_{v}^{\prime\prime}$ represent the mirror-reflection operations whose reflection planes, which are perpendicular to the monolayer plane, include in-plane vectors $\left( \vec{a}_{1} - \vec{a}_{2} \right)$, $\left( 2\vec{a}_{1} + \vec{a}_{2} \right)$, $\left(-\vec{a}_{1} - 2\vec{a}_{2}\right)$, respectively.  
Here $\vec{a}_{1}$ and $\vec{a}_{2}$ are in-plane unit vectors as shown in figure~\ref{CONFIG}(b). 

Note that the point group $C_{3v}$ has three irreducible representations $A_{1}$, $A_{2}$, and $E$~\cite{Group_Theory_Dresselhaus}. $d$ orbitals of transition metals and $p$ orbitals of chalcogen atoms are categorized in association with irreducible representations $A_{1}$ and $E$ as follows:
\begin{eqnarray}
A_{1} &:& \{d_{z^2}\}, \; \{p_{z}\}\\
E     &:& \{d_{x^2-y^2},\;d_{xy}\}, \; \{d_{zx}, \; d_{yz} \}, \; \{p_{x},\; p_{y}\}
\end{eqnarray}
Here $\{\cdots\}$ indicates basis states related to one another via irreducible representations of symmetric operations. 
When a basis state is denoted by $|\phi^{\alpha}_{i\mu}\rangle$, where $\alpha=d, p$ represents whether the state is $d$ or $p$ orbital, and $\mu i$ means that $i$th basis state belonging to the $\mu$ irreducible representation, 
basis states are written as 
\begin{eqnarray}
|\phi^{d}_{1A_{1}}\rangle &=& |d_{z^2}\rangle \\
|\phi^{d}_{1E}\rangle &=& |d_{x^2-y^2}\rangle, \; |\phi^{d}_{2E}\rangle = |d_{xy}\rangle \\
|\phi^{d}_{3E}\rangle &=& |d_{yz}\rangle,      \; |\phi^{d}_{4E}\rangle = |d_{zx}\rangle \\
|\phi^{p}_{1A_{1}}\rangle &=& |p_{z}\rangle   \\
|\phi^{p}_{1E}\rangle &=& |p_{x}\rangle,       \; |\phi^{p}_{2E}\rangle = |p_{y}\rangle
\end{eqnarray}
Energy integrals $E_{m,n}(\mathbf{R})=E^{\alpha\beta}_{\mu i,\nu j}(\mathbf{R})$ are related to one another throughout symmetric operations $\hat{\mathcal{O}}=\{ \hat{E}, \hat{C}_{3}, \hat{C}_{3}^{2}, \hat{\sigma}_{v}, \hat{\sigma}_{v}^{\prime},\hat{\sigma}_{v}^{\prime\prime} \}$ as follows~\cite{PhysRevB_88_085433_Liu}:
\begin{equation}
\label{Eq:energy_integral_symmetry}E^{\alpha\beta}_{i\mu, j\nu}\left(\hat{\mathcal{O}} \mathbf{R} \right) = \sum_{i^{\prime}}\sum_{j^{\prime}} \mathcal{D}^{\mu}_{ii^{\prime}} \left( \hat{\mathcal{O}}\right) E^{\alpha\beta}_{i^{\prime}\mu, j^{\prime}\nu}\left(\mathbf{R}\right) \mathcal{D}^{\nu*}_{j^{\prime} j}\left( \hat{\mathcal{O}}\right),      
\end{equation}
where $\mathcal{D}^{\mu}(\hat{\mathcal{O}})$ is the representation matrix corresponding to the $\mu$th irreducible representation of the symmetric operation $\hat{\mathcal{O}}$. 
For the irreducible representation $A_{1}$, the corresponding matrix is unity, i.e., 
\begin{equation}
\mathcal{D}^{A_{1}}(\hat{\mathcal{O}}) = 1,   
\end{equation}
where $\hat{\mathcal{O}}=\{ \hat{E}, \hat{C}_{3}, \hat{C}_{3}^{2}, \hat{\sigma}_{v}, \hat{\sigma}_{v}^{\prime},\hat{\sigma}_{v}^{\prime\prime} \}$. 
Irreducible representations $E$ correspond to the following matrices depending on symmetric operations of the $C_{3v}$ point group:
\begin{eqnarray}
\label{Eq:Irrep01}\mathcal{D}^{E}(\hat{E}) &=& \left[\begin{array}{cc} 1 & 0\\ 0 & 1 \end{array}\right], 
\mathcal{D}^{E}(\hat{C}_{3}) = \left[\begin{array}{cc} -\frac{1}{2} & -\frac{\sqrt{3}}{2}\\ \frac{\sqrt{3}}{2} & -\frac{1}{2} \end{array}\right] \\ 
\label{Eq:Irrep02}\mathcal{D}^{E}(\hat{C}_{3}^{2}) &=& \left[\begin{array}{cc} -\frac{1}{2} & \frac{\sqrt{3}}{2}\\ -\frac{\sqrt{3}}{2} & -\frac{1}{2} \end{array}\right], \mathcal{D}^{E}(\hat{\sigma}_{v}) = \left[\begin{array}{cc} \frac{1}{2} & \frac{\sqrt{3}}{2}\\ \frac{\sqrt{3}}{2} & -\frac{1}{2} \end{array}\right] \\
\label{Eq:Irrep03}\mathcal{D}^{E}(\hat{\sigma}^{\prime}_{v}) &=& \left[\begin{array}{cc} \frac{1}{2} & -\frac{\sqrt{3}}{2}\\ -\frac{\sqrt{3}}{2} & -\frac{1}{2}\end{array}\right], 
\mathcal{D}^{E}(\hat{\sigma}_{v}^{\prime\prime}) = \left[\begin{array}{cc} -1 & 0\\ 0 & 1 \end{array}\right]. 
\end{eqnarray}
Using equations~(\ref{Eq:energy_integral_symmetry})-(\ref{Eq:Irrep03}) one can determine relations among energy integrals and a set of independent energy integrals. 
For example, considering $\mathbf{R}_{1}=-\vec{a}_{1}$ and $\mathbf{R}_{3}=-\vec{a}_{2}$ where are related via $\mathbf{R}_{3} = \hat{C}_{3} \mathbf{R}_{1}$, 
\begin{eqnarray}
E^{dd}_{1E,1E}(\mathbf{R}_{3}) &=&   \frac{1}{4}E^{dd}_{1E,1E}(\mathbf{R}_{1})-\frac{\sqrt{3}}{4}E^{dd}_{1E,2E}(\mathbf{R}_{1}) \nonumber \\
                               & &  -\frac{\sqrt{3}}{4}E^{dd}_{2E,1E}(\mathbf{R}_{1})+\frac{3}{4}E^{dd}_{2E,2E}(\mathbf{R}_{1})
\end{eqnarray}
In this way one can derive relations among energy integrals. 
The full list of energy integrals are summarized in the supplementary data (available online). 

Energy integrals in equation~(\ref{Eq:TB}) can be constructed by performing DFT calculations and the Wannier function (WF) transformation. While the TB model based on atomic orbitals precisely satisfies the symmetric relations described in equation~(\ref{Eq:energy_integral_symmetry}), the Wannier Hamiltonian obtained by maximizing localization does not necessarily guarantee the symmetry relations. 

Symmetry constraints dictated by the crystal symmetry can be directly imposed to the maximal localization procedure of the Wannier transformation, which is called as symmetry-adapted Wannier functions~\cite{PhysRevB_87_235109_2013_Sakuma, KORETSUNE2023108645, SymWannier}. When the constructed Wannier functions are close to atomic-like orbitals, the corresponding Wannier Hamiltonian approximately respects the symmetry relations, equation~(\ref{Eq:energy_integral_symmetry}). 
In this case one can symmetrize the Wannier Hamiltonian by using equation~(\ref{Eq:energy_integral_symmetry}) in order to recover perfect crystal symmetries of the Wannier Hamiltonian. 
See references~\cite{WU2018405, ZHI2022108196,WannierTools,WannSymm} for details of symmetrization. For our case, we constructed five $d$-orbital shaped Wannier functions of the transition metal atom and six $p$-orbital shaped Wannier functions of chalcogen atoms, and then symmetrized the corresponding Wannier Hamiltonian in order to make the Wannier Hamiltonian satisfy the symmetry relations of equation~(\ref{Eq:energy_integral_symmetry}). 

Note that the symmetry conditions respected by TB models with atomic orbitals are maintained throughout the optimization process discussed in the subsequent subsections. Thus, the final hopping parameters as a result of the optimization procedure also satisfy the symmetry condition, equation~(\ref{Eq:energy_integral_symmetry}). 

\subsection{\label{subsec:truncation}Truncation}
\begin{figure}[t]
\begin{center}
\includegraphics[width=1.0\columnwidth, clip=true]{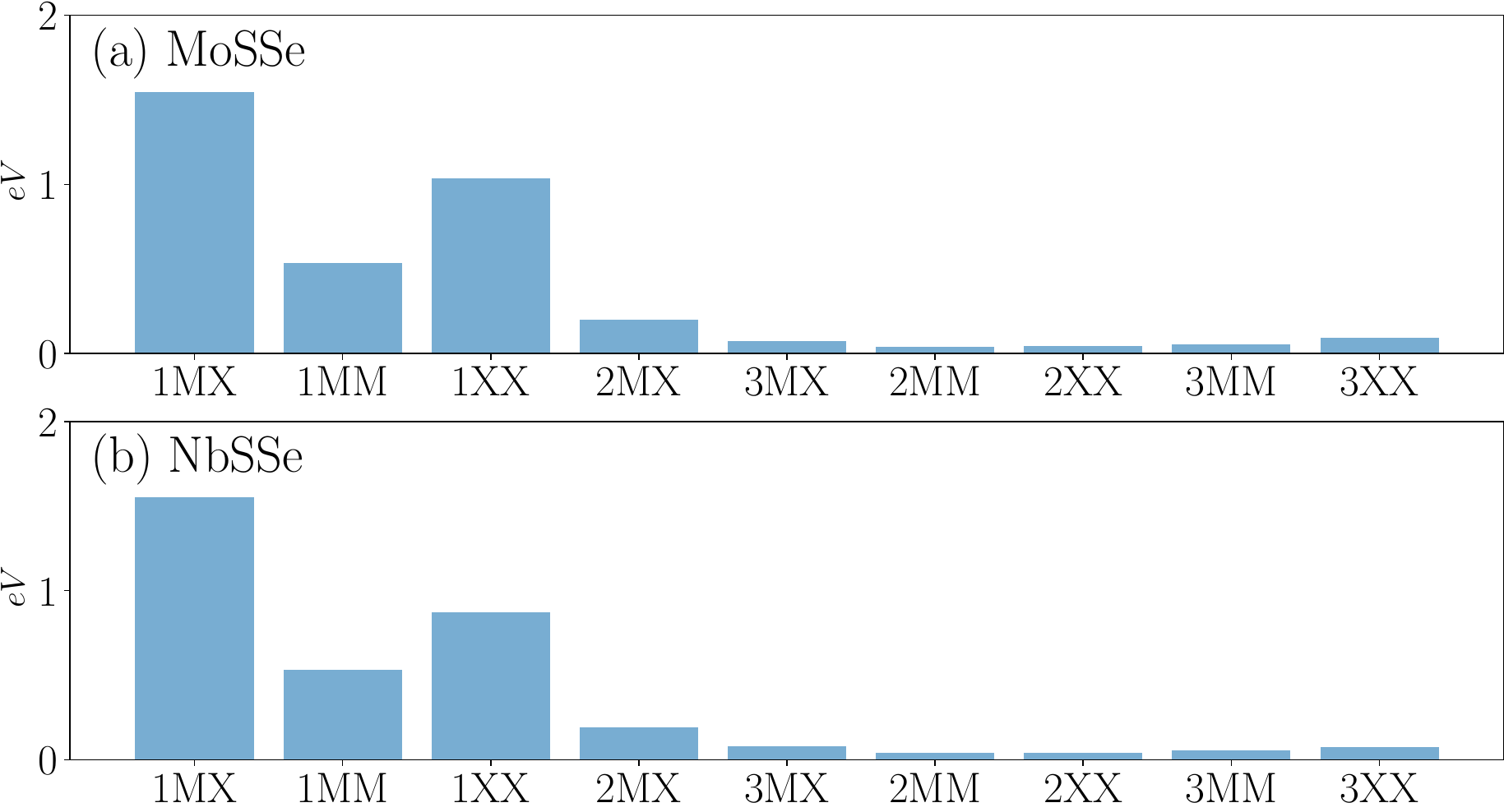} 
\end{center}
\caption{\label{Fig:WF_Truncation} (Color online) Maximum magnitudes of WF Hamiltonian for sets of nearest neighbors from 1MX to 3XX (a) for MoSSe and (b) for NbSSe.}
\end{figure}
Magnitudes of energy integrals in the WF construction decrease as the inter-site distance $\left|\mathbf{R}\right|$ increases, but a very large distance is required to make energy integrals absolutely zero. 
Thus, instead of keeping all WFs over the whole set of $\mathbf{R}$, one can construct TB models including only energy integrals within a certain cut-off radius $\left|\mathbf{R}\right|_{c}$:
\begin{equation}
\label{Eq:TB02} \mathcal{H}_{\mu i, \nu j}(\mathbf{k}) = \sum_{\left|\mathbf{R}\right|\leq \left|\mathbf{R}\right|_{c}}e^{i\mathbf{k}\cdot\mathbf{R}} E_{\mu i, \nu j}(\mathbf{R}),
\end{equation}
 The truncation process accompanies the loss of finite energy integrals, so the resulting band structures do not perfectly match DFT band structures. 
One can increase the cut-off radius $\left|\mathbf{R}\right|_{c}$ in order to obtain accurate band structures, but the model complexity accordingly increases. 
Thus it is obvious that the truncation radius $\left|\mathbf{R}\right|_{c}$ is needed to carefully choose so that the resulting TB Hamiltonian reproduces the DFT band structures. 
We call the TB model obtained by truncation from the WF Hamiltonian as \textit{the truncated TB (tTB) model}. 

Here we introduce the following notations in order to represent the truncation radius. 
When M and X denote transition metal and chalcogen atoms respectively, the $i$th nearest neighbors between A and B atoms are represented by $i$AB, where $\textrm{A}, \textrm{B} \in \left\{\textrm{M}, \textrm{X} \right\}$. 
For example, the second nearest neighbors between M and X are denoted by $2$MX. 
When $i$AB is the farthest nearest neighbor under consideration, the corresponding truncation range is denoted by $\langle\langle i\textrm{AB} \rangle\rangle$. 
For example, the truncation range $\langle\langle 3\textrm{XX} \rangle\rangle$ includes 1MX, 1MM, 1XX, 2MX, 3MX, 2MM, 2XX, 3MM, and 3XX. 
Note that $\left\Vert \textrm{1MX} \right\Vert < \left\Vert \textrm{1MM} \right\Vert \leq \left\Vert \textrm{1XX} \right\Vert < \left\Vert \textrm{2MX} \right\Vert < \left\Vert \textrm{3MX} \right\Vert < \left\Vert \textrm{2MM} \right\Vert \leq \left\Vert \textrm{2XX} \right\Vert < \left\Vert \textrm{3MM} \right\Vert \leq \left\Vert \textrm{3XX} \right\Vert$, where $\left\Vert \cdots \right\Vert$ denotes the distance between neighbors. 
For $\left\Vert i\textrm{MM} \right\Vert \leq \left\Vert i\textrm{XX} \right\Vert$ the equality holds when chalcogen neighbors are on the same plane.  

\subsection{\label{subsec:optimization}Optimization}

The deviation of the tTB model from the WF Hamiltonian can be reduced by constructing the optimized TB model with atomic orbitals. 
Despite the deviation between the tTB model and the WF Hamiltonian due to the truncation, the tTB model is still physically close to the DFT Hamiltonian, so the tTB model can be a good initial start for the optimization procedure to guarantee a physically reliable solution. 

Optimizing TB models can be done by adopting the least-squares fitting method commonly used in regression analysis and optimization problems~\cite{doi:10.1137/1.9781611971484}. 
Here we present the least-squares fitting procedure reformulated for constructing the optimized TB models. 
A cost function minimized in the least-squares fitting procedure is chosen to be a sum of squared errors (SSE) between the DFT band energies $\varepsilon^{\tiny \textrm{DFT}}_{i}(\mathbf{k}_{j})$ and TB ones $\varepsilon^{\tiny \textrm{TB}}_{i}(\mathbf{k}_{j})$  at sampling points $\{\mathbf{k}_{j}\}$ on the Brillouin zone (BZ):
\begin{equation}
\label{Eq:SSE}\textrm{(SSE)} = \sum_{i=1}^{N_{b}}\sum_{j=1}^{N_{\mathbf{k}}}\left[\varepsilon_{i \mathbf{k}_{j}}^{\tiny \textrm{DFT}} - \varepsilon_{i \mathbf{k}_{j}}^{\tiny \textrm{TB}}\right]^2,
\end{equation} 
where $N_{b}$ and $N_{\mathbf{k}}$ are numbers of bands and $\mathbf{k}$ sampling points, respectively. 
In this work we use the Nelder–Mead method~\cite{1965Nelder-Mead_Comp_Journal_7_308} implemented in the SciPy package~\cite{2020SciPy-NMeth}. 
Here the resulting TB model after finishing the optimization process is named \textit{the optimized TB (oTB) model}.

Since every band and every $\mathbf{k}$ point equally contributes to the cost function equation~(\ref{Eq:SSE}), minimizing equation~(\ref{Eq:SSE}) can lead to the oTB model fitted to overall DFT band structures. 
When it is needed to precisely optimize a subset of bands $S_{b}$ at some $\mathbf{k}$-point regions $S_{\mathbf{k}}$, a penalty function, which is also called as regularization, can be added to the cost function equation~(\ref{Eq:SSE}):
\begin{eqnarray}
\label{Eq:SSE2}\textrm{(SSE)} &=& \sum_{i\mathbf{k} \notin S_{b} \cap S_{\mathbf{k}}} \left[\varepsilon_{i \mathbf{k}}^{\tiny \textrm{DFT}} - \varepsilon_{i \mathbf{k}}^{\tiny \textrm{TB}}\right]^2 + \lambda \sum_{i\mathbf{k} \in S_{b} \cap S_{\mathbf{k}}} \left[\varepsilon_{i \mathbf{k}}^{\tiny \textrm{DFT}} - \varepsilon_{i \mathbf{k}}^{\tiny \textrm{TB}}\right]^2, 
\end{eqnarray}
where the second term is the penalty term emphasizing band structures of $S_{b}$ at $S_{\mathbf{k}}$. Here $\lambda$ is the penalty weight factor to control the relative importance of the penalty term to equation~(\ref{Eq:SSE}). The cost function corrected by the penalty term equation~(\ref{Eq:SSE2}) can be re-written as a weighted sum of squared errors (wSSE):
\begin{eqnarray}
\label{Eq:wSSE2}\textrm{(SSE)} &=& \sum_{i=1}^{N_{b}}\sum_{j=1}^{N_{\mathbf{k}}} \omega_{ij}\left[\varepsilon_{i \mathbf{k}_{j}}^{\tiny \textrm{DFT}} - \varepsilon_{i \mathbf{k}_{j}}^{\tiny \textrm{TB}}\right]^2,
\end{eqnarray}
where a weight factor $\omega_{ij}=1$ if $i\mathbf{k}_{j}\notin S_{b}\cap S_{\mathbf{k}}$ and otherwise $\omega_{ij}=1+\lambda$. The penalty weight factor $\lambda$ is needed to appropriately choose in order to achieve the accuracy of both overall energy bands and local electronic structures of the subset $S_{b}\cap S_{\mathbf{k}}$. If the penalty weight factor $\lambda$ is too small, it is likely to fail to have accurate local structures of the subset $S_{b}\cap S_{\mathbf{k}}$. On the other hand, if  the penalty weight factor $\lambda$ is chosen to be too large, one can obtain local structures of the subset $S_{b}\cap S_{\mathbf{k}}$ best-fitted to DFT calculations, but the overall band structures can be significantly deviated from DFT energy bands. 
\begin{figure}[t]
\begin{center}
\includegraphics[width=1.0\columnwidth, clip=true]{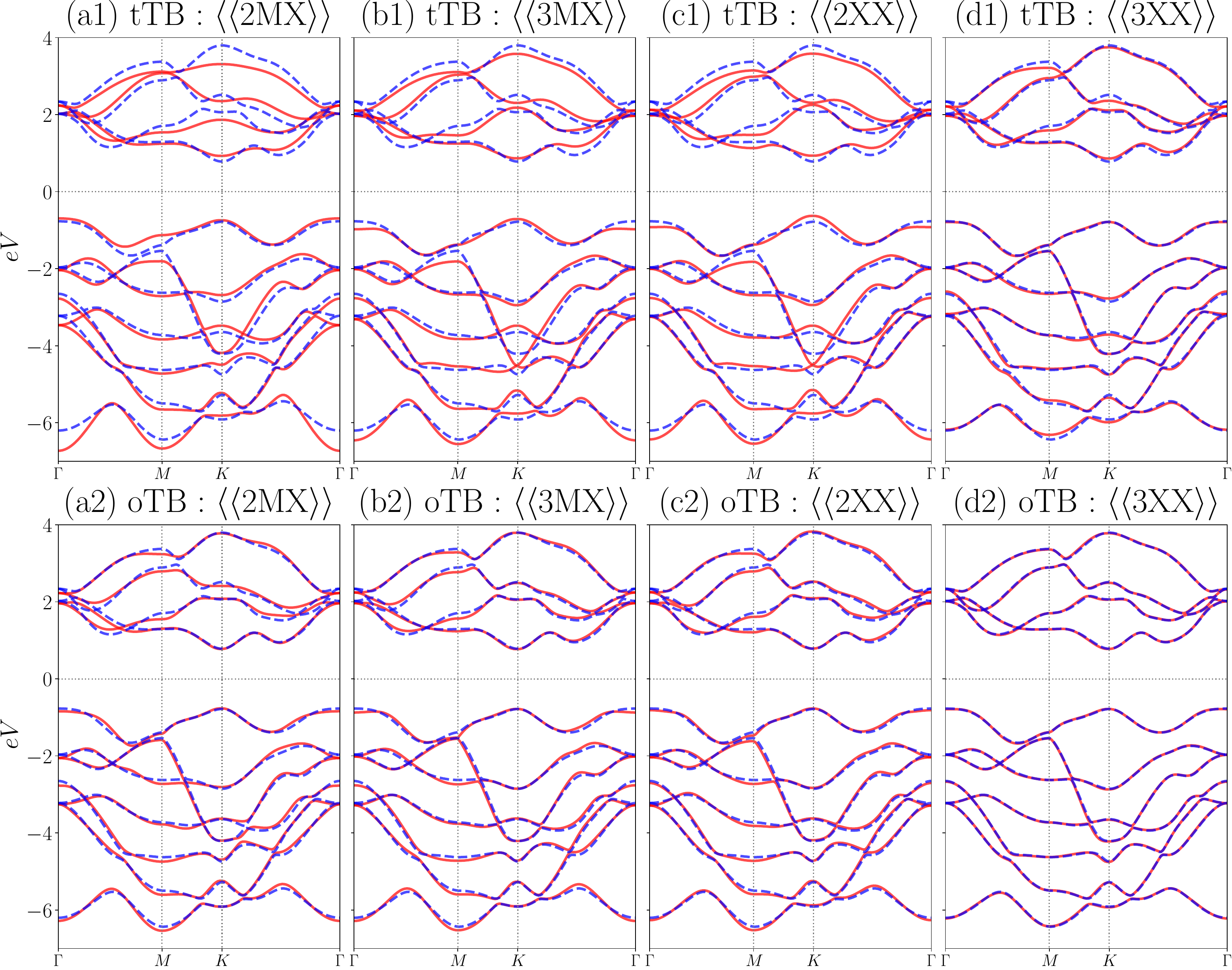} 
\end{center}
\caption{\label{Fig:MoSSe} (Color online) Band structures (red solid lines) of tTB models (upper panel) and oTB models (lower panel) for MoSSe in comparison with DFT band structures (blue dashed lines). Truncation range increases from $\langle\langle \textrm{2MX} \rangle\rangle$ [(a1) and (a2)] to $\langle\langle \textrm{3MX} \rangle\rangle$ [(b1) and (b2)] to $\langle\langle \textrm{2XX} \rangle\rangle$ [(c1) and (c2)] to $\langle\langle \textrm{3XX} \rangle\rangle$ [(d1) and (d2)]. Note that Fermi level is set to be zero in this work.}  
\end{figure}

\begin{figure}[t]
\begin{center}
\includegraphics[width=1.0\columnwidth, clip=true]{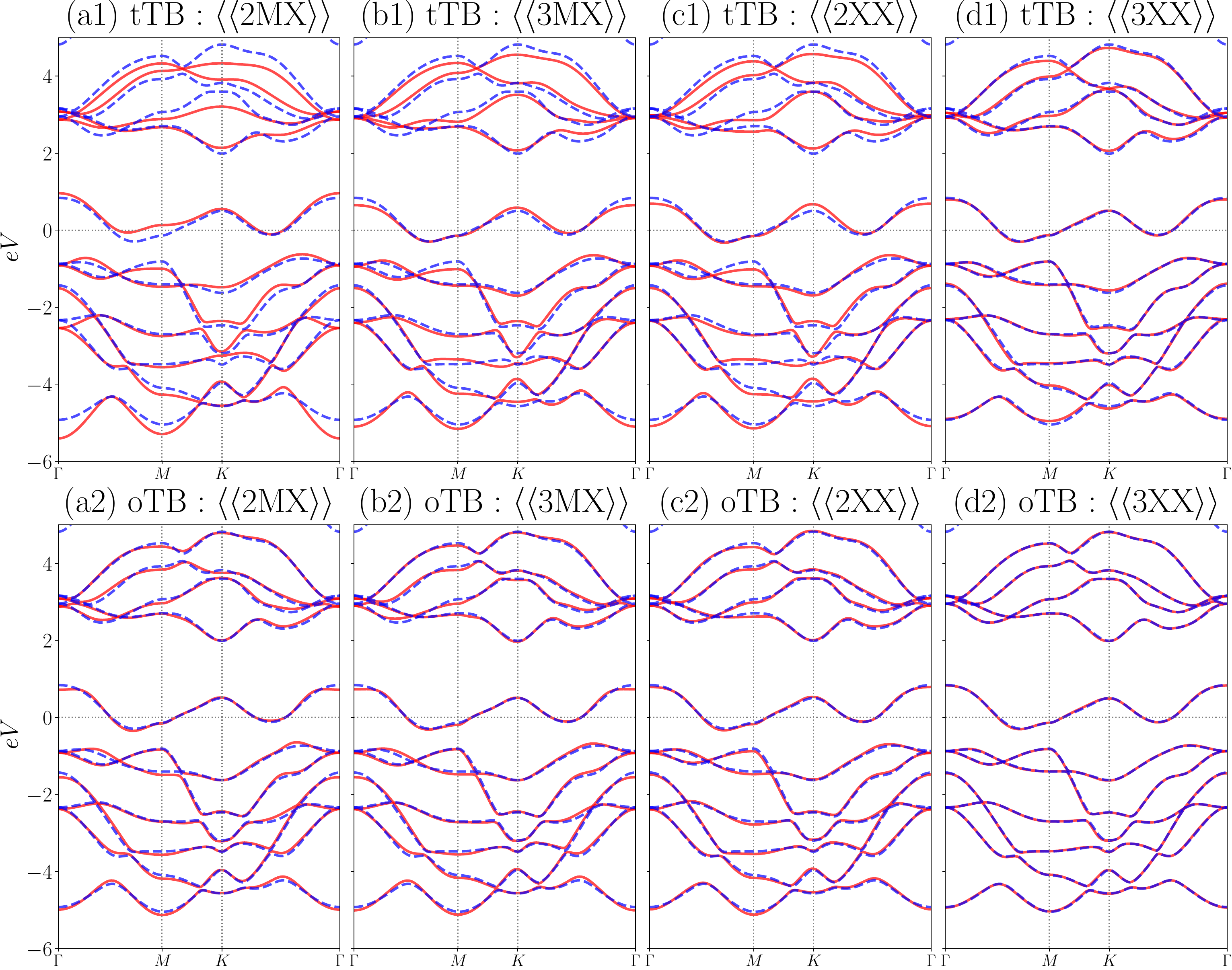} 
\end{center}
\caption{\label{Fig:NbSSe} (Color online) Band structures (red solid lines) of tTB models (upper panel) and oTB models (lower panel) for NbSSe in comparison with DFT band structures (blue dashed lines). Truncation range increases from $\langle\langle \textrm{2MX} \rangle\rangle$ [(a1) and (a2)] to $\langle\langle \textrm{3MX} \rangle\rangle$ [(b1) and (b2)] to $\langle\langle \textrm{2XX} \rangle\rangle$ [(c1) and (c2)] to $\langle\langle \textrm{3XX} \rangle\rangle$ [(d1) and (d2)].}  
\end{figure}

\section{\label{sec:discussion}Discussion}
\subsection{\label{subsec:tTB}Truncated TB Models}
Figures~\ref{Fig:MoSSe} and \ref{Fig:NbSSe} compare band structures of tTB and oTB models 
by increasing the truncation radius from $\langle\langle 2\textrm{MX} \rangle\rangle$ to $\langle\langle 3\textrm{XX} \rangle\rangle$ for MoSSe and NbSSe, respectively. 
In particular, the truncation range $\langle\langle 2\textrm{MX} \rangle\rangle$ includes MX1, MM1, XX1, and MX2, which are the same with nearest neighbors considered in the tTB model of Ref.~\cite{PhysRevB_92_205108_Fang}. Here band structures of TB models are compared with those of DFT calculations, which are in a good agreement with the well-established database of two-dimensional materials~\cite{Haastrup_2018,Gjerding_2021,C2DB}. We present how much electronic structures of tTB models are deviated from those of DFT calculations by choosing four levels of the truncation radii, and in the next section we will discuss that such deviations can be reduced by constructing oTB models for all of the truncation radii.  

In the upper panels of figures~\ref{Fig:MoSSe} and \ref{Fig:NbSSe}, it is clearly shown that band structures of tTB models approach those of the DFT calculations as the truncation range increases.
Especially the tTB model with $\langle\langle 3\textrm{XX}\rangle\rangle$ well reproduces DFT band structures not only qualitatively but also quantitatively.
There are some discrepancies, for example, for four topmost unoccupied bands and two lowest occupied ones, but they are very small. 

When the truncation radius is smaller than $\langle\langle 3\textrm{XX}\rangle\rangle$, it is noticed that deviations between tTB bands and DFT ones are evidently observed over the whole band manifolds. 
Although the truncation range $\langle\langle 2\textrm{MX} \rangle\rangle$ fairly reproduces DFT band structures as discussed in Ref.~\cite{PhysRevB_92_205108_Fang}, it does not provide accurate energy bands, especially five highest bands mostly composed of $d$-orbitals, which include conduction and valence bands around the Fermi energy $\varepsilon_{F}$. 
The deviations can be quantified by using the root mean squared error (RMSE),
\begin{equation}
\label{Eq:RMSE}\textrm{(RMSE)} = \sqrt{\frac{1}{N_{b}N_{\mathbf{k}}}\left(\textrm{SSE}\right)}.
\end{equation}
For MoSSe (NbSSe) the RMSE in equation~(\ref{Eq:RMSE}) gives 0.157 (0.155), 0.114 (0.099), 0.111 (0.093), and 0.048 (0.040) when the truncation range is $\langle\langle \textrm{2MX} \rangle\rangle$, $\langle\langle \textrm{3MX} \rangle\rangle$, $\langle\langle \textrm{2XX} \rangle\rangle$, and $\langle\langle \textrm{3XX} \rangle\rangle$, respectively. 
Note that the RMSE is in a unit of eV.

For the semiconducting monolayer MoSSe, main physical properties are determined by valence and conduction bands in the vicinity of the $K$ point, where the semiconducting energy gap is located. 
As seen in figures~\ref{Fig:MoSSe}(a1)-(d1), the tTB model predicts valence band energy $E_{v}(K)$ and conduction band one $E_{c}(K)$ at the $K$ point that are deviated from those energies from DFT calculations. 
In addition, curvatures of valence and conduction bands around the $K$ point are different between the tTB model and DFT calculations. 

For the metallic TMDC monolayer NbSSe, the resulting energy band at the Fermi level $\varepsilon_{F}$ of the tTB model with $\langle\langle \textrm{2MX} \rangle\rangle$ is problematic as shown in figure~\ref{Fig:NbSSe}(a1). 
The difference between the partially occupied band of the tTB model and that of DFT calculations is very large except for the symmetric line from the $K$ point to the saddle point on the $\overline{K \Gamma}$. 
When the truncation range of the tTB Hamiltonian increases beyond $\langle\langle 2\textrm{MX} \rangle\rangle$, the energy band of the tTB model at the Fermi level gets closer to that of DFT calculations, but there are still deviations around $\Gamma$ or $K$ points as shown in figures~\ref{Fig:NbSSe}(b1)-(d1). 

The aforementioned discrepancies of the metallic energy bands between tTB models and DFT calculations are also observed when Fermi surfaces are calculated as shown in figures~\ref{Fig:Fermi-Surface-NbSSe}(a1) to (d1). For the tTB model with $\langle\langle \textrm{2MX} \rangle\rangle$, the resulting Fermi surfaces are completely different from those of DFT calculations; While DFT Fermi surfaces exhibit two hole pockets around $\Gamma$ and $K$, the tTB model leads to two circles surrounding the $\Gamma$ point as illustrated in figure~\ref{Fig:Fermi-Surface-NbSSe}(a1). 
This observation is consistent with the fact that the metallic band of the tTB model with $\langle\langle \textrm{2MX} \rangle\rangle$ has no crossing point with the Fermi level along $\overline{MK}$, but there are two crossing points both on $\overline{\Gamma M}$ and $\overline{K\Gamma}$. 

For the cases of tTB models with $\langle\langle \textrm{3MX} \rangle\rangle$, $\langle\langle \textrm{2XX} \rangle\rangle$, and $\langle\langle \textrm{3XX} \rangle\rangle$, resulting Fermi surfaces share the same topology with those of DFT calculations: hole pockets at $\Gamma$ and $K$. 
The hexagon-like hole pocket around $\Gamma$ is well matched with that of DFT calculations. 
Hole pockets at $K$ have different shapes against DFT calculations, when truncation ranges $\langle\langle \textrm{3MX} \rangle\rangle$ and $\langle\langle \textrm{2XX} \rangle\rangle$ are considered. 
The tTB model with $\langle\langle \textrm{3XX} \rangle\rangle$ perfectly reproduces DFT Fermi surfaces.

\subsection{\label{subsec:oTB}Optimized TB Models}
\begin{figure}[t]
\begin{center}
\includegraphics[width=1.0\columnwidth, clip=true]{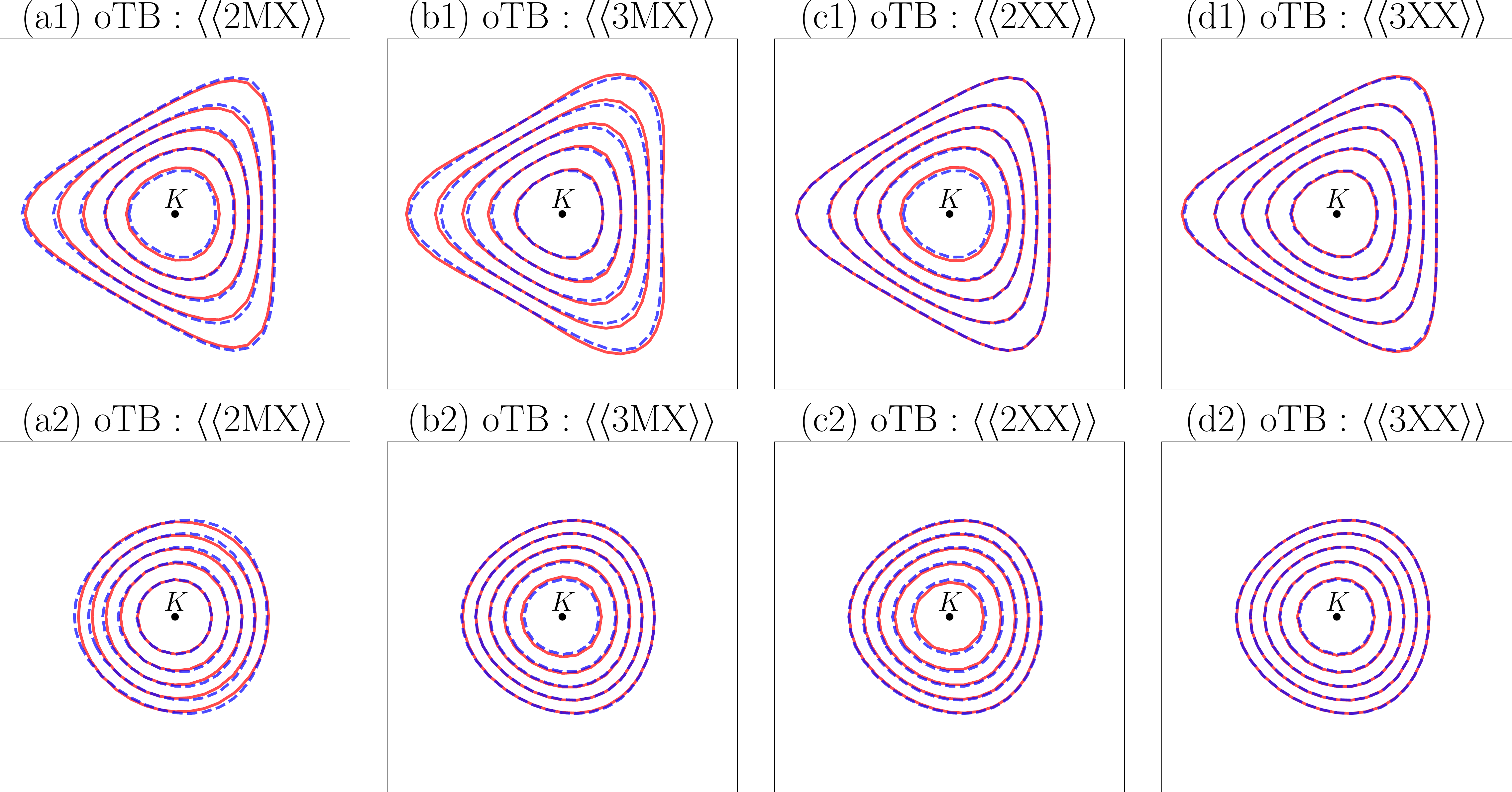} 
\end{center}
\caption{\label{Fig:Fermi-Surface-MoSSe} (Color online) Energy contours of oTB models around valence band tops (upper panel) and conduction band bottoms (lower panel) of MoSSe at the $K$ point by changing truncation range from $\langle\langle \textrm{2MX} \rangle\rangle$ to $\langle\langle \textrm{3XX} \rangle\rangle$. Red solid lines and blue dashed ones represent energy contours of oTB models and DFT calculations, respectively. Five energy contours are chosen at intervals of 0.05 eV from valence band tops or conduction band bottoms.}
\end{figure}
Next we perform the optimization process in order to show that deviations between TB bands and DFT ones are minimized by constructing oTB models. 
As shown on lower panels of figures~\ref{Fig:MoSSe} and \ref{Fig:NbSSe}, band energies from oTB models are in a very good agreement with DFT band structures. 
In particular, the five topmost bands are noticeably improved throughout the optimization process. 
When the optimization is performed, the penalty function discussed in equation~(\ref{Eq:SSE2}) is included in order to precisely reproduce structures of conduction and valence bands around $K$ for semiconducting MoSSe and energy bands around the Fermi level for metallic NbSSe monolayers.

For the semiconducting monolayer MoSSe, the penalty function with $\lambda=10$ in equation~(\ref{Eq:SSE2}) is applied to the subset $S_{b} \cap S_{\mathbf{k}}$, where the band subset $S_{b}$ and the $\mathbf{k}$-point subset are as follows:
\begin{eqnarray}
 S_{b} &=& \{i|i=\textrm{(conduction) or (valence)}\} \\Note
 S_{\mathbf{k}} &=& \{\mathbf{k}| |\mathbf{k}_{K} - \mathbf{k}| < \varepsilon \}
\end{eqnarray}
where $\mathbf{k}_{K}$ is the $\mathbf{k}$ vector corresponding to the $K$ symmetric point and $\varepsilon$ is a small radius around $K$. 
For the metallic TMDC monolayer NbSSe, it is desirable to optimize the TB model that enables to reproduce the partially occupied energy band at the Fermi level accurately. 
When equation~(\ref{Eq:SSE}) is chosen as the cost function, the resulting energy band at Fermi level does not match that of DFT calculations. For this reason, the penalty function with $\lambda=10$ is added to the cost function as discussed in equation~(\ref{Eq:SSE2}). The penalty term is applied to the partially occupied band at $\mathbf{k}$ points lying on the Fermi surface:
\begin{eqnarray}
S_{b}&=&\{i|i=\textrm{(partially occupied bands)}\} \\   
S_{\mathbf{k}}&=&\{\mathbf{k}|\varepsilon_{\mathbf{k}}=\varepsilon_{F}\}.
\end{eqnarray} 

\begin{figure}[t]
\begin{center}
\includegraphics[width=1.0\columnwidth, clip=true]{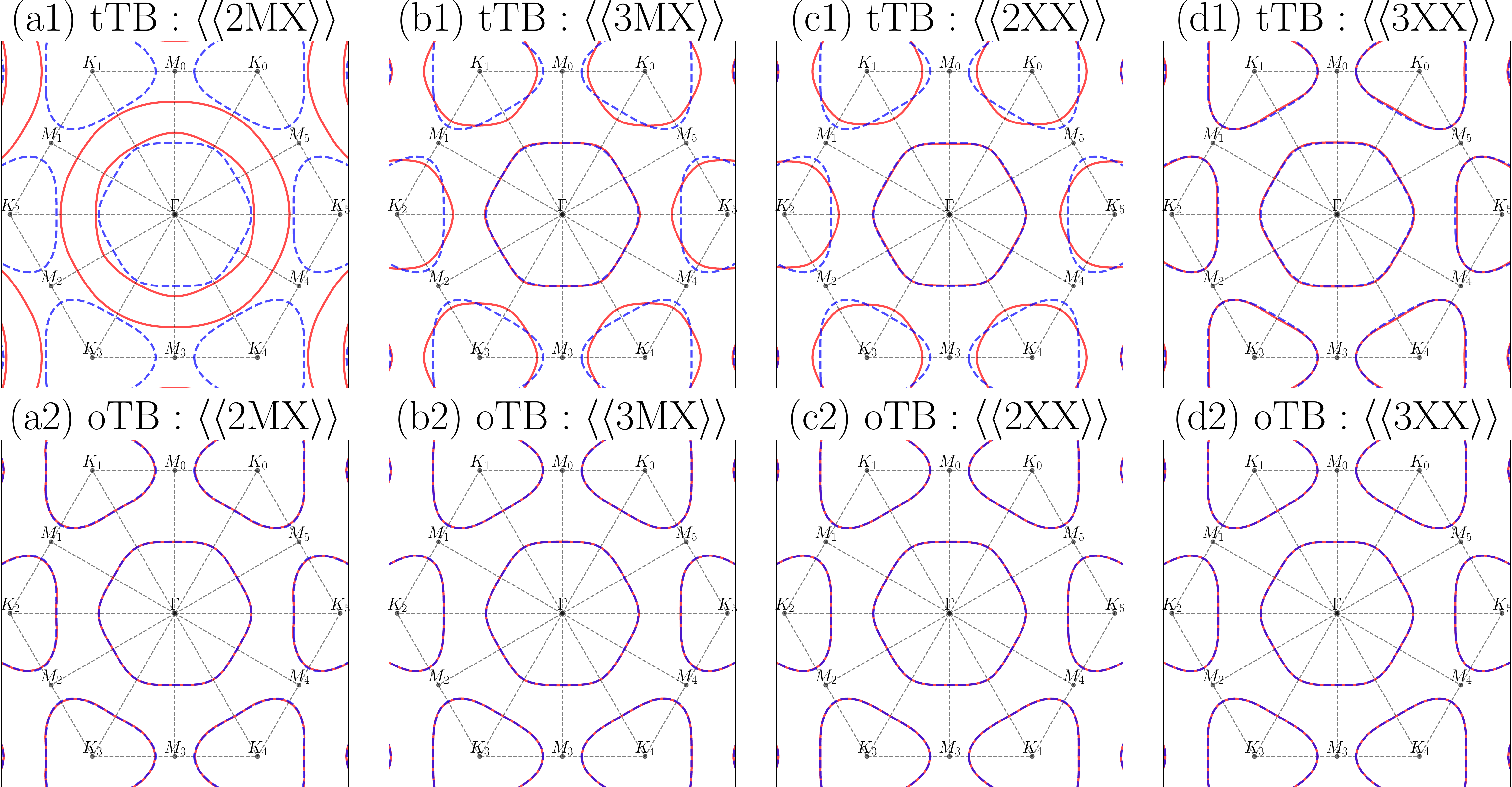} 
\end{center}
\caption{\label{Fig:Fermi-Surface-NbSSe} (Color online) Fermi surfaces (red solid lines) of NbSSe calculated by tTB models (upper panel) and oTB ones (lower panel). DFT Fermi surfaces are denoted by blue dashed lines.}
\end{figure}

\begin{figure}[h]
\begin{center}
\includegraphics[width=1.0\columnwidth, clip=true]{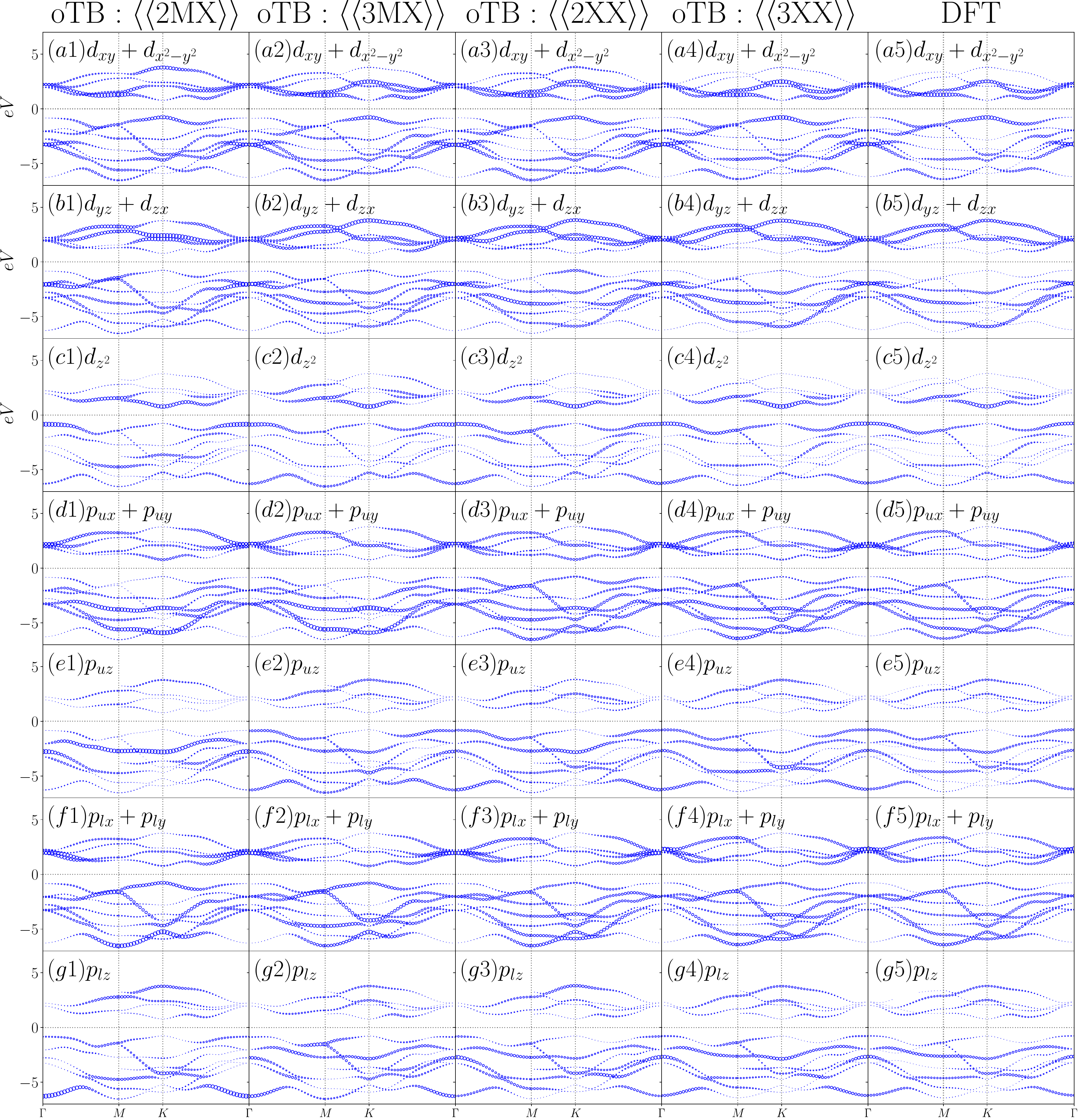} 
\end{center}
\caption{\label{Fig:MoSSe-orbitals} (Color online) Orbital-projected band structures of TB models for MoSSe in comparison with DFT calculations in the rightmost panel. Orbitals $d_{xy}+d_{x^2-y^2}$, $d_{yz}+d_{zx}$, $d_{z^2}$, $p_{ux}+p_{uy}$, $p_{uz}$, $p_{lx}+p_{ly}$, and $p_{lz}$ are projected onto energy bands. Here superscripts $u$ and $l$ represent upper and lower chalcogen layers respectively, The size of blue open circle is proportional to orbital contributions to bands.}
\end{figure}

\begin{figure}[h]
\begin{center}
\includegraphics[width=1.0\columnwidth, clip=true]{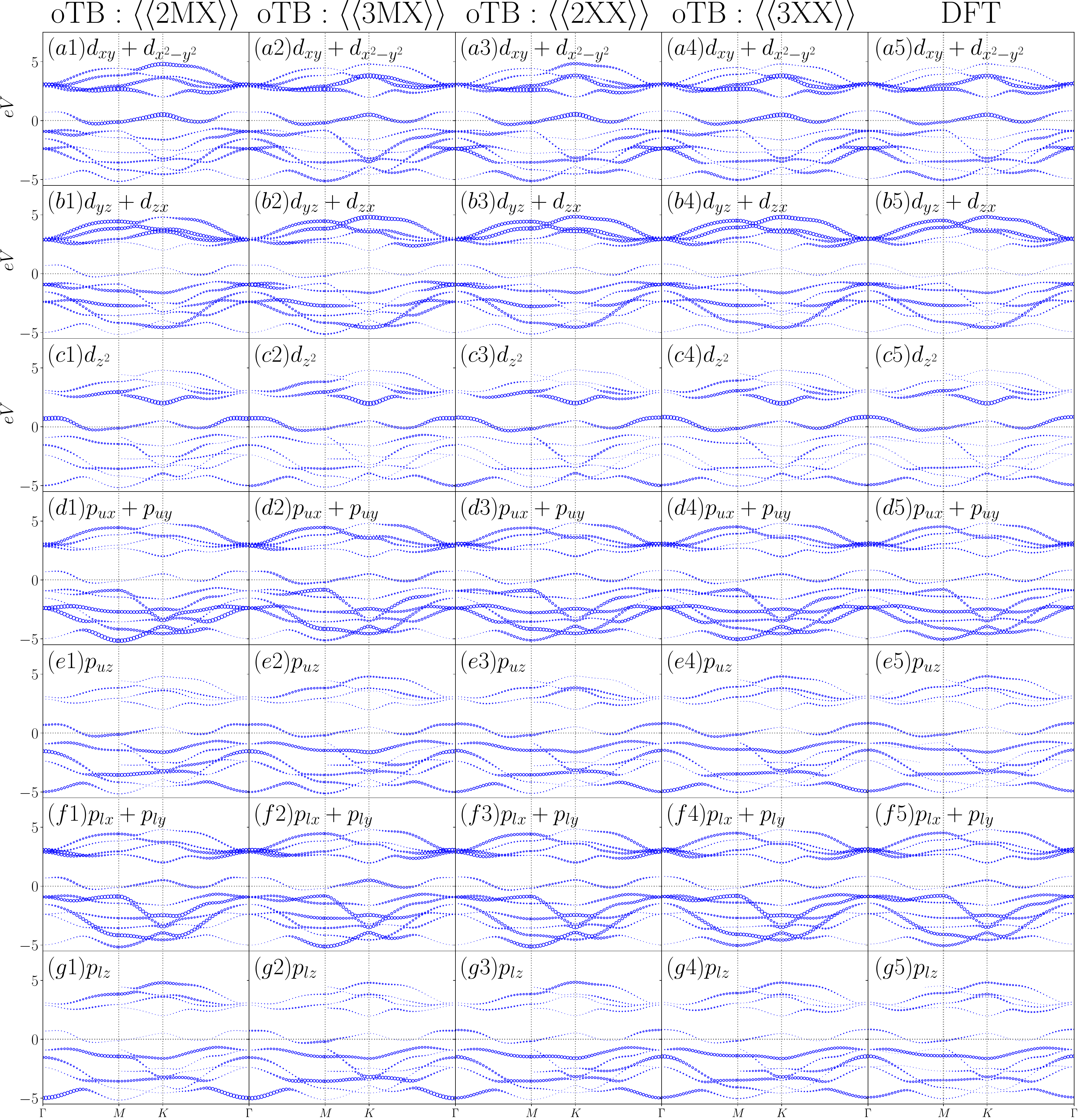} 
\end{center}
\caption{\label{Fig:NbSSe-orbitals} (Color online) Orbital-projected band structures of TB models for NbSSe in comparison with DFT calculations in the rightmost panel. Orbitals $d_{xy}+d_{x^2-y^2}$, $d_{yz}+d_{zx}$, $d_{z^2}$, $p_{ux}+p_{uy}$, $p_{uz}$, $p_{lx}+p_{ly}$, and $p_{lz}$ are projected onto energy bands. Here superscripts $u$ and $l$ represent upper and lower chalcogen layers respectively, The size of blue open circle is proportional to orbital contributions to bands.}
\end{figure}

Figure~\ref{Fig:Fermi-Surface-MoSSe} shows energy contours of MoSSe around the valence band top [figures~\ref{Fig:Fermi-Surface-MoSSe}(a1)-(d1)] and the conduction band bottom [figures~\ref{Fig:Fermi-Surface-MoSSe}(a2)-(d2)] at $K$. 
The oTB models tuned by equation~(\ref{Eq:SSE2}) and DFT calculations produce almost the same contour lines, triangle-shaped valence contours and circular conduction ones for all the truncation ranges from $\langle\langle \textrm{2MX} \rangle\rangle$ to $\langle\langle \textrm{3XX} \rangle\rangle$. 

For monolayer NbSSe, when we perform the optimization without imposing the penalty function onto the partial occupied band around the Fermi level, it is found that Fermi surfaces constructed by the oTB model do not match those of DFT calculations, as illustrated in figures~\ref{Fig:Fermi-Surface-NbSSe}(a1)-(d1).
In contrast, the inclusion of the penalty function in equation~\ref{Eq:SSE2} significantly resolves the discrepancy observed in Fermi surfaces. The partially occupied bands of the oTB model and DFT calculations cross the Fermi level at the same $\mathbf{k}$-points as shown in figures~\ref{Fig:NbSSe}(a2)-(d2). 
In fact, as illustrated in figure~\ref{Fig:Fermi-Surface-NbSSe}, Fermi surfaces of DFT energy bands are excellently replicated by the oTB models optimized by equation~(\ref{Eq:SSE2}) from $\langle\langle \textrm{2MX} \rangle\rangle$ to $\langle\langle \textrm{3XX} \rangle\rangle$. 
These excellent agreements demonstrated in figure~\ref{Fig:Fermi-Surface-NbSSe} show that the penalty function imposed in equation~(\ref{Eq:SSE2}) is very useful to fine-tune the TB model so that the TB model can reproduce specific features of band structures.

For the case where the penalty function is added to the cost function in equation~(\ref{Eq:SSE2}), one can separately calculate RMSEs for sampling points not included in the penalty function and for those included in the penalty function,
\begin{eqnarray}
\label{Eq:RMSE1}\textrm{(RMSE)}_{1} &=& \sqrt{\frac{1}{n\left(S_{b}^{c} \cup S_{\mathbf{k}}^{c}\right)} \sum_{i\mathbf{k} \notin S_{b} \cap S_{\mathbf{k}}} \left[\varepsilon_{i \mathbf{k}}^{\tiny \textrm{DFT}} - \varepsilon_{i \mathbf{k}}^{\tiny \textrm{TB}}\right]^2} \\ 
\label{Eq:RMSE2}\textrm{(RMSE)}_{2} &=& \sqrt{\frac{1}{n\left(S_{b} \cap S_{\mathbf{k}}\right)} \sum_{i\mathbf{k} \in S_{b} \cap S_{\mathbf{k}}} \left[\varepsilon_{i \mathbf{k}}^{\tiny \textrm{DFT}} - \varepsilon_{i \mathbf{k}}^{\tiny \textrm{TB}}\right]^2}, 
\end{eqnarray}
where $n\left(S\right)$ stands for the number of elements included in a set $S$ and $S^{c}$ is the complement set of $S$. 
Equation~(\ref{Eq:RMSE1}) represents standard deviations of overall band structures between DFT calculations and oTB models, while equation~(\ref{Eq:RMSE2}) quantifies how well the oTB model reproduces local features of interest such as valence and conduction bands and Fermi surfaces, imposed in the penalty function. 

For optimized TB bands of MoSSe (NbSSe), $(\textrm{RMSE})_{1}$ in equation~(\ref{Eq:RMSE1}) leads to 0.064 (0.053), 0.055 (0.047), 0.050 (0.040), and 0.010 (0.009) for $\langle\langle \textrm{2MX} \rangle\rangle$, $\langle\langle \textrm{3MX} \rangle\rangle$, $\langle\langle \textrm{2XX} \rangle\rangle$, $\langle\langle \textrm{3XX} \rangle\rangle$, respectively. 
For the penalty function imposed onto equation~(\ref{Eq:SSE2}), $(\textrm{RMSE})_{2}$ of MoSSe (NbSSe) is 0.008 (0.014), 0.009 (0.009), 0.009 (0.006), and 0.001 (0.004) for $\langle\langle \textrm{2MX} \rangle\rangle$, $\langle\langle \textrm{3MX} \rangle\rangle$, $\langle\langle \textrm{2XX} \rangle\rangle$, $\langle\langle \textrm{3XX} \rangle\rangle$, respectively. 

It is also needed to confirm that the resulting TB model not only correctly reproduces energy bands of DFT calculations, but also retrains physical properties of the system. 
Technically the TB model construction based on fitting of energy bands is the minimization process, so it is possible that the construction might reach some undesirable local minimum, which is not a physically correct Hamiltonian. 
Since the optimization described here starts with tTB models originated from DFT calculations, it is highly likely to obtain a physically reliable TB model. 
Here we compare orbital characters of the oTB models and DFT calculations, in order to confirm whether the oTB models are physically reliable. 
Figures~\ref{Fig:MoSSe-orbitals} and \ref{Fig:NbSSe-orbitals} illustrate $d$ and $p$ orbital-projected band structures of oTB models of MoSSe and NbSSe, compared with those of DFT calculations. 
Although there are some differences, for example, $d_{xy}+d_{x^2-y^2}$ orbital contributions to the topmost unoccupied band of the oTB model with $\langle\langle\textrm{2MX}\rangle\rangle$, oTB models reproduces overall orbital distributions on band structures, which resemble those of DFT calculations.

In summary, the oTB models constructed here provide electronic structure properties well matched with DFT calculations for all of the chosen truncation ranges. It is in a sharp contrast to tTB models whose deviations against DFT calculations increase as the truncation radius decreases. It means that one can use the oTB model with the smaller number of hopping parameters to well represent electronic properties, instead of using the \textit{ab initio} Hamiltonian truncated at the larger truncation radius. For example, while the tTB model with $\langle\langle \textrm{3XX} \rangle\rangle$ consists of 178 parameters, the oTB model with $\langle\langle \textrm{2MX} \rangle\rangle$ has 78 parameters. Despite the difference in the number of parameters, electronic properties reproduced by the two models are comparable as shown above. 

\subsection{\label{subsec:SOC}Spin-orbit coupling}
Here we introduce the spin-orbit coupling (SOC) to monolayers TMDC with the broken mirror-reflection symmetry. 
Atomic SOC terms for $d$ and $p$ orbitals added to the TB model read as follows:
\begin{equation}
\label{Eq:SOC-M}
\fl\mathcal{H}_{\tiny \textrm{SOC}}^{\tiny \textrm{M}}=\lambda_{\tiny \textrm{M}}\hat{S}\cdot\hat{L}\dot{=}\lambda_{\textrm{M}}\left[\begin{array}{cccccccccc}
0 & 0 & 0 & 0 & 0 & 0 & 0 & 0 & -\frac{\sqrt{3}}{2} & i\frac{\sqrt{3}}{2}\\
0 & 0 & -i & 0 & 0 & 0 & 0 & 0 & \frac{1}{2} & \frac{i}{2}\\
0 & i & 0 & 0 & 0 & 0 & 0 & 0 & -\frac{i}{2} & \frac{1}{2}\\
0 & 0 & 0 & 0 & -\frac{i}{2} & \frac{\sqrt{3}}{2} & -\frac{1}{2} & \frac{i}{2} & 0 & 0\\
0 & 0 & 0 & \frac{i}{2} & 0 & -i\frac{\sqrt{3}}{2} & -\frac{i}{2} & -\frac{1}{2} & 0 & 0\\
0 & 0 & 0 & \frac{\sqrt{3}}{2} & i\frac{\sqrt{3}}{2} & 0 & 0 & 0 & 0 & 0\\
0 & 0 & 0 & -\frac{1}{2} & \frac{i}{2} & 0 & 0 & i & 0 & 0\\
0 & 0 & 0 & -\frac{i}{2} & -\frac{1}{2} & 0 & -i & 0 & 0 & 0\\
-\frac{\sqrt{3}}{2} & \frac{1}{2} & \frac{i}{2} & 0 & 0 & 0 & 0 & 0 & 0 & \frac{i}{2}\\
-i\frac{\sqrt{3}}{2} & -\frac{i}{2} & \frac{1}{2} & 0 & 0 & 0 & 0 & 0 & -\frac{i}{2} & 0
\end{array}\right],   
\end{equation}
which is written with the basis of $|z^{2},\uparrow\rangle$, $|x^{2}-y^{2},\uparrow\rangle$, $|xy,\uparrow\rangle$, $|zx,\uparrow\rangle, |yz,\uparrow\rangle$, $|z^{2},\downarrow\rangle$, $|x^{2}-y^{2},\downarrow\rangle$, $|xy,\downarrow\rangle$, $|zx,\downarrow\rangle$, and $|yz,\downarrow\rangle$, and here $\uparrow$ and $\downarrow$ stand for up and down spins, and 
\begin{equation}
\label{Eq:SOC-X}\mathcal{H}_{\tiny \textrm{SOC}}^{\tiny \textrm{X}}=\lambda_{\tiny \textrm{X}}\hat{S}\cdot\hat{L}\dot{=}\frac{1}{2}\lambda_{\textrm{X}}\left[\begin{array}{cccccc}
0 & 0 & 0 & 0 & i & -1\\
0 & 0 & i & -i & 0 & 0\\
0 & -i & 0 & 1 & 0 & 0\\
0 & i & 1 & 0 & 0 & 0 \\
-i & 0 & 0 & 0 & 0 & -i\\
-1 & 0 & 0 & 0 & i & 0
\end{array}\right],   
\end{equation}
which is written in the basis of  $|p_{z},\uparrow\rangle$, $|p_{y},\uparrow\rangle$, $|p_{x},\uparrow\rangle$, $|p_{z},\downarrow\rangle$, $|p_{y},\downarrow\rangle$, and $|p_{x},\downarrow\rangle$~\cite{PhysRevB_88_085433_Liu, PhysRevB_96_155439_Kim, PhysRevB_104_045426_Kim}. 
Here $\lambda_{\textrm{M}}$ and $\lambda_{\textrm{X}}$ are atomic spin-orbit coupling parameters for transition metal and chalcogen atoms, respectively. 

First we calculate band structures by inserting bare atomic SOC (bSOC) terms into the tTB model. 
Let us call this model as \textit{tTB+bSOC}. 
Here \textit{bare} means the atomic SOC calculated from DFT calculations when an isolated atom is placed inside a very large cubic unit cell, whose side is 20 $\AA$ in our case~\cite{PhyRevB_98_075106_2018_Fang}.The bare SOC strength can be calculated by comparing atomic energy levels when the SOC term is turned on and off in the DFT calculations.

Note that here we do not construct Wannier functions including the SOC. The tTB model is obtained by truncating the Wannier Hamiltonian constructed without inclusion of the SOC. The SOC terms, equations~(\ref{Eq:SOC-M}) and (\ref{Eq:SOC-X}) are added to the oTB model. The initial values of atomic SOC are computed by using DFT calculations on isolated atoms as mentioned before.   

\begin{figure}[t]
\begin{center}
\includegraphics[width=1.0\columnwidth, clip=true]{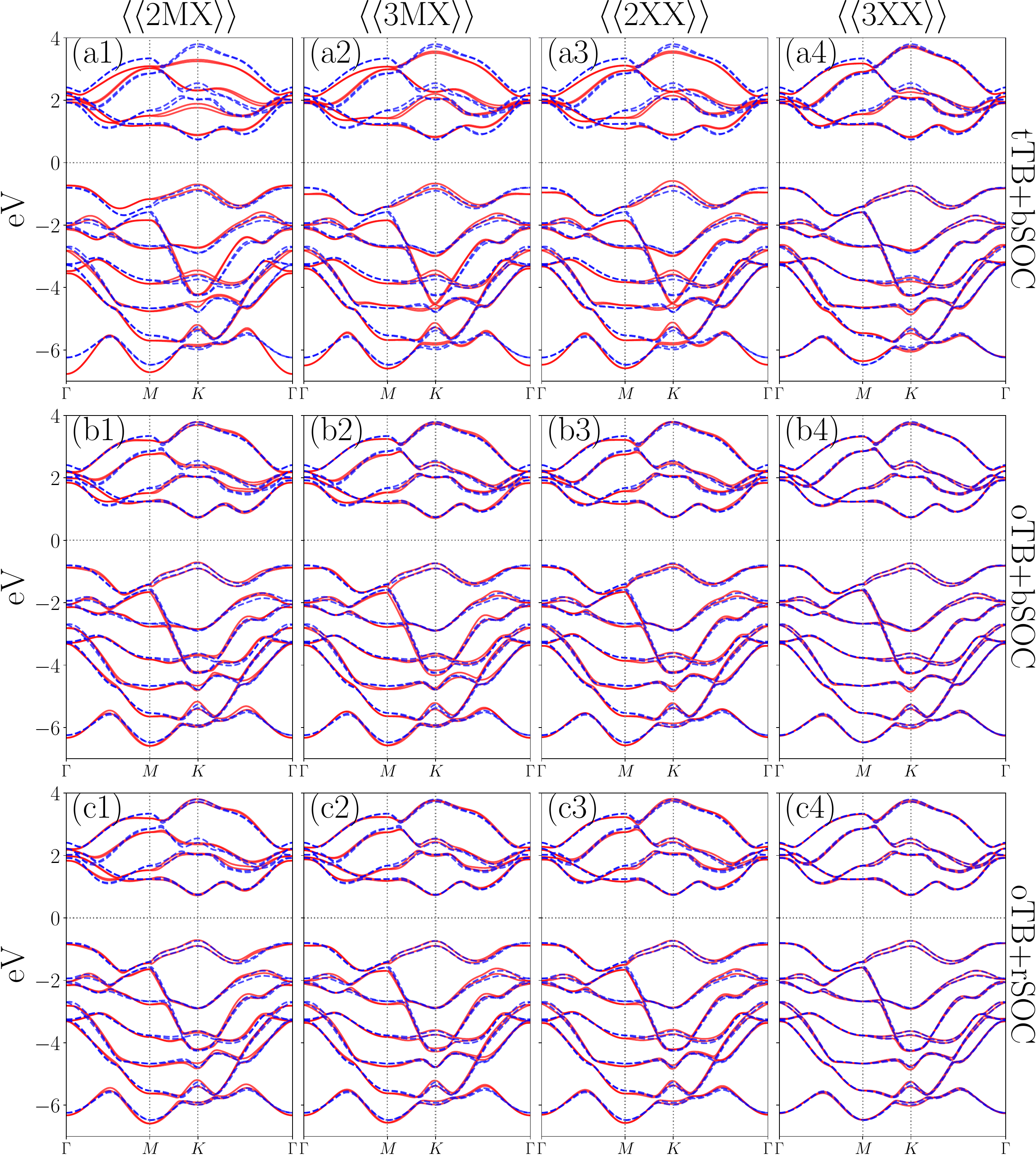} 
\end{center}
\caption{\label{Fig:MoSSe-SOC} (Color online) Band structures (red solid lines) of TB models including SOC for MoSSe when the truncation range are $\langle\langle \textrm{2MX} \rangle\rangle$, $\langle\langle \textrm{3MX} \rangle\rangle$, $\langle\langle \textrm{2XX} \rangle\rangle$, and $\langle\langle \textrm{3XX} \rangle\rangle$. DFT band structures (blue dashed lines) are compared. Top [(a1) to (a4)], middle [(b1) to (b4)], and bottom [(c1) to (c4)] panels are band structures calculated by tTB+bSOC, oTB+bSOC, and oTB+rSOC, respectively.}
\end{figure}

\begin{figure}[t]
\begin{center}
\includegraphics[width=1.0\columnwidth, clip=true]{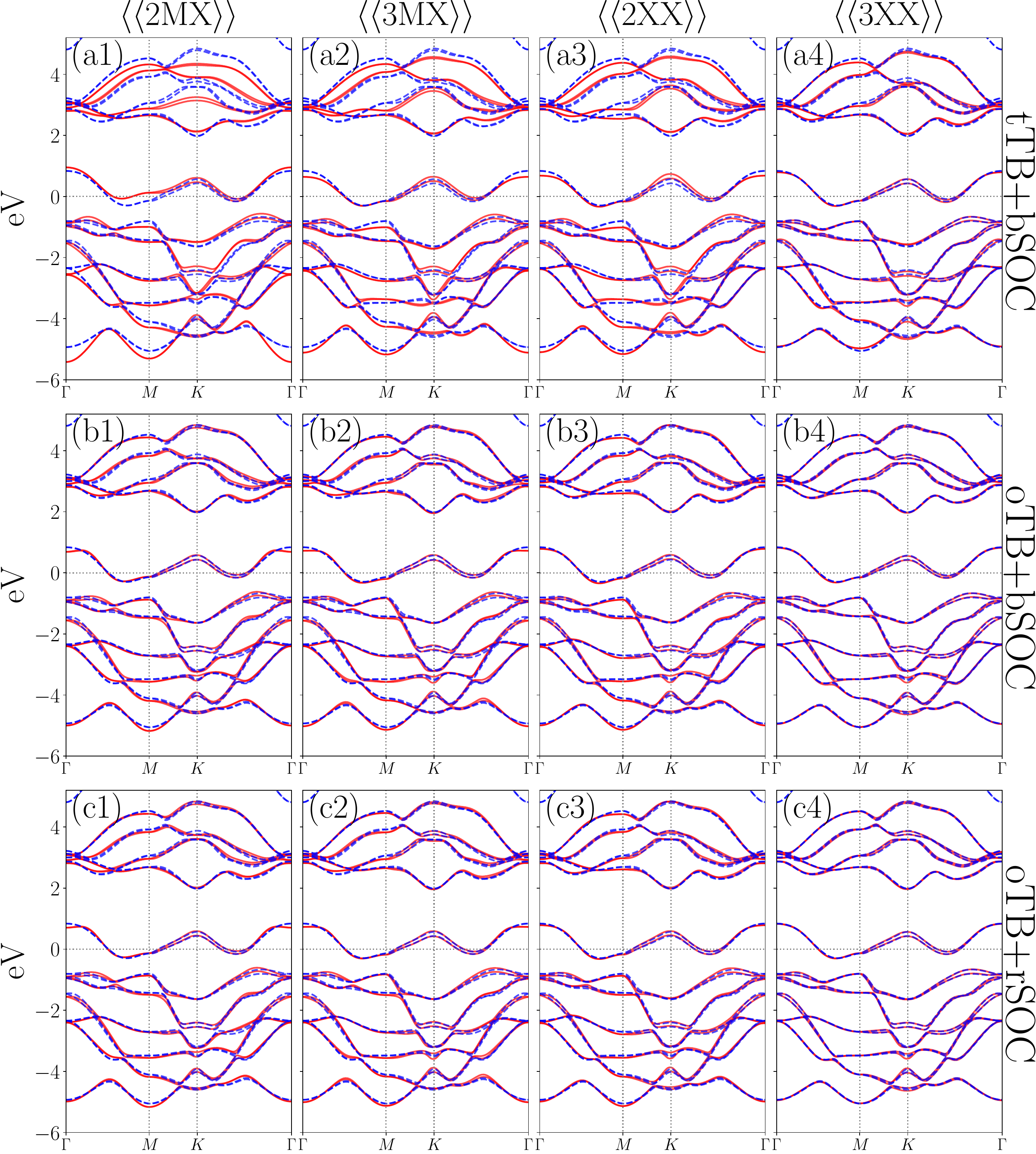} 
\end{center}
\caption{\label{Fig:NbSSe-SOC} (Color online) Band structures (red solid lines) of TB models including SOC for NbSSe when the truncation range are $\langle\langle \textrm{2MX} \rangle\rangle$, $\langle\langle \textrm{3MX} \rangle\rangle$, $\langle\langle \textrm{2XX} \rangle\rangle$, and $\langle\langle \textrm{3XX} \rangle\rangle$. DFT band structures (blue dashed lines) are compared. Top [(a1) to (a4)], middle [(b1) to (b4)], and bottom [(c1) to (c4)] panels are band structures calculated by tTB+bSOC, oTB+bSOC, and oTB+rSOC, respectively.}
\end{figure}

\begin{table}
\caption{\label{TABLE1} bSOC and rSOC parameters for MoSSe and NbSSe. SOC values are in the unit of eV/$\hbar^2$.}
\begin{indented}
\item[]
\begin{tabular}{@{}lllllll}
\br
&&MoSSe&&&NbSSe&\\
\ns\ns
&\crule{3}&\crule{3}\\
\ns
              & \textrm{Mo} & \textrm{S} & \textrm{Se} & \textrm{Nb} & \textrm{S} & \textrm{Se} \\
\mr
\textrm{bSOC} &  0.0842  &  0.0556  &  0.2491  &  0.0690  &  0.0556  &  0.2491 \\
\textrm{rSOC}, $\langle\langle \textrm{2MX}\rangle\rangle$ &  0.0891  &  0.0563  &  0.2416 &  0.0667  &  0.0535  & 
 0.2398  \\
\textrm{rSOC}, $\langle\langle \textrm{3MX}\rangle\rangle$  &    0.0932   &   0.0823   &   0.2340 &    0.0686   &   0.0554   &   0.2479   \\ 
\textrm{rSOC}, $\langle\langle \textrm{2XX}\rangle\rangle$  &    0.0713   &   0.0710   &   0.2728 &    0.0675   &   0.0552   &   0.2490   \\   
\textrm{rSOC}, $\langle\langle \textrm{3XX}\rangle\rangle$  &    0.0889   &   0.0666   &   0.2607 &    0.0711   &   0.0571   &   0.2560   \\
\br
\end{tabular}    
\end{indented}
\end{table}

Band structures calculated by tTB+bSOC models are shown in figures~\ref{Fig:MoSSe-SOC}(a1)-(d1) and \ref{Fig:NbSSe-SOC}(a1)-(d1) for MoSSe and NbSSe, respectively. 
Since the tTB models with $\langle\langle \textrm{2XX}\rangle\rangle$ do not well reproduce DFT band structures as shown in figures~(\ref{Fig:MoSSe})(a1) and (\ref{Fig:NbSSe})(a1), corresponding tTB+bSOC models also do not provide quantitatively good SOC-corrected band structures compared to DFT ones. 
In contrast the tTB+bSOC models with $\langle\langle \textrm{3XX}\rangle\rangle$ maintain the quality of the tTB models despite discrepancies in four topmost bands around $M$ and $K$ points. See figures~\ref{Fig:MoSSe-SOC}(a4) and \ref{Fig:NbSSe-SOC}(a4).   

We can also calculate band structures by combining oTB models developed in Sec.~\ref{sec:discussion} with bare SOC terms, equations~(\ref{Eq:SOC-M}) and (\ref{Eq:SOC-X}). 
This model is named \textit{oTB+bSOC}.  
Figures~\ref{Fig:MoSSe-SOC}(b1)-(b4) and \ref{Fig:NbSSe-SOC}(b1)-(b4) show band structures calculated by oTB+bSOC models.  
Compared with tTB+bSOC models, band structures of oTB+bSOC models exhibit smaller deviations from DFT band energies. 

Despite that oTB+bSOC models provides better SOC-corrected band structures than tTB+bSOC models, there are still inaccurate features compared with DFT band structures, for example, valence band top splitting due to SOC at $K$ as shown in figure~\ref{Fig:MoSSe-SOC}(b3). 
Therefore, in order to obtain the best-fitted band structures including correct local features, we can perform the second optimization process, where not only TB hopping parameters, but also SOC ones are optimized. 
In this optimization, bare SOC parameters from DFT calculations are used as the initial input values of the optimization. 
The SOC parameters determined by the optimization are regarded as renormalized SOC (rSOC) strengths influenced by change in the valence of the atom in solids~\cite{PhysRevB_102_045109_2020_Kurita, PhysRevB_104_195104_2021_Cuadrado}. 
For this reason we call the TB model obtained in the second optimization scheme as the \textit{oTB+rSOC} model.  
Table~\ref{TABLE1} compares bSOC parameters and rSOC ones for MoSSe and NbSSe. 
Figures~\ref{Fig:MoSSe-SOC}(c1)-(c4) and \ref{Fig:NbSSe-SOC}(c1)-(c4) show band structures of MoSSe and NbSSe, respectively, obtained by oTB+rSOC models where both TB hopping parameters and SOC strengths are optimized so that resulting TB bands are best-fitted to DFT band structures.  

\begin{figure}[t]
\begin{center}
\includegraphics[width=1.0\columnwidth, clip=true]{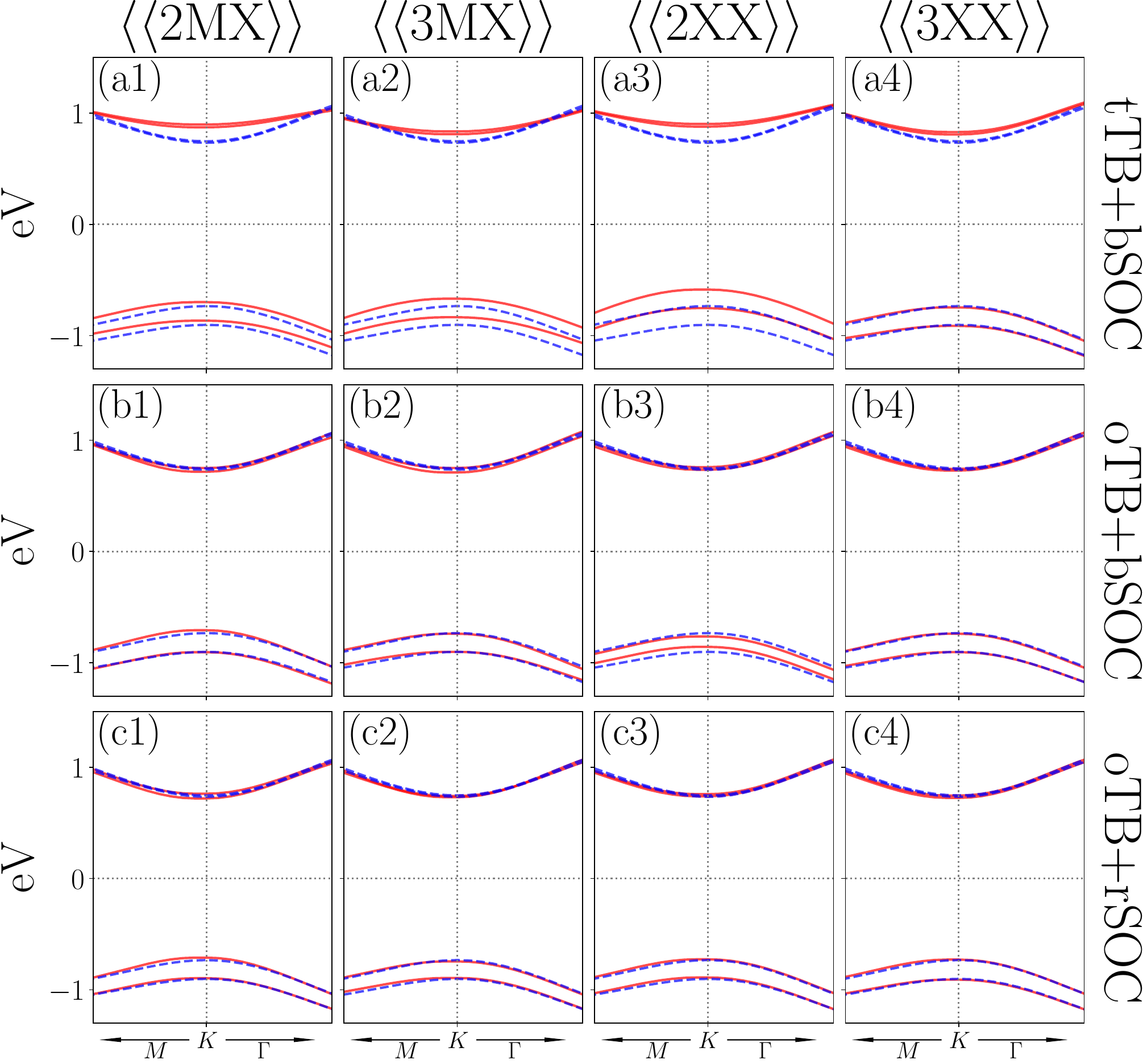} 
\end{center}
\caption{\label{Fig:MoSSe-SOC-KPT} (Color online) Valence bands and conduction ones around $K$ point when SOC is considered. The bands (red solid lines) calculated by tTB+bSOC (top panel), oTB+bSOC (middle panel), and oTB+rSOC (bottom panel) are compared with DFT calculations (blue dashed lines).}
\end{figure}

Note that penalty functions in equation~(\ref{Eq:SSE2}) are used when oTB+rSOC models are constructed. 
For MoSSe, penalty functions are imposed on valence band tops and conduction band bottoms around $K$ in order to capture the SOC-induced valence band splitting. 
Figure~\ref{Fig:MoSSe-SOC-KPT} compares SOC-corrected valance band tops and conduction band bottoms calculated by tTB+bSOC, oTB+bSOC, and oTB+rSOC models. It is clearly shown that the oTB+rSOC model provides valence and conduction band edges best-fitted to DFT results. 

\begin{figure}[t]
\begin{center}
\includegraphics[width=1.0\columnwidth, clip=true]{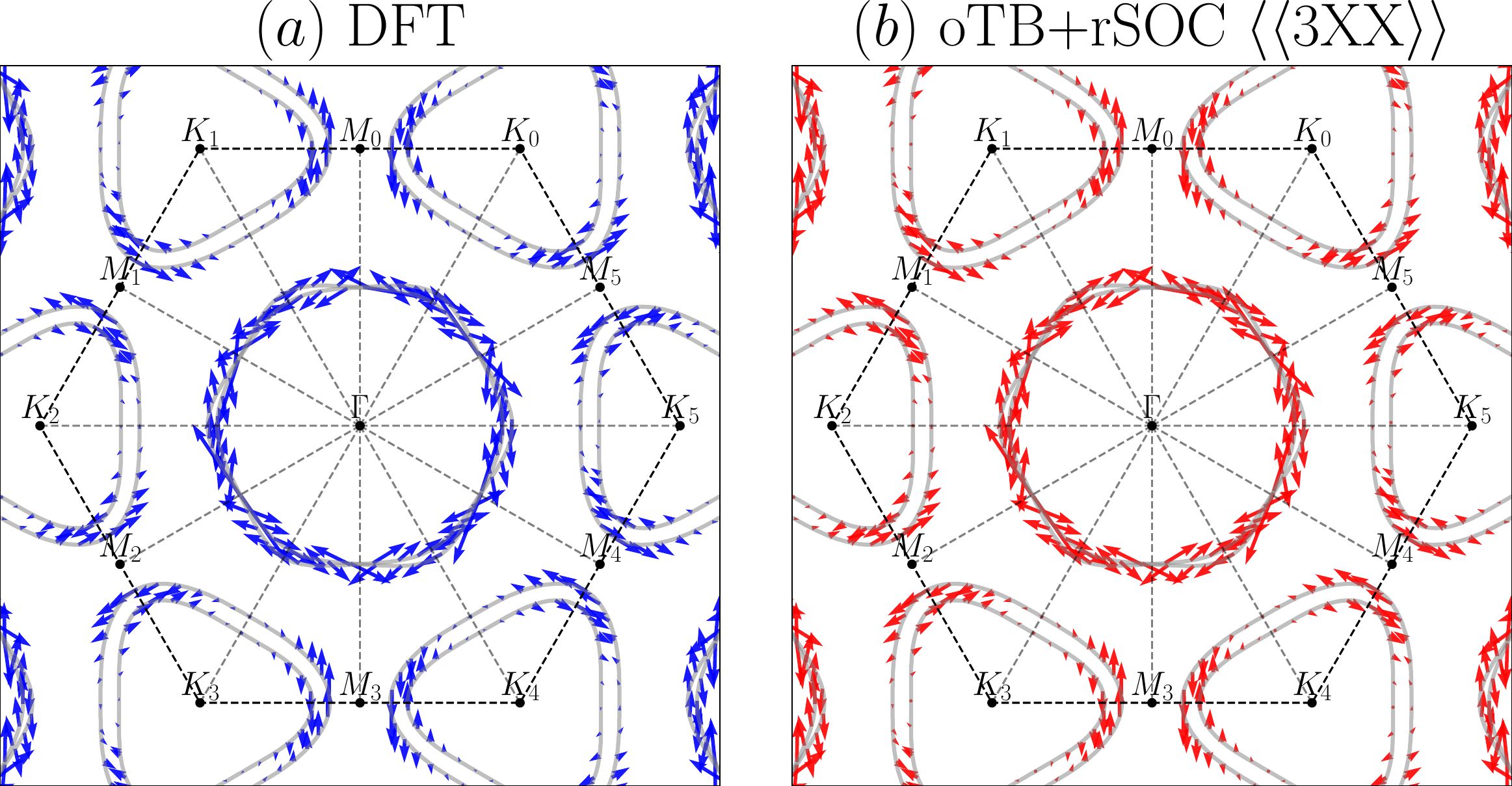} 
\end{center}
\caption{\label{Fig:SPIN-TEXTURE-DFT-3XX} (Color online) Spin textures derived from (a) DFT calculations (blue) and (b) oTB+rSOC model with $\langle\langle \textrm{3XX} \rangle\rangle$ (red). Fermi surfaces are indicated by gray solid lines.}
\end{figure}

For NbSSe we impose not only constraints on Fermi surfaces as done in the subsection~\ref{subsec:optimization}, but also another penalty term to the cost function to elaborate spin textures along Fermi surfaces. 
The $2H$ structure of monolayer TMDC has no inversion symmetry, so the monolayer generally has spin-split band structures~\cite{PRL2012Xiao}. 
When the monolayer respects the mirror-reflection symmetry $\mathcal{M}_{z}$, energy bands along the $\overline{\Gamma M}$ line are not split by inclusion of the SOC, and spin orientations induced by the SOC are Ising-type~\cite{PRL2012Xiao, PhysRevB_93_180501_Zhou}.
When the mirror symmetry $\mathcal{M}_{z}$ is broken, energy bands on $\overline{\Gamma M}$ are spin-split and planar spin textures are induced~\cite{PhysRevB_104_045426_Kim}. 
For this reason, NbSSe, where the mirror symmetry is broken due to different chalcogen layers, possesses planar spin textures on Fermi surfaces as illustrated in figure~\ref{Fig:SPIN-TEXTURE-DFT-3XX}(a). 
An additional penalty function can be added to the cost function equation~(\ref{Eq:SSE2}) in order to preserve spin orientation properties of DFT calculations as follows:
\begin{eqnarray}
\label{Eq:SSE3}\textrm{(SSE)} &=& \sum_{i\mathbf{k} \notin S_{b} \cap S_{\mathbf{k}}} \left[\varepsilon_{i \mathbf{k}}^{\tiny\textrm{DFT}} - \varepsilon_{i \mathbf{k}}^{\tiny\textrm{TB}}\right]^2 \\
&& + \lambda_{1} \sum_{i\mathbf{k} \in S_{b} \cap S_{\mathbf{k}}} \left[\varepsilon_{i \mathbf{k}}^{\tiny\textrm{DFT}} - \varepsilon_{i \mathbf{k}}^{\tiny\textrm{TB}}\right]^2 \\
&& + \lambda_{2} \sum_{i\mathbf{k} \in S_{b} \cap S_{\mathbf{k}}}\sum_{j=x,y,z} \left[\langle \hat{S}_{j}\rangle^{\tiny \textrm{DFT}}_{i\mathbf{k}} - \langle \hat{S}_{j}\rangle^{\tiny \textrm{TB}}_{i\mathbf{k}} \right]^2, 
\end{eqnarray}
where $\hat{S}_{j}$ is the spin operator for $j=x,y,z$, and $\langle \hat{S}_{j}\rangle^{\tiny \textrm{DFT(TB)}}_{i\mathbf{k}}$ is the spin expectation value of $\hat{S}_{j}$ with respect to DFT (TB) Bloch state $|\Psi^{\tiny \textrm{DFT(TB)}}(i\mathbf{k})\rangle$. 

\begin{figure}[t]
\begin{center}
\includegraphics[width=1.0\columnwidth, clip=true]{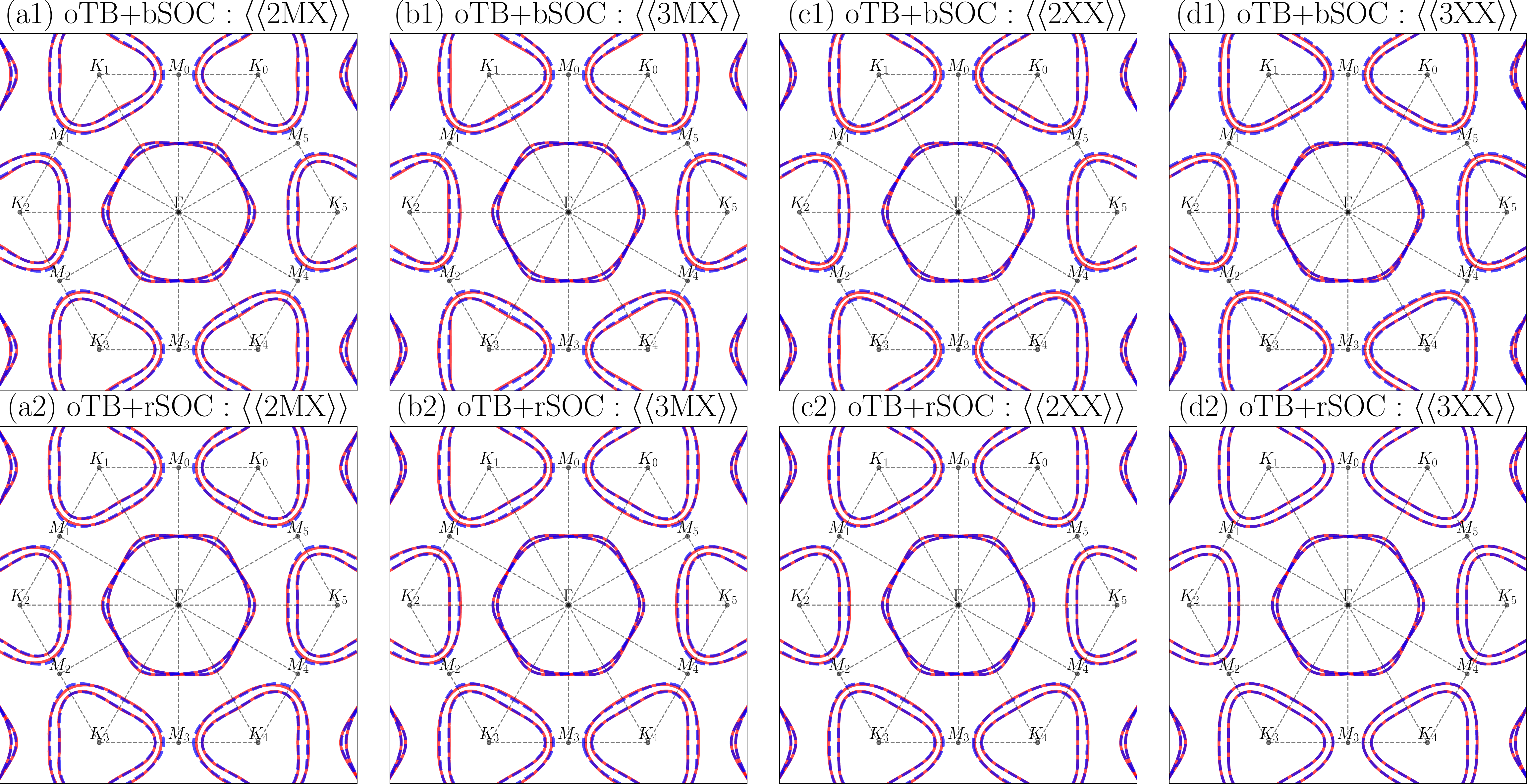} 
\end{center}
\caption{\label{Fig:SOC-Fermi-Surfaces} (Color online) Fermi surfaces (red solid lines) calculated by oTB+bSOC (top panel) and oTB+rSOC (bottom panel). Fermi surfaces (blue dashed lines) from DFT calculations are compared. Truncation ranges are indicated in subplot titles.}
\end{figure}

Varying truncation ranges from $\langle\langle \textrm{2MX} \rangle\rangle$ to $\langle\langle \textrm{3XX} \rangle\rangle$, Fermi surfaces calculated by oTB+bSOC and oTB+rSOC models are compared by DFT calculations in figure~\ref{Fig:SOC-Fermi-Surfaces}. It is shown that oTB+bSOC and oTB+rSOC models well reproduce two hexagon hole pockets around $\Gamma$ and two triangular hole pockets around $K$ for all the truncation ranges. 

\begin{figure}[t]
\begin{center}
\includegraphics[width=1.0\columnwidth, clip=true]{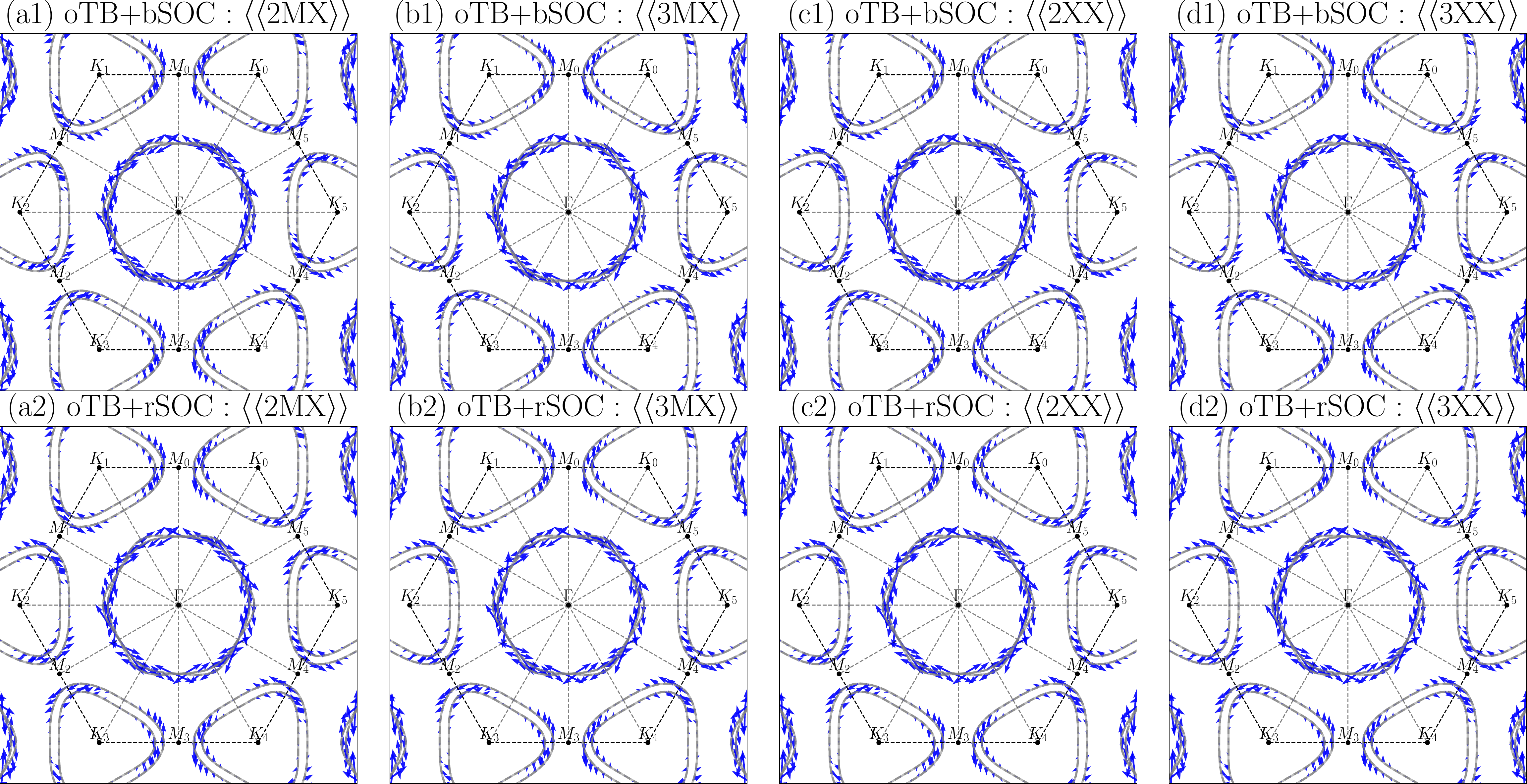} 
\end{center}
\caption{\label{Fig:SPIN-TEXTURE-NbSSe} (Color online) Spin textures (blue arrows) computed from oTB+bSOC [top panel, (a1) to (d1)] and (b) oTB+rSOC [bottom panel, (a2) to (d2)]. Fermi surfaces are drawn in gray solid lines. Truncation ranges are indicated in subplot titles.}
\end{figure}

Furthermore we calculate spin textures on Fermi surfaces by using oTB+bSOC and oTB+rSOC models, and compare them with DFT calculations in figures~\ref{Fig:SPIN-TEXTURE-DFT-3XX} and \ref{Fig:SPIN-TEXTURE-NbSSe}. 
According to DFT calculations in figure~\ref{Fig:SPIN-TEXTURE-DFT-3XX}(a), spin textures on Fermi surfaces have following properties. 
For two hexagon hole pockets around $\Gamma$, planar spin vectors on the outer hexagon rotate clockwise, while those of the inner hexagon orients counterclockwise. 
For two triangular hole pockets around $K$, planer spins have maximum magnitude around the $\overline{MK}$ line, and their orientations are perpendicular to the $\overline{MK}$ line. 
As like hole pockets around $\Gamma$, the two triangular hole pockets around $K$ show opposite spin orientations to each other. 
As shown in figure~\ref{Fig:SPIN-TEXTURE-DFT-3XX}, these key features of spin textures, not only spin orientations but also spin magnitudes, are well replicated by the oTB+rSOC model with $\langle\langle \textrm{3XX} \rangle\rangle$.

We also compute spin textures of oTB+bSOC and oTB+rSOC models by extending truncation ranges from $\langle\langle \textrm{2MX} \rangle\rangle$ to $\langle\langle \textrm{3XX} \rangle\rangle$ in figure~\ref{Fig:SPIN-TEXTURE-NbSSe}. Note that spin textures in figure~\ref{Fig:SPIN-TEXTURE-NbSSe} are all in an excellent agreement with DFT results [figure~\ref{Fig:SPIN-TEXTURE-DFT-3XX}(a)]. When the root-mean-squared error for spin vectors is defined as
\begin{equation}
\left(\textrm{RMSE}\right)_{3} = \sqrt{\frac{1}{n\left(S_{b} \cap S_{\mathbf{k}}\right)} \sum_{ i\mathbf{k} \in S_{b} \cap S_{\mathbf{k}}}  \sum_{j=x,y,z}  \left[\langle \hat{S}_{j}\rangle^{\tiny \textrm{DFT}}_{i\mathbf{k}} - \langle \hat{S}_{j}\rangle^{\tiny \textrm{TB}}_{i\mathbf{k}} \right]^2 }, 
\end{equation}
$\left(\textrm{RMSE}\right)_{3}$ of oTB+bSOC (oTB+rSOC) corresponds to 0.011 (0.006), 0.011 (0.005), 0.005 (0.003), and 0.004 (0.002) for $\langle\langle \textrm{2MX} \rangle\rangle$, $\langle\langle \textrm{3MX} \rangle\rangle$, $\langle\langle \textrm{2XX} \rangle\rangle$, and $\langle\langle \textrm{3XX} \rangle\rangle$, respectively. 

\section{\label{sec:conclusions}Conclusions}
In this work we have presented the detailed process to construct the optimized tight-binding model with atomic orbitals by using the truncated \textit{ab initio} tight-binding models. Applying to semiconducting and metallic Janus TMDC monolayers, it is shown that the optimization process improves the accuracy of overall band energies when the optimized tight-binding models are compared with the full DFT calculation where the \textit{ab initio} tight-binding models are derived. Furthermore we have shown that the optimization well reproduces detailed local band structures of interest such as valence and conduction band edges for semiconductor or Fermi surfaces and spin textures for metallic systems, by imposing additional constraints to the minimization process in terms of penalty functions. The resulting fine-tuned tight-binding model can be used to develop further simplified theories such as the $\mathbf{k}\cdot\mathbf{p}$ expansion. 

Here we have used planewave-based DFT calculations accompanied by the Wannier function transformation, but the optimization process discussed here is not limited to the construction method of the fully \textit{ab initio} Hamiltonian. The optimization process can be applied to DFT calculations based on atomic localized orbitals. Different from Wannier functions that are by construction orthogonal to one another, atomic localized orbitals are generally non-orthogonal to those at different sites, so the \textit{ab initio} tight-binding Hamiltonian based on atomic localized orbitals requires to use the overlap matrix~\cite{PhysRevB_52_14531_1995_McKinnon, PhysRevB_47_12995_1993_Dorantes_Davila, Electronic_structure_and_the_properties_of_solids_Harrison}. Nevertheless, the introduction of the overlap matrix does not change the overall optimization procedure presented in this work. Alternatively, even with DFT calculations based on atomic localized orbitals, the Wannier function transformation can be further applied to construct the orthogonal \textit{ab initio} tight-binding Hamiltonian choosing orbitals relevant to the energy window of our interest around the Fermi energy. Considering above discussions, the optimized tight-binding models demonstrated here can be generally constructed with various DFT calculations.

\section*{Acknowledgement}
We thank Hosub Jin, Seoung-Hun Kang, and Pilkyung Moon for fruitful discussions. This work was supported by the National Research Foundation of Korea (NRF) grant funded by the Korea government (MSIT) (Grant No. 2018R1C1B6007233) and by the Open KIAS Center at Korea Institute for Advanced Study. Computational resources were provided by the Center for Advanced Computation (CAC) of KIAS. 

\section*{References}
\bibliographystyle{unsrt}
\bibliography{bibl}

\begin{thebibliography}{10}

\bibitem{PRB2014Pike}
N.~A. Pike and D.~Stroud.
\newblock Tight-binding model for adatoms on graphene: Analytic density of states, spectral function, and induced magnetic moment.
\newblock {\em Phys. Rev. B}, 89:115428, 2014.

\bibitem{PRB2004XiaoDong}
Xiao Dong, Guan~Ming Wang, and Estela Blaisten-Barojas.
\newblock Tight-binding model for calcium nanoclusters: Structural, electronic, and dynamical properties.
\newblock {\em Phys. Rev. B}, 70:205409, 2004.

\bibitem{PRB2005Schulz}
S.~Schulz and G.~Czycholl.
\newblock Tight-binding model for semiconductor nanostructures.
\newblock {\em Phys. Rev. B}, 72:165317, 2005.

\bibitem{PRB1976Pandey}
K.~C. Pandey.
\newblock Realistic tight-binding model for chemisorption: H on si and ge (111).
\newblock {\em Phys. Rev. B}, 14:1557, 1976.

\bibitem{PRB1994Davidson}
B.~N. Davidson and W.~E. Pickett.
\newblock Tight-binding study of hydrogen on the c(111), c(100), and c(110) diamond surfaces.
\newblock {\em Phys. Rev. B}, 49:11253, 1994.

\bibitem{JChemPhys1993}
Hui Ou-Yang, Bruno K\"{a}llebring, and FL~A. Marcus.
\newblock Surface properties of solids using a semi-infinite approach and the tight-binding approximation.
\newblock {\em J. Chem. Phys., Vol. 98, No. 9, 1 May 1993}, 98:7405, 1993.

\bibitem{PRB2009Pereira}
Vitor~M. Pereira, A.~H.~Castro Neto, and N.~M.~R. Peres.
\newblock Tight-binding approach to uniaxial strain in graphene.
\newblock {\em Phys. Rev. B}, 80:045401, 2009.

\bibitem{PhyRevB_98_075106_2018_Fang}
Shiang Fang, Stephen Carr, Miguel~A. Cazalilla, and Efthimios Kaxiras.
\newblock Electronic structure theory of strained two-dimensional materials with hexagonal symmetry.
\newblock {\em Phys. Rev. B}, 98:075106, 2018.

\bibitem{PhyRevB_76_115202_2007_Jancu}
J.-M. Jancu and P.~Voisin.
\newblock Tetragonal and trigonal deformations in zinc-blende semiconductors: A tight-binding point of view.
\newblock {\em Phys. Rev. B}, 76:115202, 2007.

\bibitem{PhyRevB_94_155416_2016_Pearce}
Alexander~J. Pearce, Eros Mariani, and Guido Burkard.
\newblock Tetragonal and trigonal deformations in zinc-blende semiconductors: A tight-binding point of view.
\newblock {\em Phys. Rev. B}, 94:155416, 2016.

\bibitem{PhyRevB_79_245201_2009_Niquet}
Y.~M. Niquet, D.~Rideau, C.~Tavernier, H.~Jaouen, and X.~Blase.
\newblock Onsite matrix elements of the tight-binding hamiltonian of a strained crystal: Application to silicon, germanium, and their alloys.
\newblock {\em Phys. Rev. B}, 79:245201, 2009.

\bibitem{PRB_85_194558_2012_Moon}
Pilkyung Moon and Mikito Koshino.
\newblock Energy spectrum and quantum hall effect in twisted bilayer graphene.
\newblock {\em Phys. Rev. B}, 85:194558, 2012.

\bibitem{PRB_90_155406_2014_Moon}
Pilkyung Moon and Mikito Koshino.
\newblock Electronic properties of graphene/hexagonal-boron-nitride moiré superlattice.
\newblock {\em Phys. Rev. B}, 90:155406, 2014.

\bibitem{PRB_99_165430_2019_Moon}
Pilkyung Moon, Mikito Koshino, and Young-Woo Son.
\newblock Quasicrystalline electronic states in $30\circ$ rotated twisted bilayer graphene.
\newblock {\em Phys. Rev. B}, 99:165430, 2019.

\bibitem{PhysRev_94_1498_1954_Slater_Koster}
J.~C. Slater and G.~F. Koster.
\newblock Simplified lcao method for the periodic potential problem.
\newblock {\em Phys. Rev.}, 94:1498, 1954.

\bibitem{JPhysCondensMatter_2003_Papaconstantopoulos_Mehl}
D.~A. Papaconstantopoulos and M.~J. Mehl.
\newblock The slater–koster tight-binding method: a computationally efficient and accurate approach.
\newblock {\em J. Phys.: Condens. Matter}, 15:R413, 2003.

\bibitem{Handbook_TB_Papaconstantopoulos}
Dimitris~A. Papaconstantopoulos.
\newblock {\em Handbook of the Band Structure of Elemental Solids}.
\newblock Springer New York, NY, 1 edition, 2014.

\bibitem{PhysRev_136_B864_1964}
P.~Hohenberg and W.~Kohn.
\newblock Inhomogeneous electron gas.
\newblock {\em Phys. Rev.}, 136:B864, 1964.

\bibitem{PhysRev_140_A1133_1965}
W.~Kohn and L.~J. Sham.
\newblock Self-consistent equations including exchange and correlation effects.
\newblock {\em Phys. Rev.}, 1965:A1133, 1965.

\bibitem{JPhysCondMatter2015Ridolfi}
E.~Ridolfi, D.~Le, T.~S. Rahman, E.~R. Mucciolo, and C.~H. Lewenkopf.
\newblock A tight-binding model for mos$_{2}$ monolayers.
\newblock {\em J. Phys.: Condens. Matter}, 27:365501, 2015.

\bibitem{PhysRev_52_191_1937_Wannier}
Gregory~H. Wannier.
\newblock The structure of electronic excitation levels in insulating crystals.
\newblock {\em Phys. Rev.}, 52:191--197, 1937.

\bibitem{RevModPhys_34_645_1962_Wannier}
Gregory~H. Wannier.
\newblock Dynamics of band electrons in electric and magnetic fields.
\newblock {\em Rev. Mod. Phys.}, 34:645, 1962.

\bibitem{PhysRevB_56_12847_1997_Marzari}
Nicola Marzari and David Vanderbilt.
\newblock Maximally localized generalized wannier functions for composite energy bands.
\newblock {\em Phys. Rev. B}, 56:12847, 1997.

\bibitem{PhysRevB_65_035109_2002_Souza}
I.~Souza, N.~Marzari, and D.~Vanderbilt.
\newblock Maximally localized wannier functions for entangled energy bands.
\newblock {\em Phys. Rev. B}, 65:035109, 2002.

\bibitem{RevModPhys_84_1419_2012_Marzari}
Nicola Marzari, Arash~A. Mostofi, Jonathan~R. Yates, Ivo Souza, and David Vanderbilt.
\newblock Maximally localized wannier functions: Theory and applications.
\newblock {\em Rev. Mod. Phys.}, 84:1419, 2012.

\bibitem{PhysRevLett_94_026405_2005_Thygesen}
K.~S. Thygesen, L.~B. Hansen, and K.~W. Jacobsen.
\newblock Partly occupied wannier functions.
\newblock {\em Phys. Rev. Lett.}, 94:026405, Jan 2005.

\bibitem{PhysRevB_104_125140_Fontana}
Pietro~F. Fontana, Ask~H. Larsen, Thomas Olsen, and Kristian~S. Thygesen.
\newblock Spread-balanced wannier functions: Robust and automatable orbital localization.
\newblock {\em Phys. Rev. B}, 104:125140, Sep 2021.

\bibitem{PhysRevB_61_10040_2000_Berghold}
Gerd Berghold, Christopher~J. Mundy, Aldo~H. Romero, J\"urg Hutter, and Michele Parrinello.
\newblock General and efficient algorithms for obtaining maximally localized wannier functions.
\newblock {\em Phys. Rev. B}, 61:10040--10048, Apr 2000.

\bibitem{Multiscale_Model_Simul_17_167_Damle}
Anil Damle, Antoine Levitt, and Lin Lin.
\newblock Variational formulation for wannier functions with entangled band structure.
\newblock {\em Multiscale Modeling \& Simulation}, 17(1):167--191, 2019.

\bibitem{PhysRevB_89_035405_Jung}
Jeil Jung and Allan~H. MacDonald.
\newblock Accurate tight-binding models for the $\pi$ bands of bilayer graphene.
\newblock {\em Phys. Rev. B}, 89:035405, 2014.

\bibitem{PhysRevB_92_205108_Fang}
S.~Fang, R.~K. Defo, S.~N. Shirodkar, S.~Lieu, G.~A. Tritsaris, and E.~Kaxiras.
\newblock \textit{Ab initio} tight-binding hamiltonian for transition metal dichalcogenides.
\newblock {\em Phys. Rev. B}, 92:205108, 2015.

\bibitem{NatNanoTech_12_744_2017_Lu}
Ang-Yu Lu, Hanyu Zhu, Jun Xiao, Chih-Piao Chuu, Yimo Han, Ming-Hui Chiu, Chia-Chin Cheng, Chih-Wen Yang, Kung-Hwa Wei, Yiming Yang, Yuan Wang, Dimosthenis Sokaras, Dennis Nordlund, Peidong Yang, David~A. Muller, Mei-Yin Chou, Xiang Zhang, and Lain-Jong Li.
\newblock Janus monolayers of transition metal dichalcogenides.
\newblock {\em Nature Nanotechnology}, 12:744--749, 2017.

\bibitem{ACSNano_11_8192_2017_Zhang}
Jing Zhang, Shuai Jia, Iskandar Kholmanov, Liang Dong, Dequan Er, Weibing Chen, Hua Guo, Zehua Jin, Vivek~B. Shenoy, Li~Shi, and Jun Lou.
\newblock Janus monolayer transition-metal dichalcogenides.
\newblock {\em ACS Nano}, 11:8192--8198, 2017.

\bibitem{JPhysCM_30_215301_2018_Shi}
Wenwu Shi and Zhiguo Wang.
\newblock Mechanical and electronic properties of janus monolayer transition metal dichalcogenides.
\newblock {\em J. Phys.:Condens. Matter}, 30:215301, 2018.

\bibitem{PhysRevB_55_4168_1997_Stiles}
M.~D. Stiles.
\newblock Generalized slater-koster method for fitting band structures.
\newblock {\em Phys. Rev. B}, 55:4168, 1997.

\bibitem{NatNano2012Wang}
Q.~H. Wang, K.~Kalantar-Zadeh, A.~Kis, J.~N. Coleman, and M.~S. Strano.
\newblock Electronics and optoelectronics of two-dimensional transition metal dichalcogenides.
\newblock {\em Nat. Nano.}, 7:699--712, 2012.

\bibitem{JPhys_CM_21_395502_2009}
P.~Giannozzi, O.~Andreussi, T.~Brumme, O.~Bunau, M.~Buongiorno Nardelli, M.~Calandra, R.~Car, C.~Cavazzoni, D.~Ceresoli, M.~Cococcioni, N.~Colonna, I.~Carnimeo, A.~Dal Corso, S.~de~Gironcoli, P.~Delugas, R.~A.~DiStasio Jr, A.~Ferretti, A.~Floris, G.~Fratesi, G.~Fugallo, R.~Gebauer, U.~Gerstmann, F.~Giustino, T.~Gorni, J~Jia, M.~Kawamura, H.-Y. Ko, A.~Kokalj, E.~K\"{u}c\"{u}kbenli, M~.Lazzeri, M.~Marsili, N.~Marzari, F.~Mauri, N.~L. Nguyen, H.-V. Nguyen, A.~Otero de-la Roza, L.~Paulatto, S.~Ponc\'{e}, D.~Rocca, R.~Sabatini, B.~Santra, M.~Schlipf, A.~P. Seitsonen, A.~Smogunov, I.~Timrov, T.~Thonhauser, P.~Umari, N.~Vast, X.~Wu, and S.~Baroni.
\newblock $\textsc{Quantum ESPRESSO}$: a modular and open-source software project for quantum simulations of materials.
\newblock {\em J. Phys.: Condens. Matter}, 21:395502, 2009.

\bibitem{JPhys_CM_29_465901_2017}
P.~Giannozzi, S.~Baroni, N.~Bonini, M.~Calandra, R.~Car, C.~Cavazzoni, D.~Ceresoli, G.~L. Chiarotti, M.~Cococcioni, I.~Dabo amd A.~Dal~Corso, S.~Fabris, G.~Fratesi, S.~de~Gironcoli, R.~Gebauer, U.~Gerstmann, C.~Gougoussis, A.~Kokalj, M.~Lazzeri, L.~Martin-Samos, N.~Marzari, F.~Mauri, R.~Mazzarello, S.~Paolini, A.~Pasquarello, L.~Paulatto, C.~Sbraccia, S.~Scandolo, G.~Sclauzero, A.~P. Seitsonen, A.~Smogunov, P.~Umari, and R.~M. Wentzcovitch.
\newblock Advanced capabilities for materials modelling with $\textsc{Quantum ESPRESSO}$.
\newblock {\em J. Phys.: Condens. Matter}, 29:465901, 2017.

\bibitem{PhysRevLett_77_3865_1996}
J.~P. Perdew, K.~Burke, and M.~Ernzerhof.
\newblock Generalized gradient approximation made simple.
\newblock {\em Phys. Rev. Lett.}, 77:3865, 1996.

\bibitem{PhysRevB_88_085117_2013}
D.~R. Hamann.
\newblock Optimized norm-conserving vanderbilt pseudopotentials.
\newblock {\em Phys. Rev. B}, 88:085117, 2013.

\bibitem{ComPhysComms_226_39_2018_Setten}
M.~J. van Setten, M.~Giantomassi, E.~Bousquet, M.~J. Verstraete, D.~R. Hamann, X.~Gonze, and G.-M. Rignanese.
\newblock The pseudodojo: Training and grading a 85 element optimized norm-conserving pseudopotential table.
\newblock {\em Comput. Phys. Commun.}, 226:39--54, 2018.

\bibitem{PhysRevB_21_2603_1980_Kleinman}
Leonard Kleinman.
\newblock Relativistic norm-conserving pseudopotential.
\newblock {\em Phys. Rev. B}, 21:2603, 1980.

\bibitem{PhysRevB_25_2103_1982_Bachelet}
Giovanni~B. Bachelet and M.~Schl\"{u}ter.
\newblock Relativistic norm-conserving pseudopotentials.
\newblock {\em Phys. Rev. B}, 25:2103, 1982.

\bibitem{PhysRevLett_82_3296_1999_Marzari}
Nicola Marzari, David Vanderbilt, Alessandro~De Vita, and M.~C. Payne.
\newblock Thermal contraction and disordering of the al(110) surface.
\newblock {\em Phys. Rev. Lett.}, 82:3296, 1999.

\bibitem{JPhysCM_32_165902_Pizzi}
Giovanni Pizzi, Valerio Vitale, Ryotaro Arita, Stefan Bl\"{u}gel, Frank Freimuth, Guillaume G\'{e}ranton, Marco Gibertini, Dominik Gresch, Charles Johnson, Takashi Koretsune, Julen Ibañez-Azpiroz, Hyungjun Lee, Jae-Mo Lihm, Daniel Marchand, Antimo Marrazzo, Yuriy Mokrousov, Jamal~I Mustafa, Yoshiro Nohara, Yusuke Nomura, Lorenzo Paulatto, Samuel Ponc\'{e}, Thomas Ponweiser, Junfeng Qiao, Florian Th\"{o}le, Stepan~S. Tsirkin, Małgorzata Wierzbowska, Nicola Marzari, David Vanderbilt, Ivo Souza, Arash~A. Mostofi, and Jonathan~R. Yates.
\newblock Wannier90 as a community code: new features and applications.
\newblock {\em J. Phys.: Condens. Matter}, 32:165902, 2020.

\bibitem{PhysRevB_88_085433_Liu}
Gui-Bin Liu, Wen-Yu Shan, Yugui Yao, Wang Yao, and Di~Xiao.
\newblock Three-band tight-binding model for monolayers of group-vib transition metal dichalcogenides.
\newblock {\em Phys. Rev. B}, 88:085433, 2013.

\bibitem{Group_Theory_Dresselhaus}
M.~S. Dresselhaus, G.~Dresselhaus, and A.~Jorio.
\newblock {\em Group Theory: Application to the Physics of Condensed Matter}.
\newblock Springer-Verlag Berlin Heidelberg, 1 edition, 2008.

\bibitem{PhysRevB_87_235109_2013_Sakuma}
R.~Sakuma.
\newblock Symmetry-adapted wannier functions in the maximal localization procedure.
\newblock {\em Phys. Rev. B}, 87:235109, Jun 2013.

\bibitem{KORETSUNE2023108645}
Takashi Koretsune.
\newblock Construction of maximally-localized wannier functions using crystal symmetry.
\newblock {\em Computer Physics Communications}, 285:108645, 2023.

\bibitem{SymWannier}
Symwannier.
\newblock \url{https://github.com/wannier-utils-dev/symWannier}.

\bibitem{WU2018405}
QuanSheng Wu, ShengNan Zhang, Hai-Feng Song, Matthias Troyer, and Alexey~A. Soluyanov.
\newblock Wanniertools: An open-source software package for novel topological materials.
\newblock {\em Computer Physics Communications}, 224:405--416, 2018.

\bibitem{ZHI2022108196}
Guo-Xiang Zhi, Chenchao Xu, Si-Qi Wu, Fanlong Ning, and Chao Cao.
\newblock Wannsymm: A symmetry analysis code for wannier orbitals.
\newblock {\em Computer Physics Communications}, 271:108196, 2022.

\bibitem{WannierTools}
Wanniertools.
\newblock \url{http://www.wanniertools.com/}.

\bibitem{WannSymm}
Wannsymm.
\newblock \url{https://github.com/ccao/WannSymm}.

\bibitem{doi:10.1137/1.9781611971484}
Åke Björck.
\newblock {\em Numerical Methods for Least Squares Problems}.
\newblock Society for Industrial and Applied Mathematics, 1996.

\bibitem{1965Nelder-Mead_Comp_Journal_7_308}
John~A. Nelder and R.~Mead.
\newblock A simplex method for function minimization.
\newblock {\em Computer Journal}, 7 (4):308–313, 1965.

\bibitem{2020SciPy-NMeth}
Pauli Virtanen, Ralf Gommers, Travis~E. Oliphant, Matt Haberland, Tyler Reddy, David Cournapeau, Evgeni Burovski, Pearu Peterson, Warren Weckesser, Jonathan Bright, St{\'e}fan~J. {van der Walt}, Matthew Brett, Joshua Wilson, K.~Jarrod Millman, Nikolay Mayorov, Andrew R.~J. Nelson, Eric Jones, Robert Kern, Eric Larson, C~J Carey, {\.I}lhan Polat, Yu~Feng, Eric~W. Moore, Jake {VanderPlas}, Denis Laxalde, Josef Perktold, Robert Cimrman, Ian Henriksen, E.~A. Quintero, Charles~R. Harris, Anne~M. Archibald, Ant{\^o}nio~H. Ribeiro, Fabian Pedregosa, Paul {van Mulbregt}, and {SciPy 1.0 Contributors}.
\newblock {{SciPy} 1.0: Fundamental Algorithms for Scientific Computing in Python}.
\newblock {\em Nature Methods}, 17:261--272, 2020.

\bibitem{Haastrup_2018}
Sten Haastrup, Mikkel Strange, Mohnish Pandey, Thorsten Deilmann, Per~S Schmidt, Nicki~F Hinsche, Morten~N Gjerding, Daniele Torelli, Peter~M Larsen, Anders~C Riis-Jensen, Jakob Gath, Karsten~W Jacobsen, Jens~Jørgen Mortensen, Thomas Olsen, and Kristian~S Thygesen.
\newblock The computational 2d materials database: high-throughput modeling and discovery of atomically thin crystals.
\newblock {\em 2D Materials}, 5(4):042002, sep 2018.

\bibitem{Gjerding_2021}
Morten~Niklas Gjerding, Alireza Taghizadeh, Asbjørn Rasmussen, Sajid Ali, Fabian Bertoldo, Thorsten Deilmann, Nikolaj~Rørbæk Knøsgaard, Mads Kruse, Ask~Hjorth Larsen, Simone Manti, Thomas~Garm Pedersen, Urko Petralanda, Thorbjørn Skovhus, Mark~Kamper Svendsen, Jens~Jørgen Mortensen, Thomas Olsen, and Kristian~Sommer Thygesen.
\newblock Recent progress of the computational 2d materials database (c2db).
\newblock {\em 2D Materials}, 8(4):044002, jul 2021.

\bibitem{C2DB}
Computational 2d materials database (c2db).
\newblock \url{https://cmr.fysik.dtu.dk/c2db/c2db.html}.

\bibitem{PhysRevB_96_155439_Kim}
Sejoong Kim and Young-Woo Son.
\newblock Quasiparticle energy bands and fermi surfaces of monolayer $\textrm{NbSe}_2$.
\newblock {\em Phys. Rev. B}, 96:155439, 2017.

\bibitem{PhysRevB_104_045426_Kim}
Sejoong Kim and Young-Woo Son.
\newblock Dichotomy of saddle points in energy bands of monolayer $\textrm{NbSe}_2$.
\newblock {\em Phys. Rev. B}, 104:045426, 2021.

\bibitem{PhysRevB_102_045109_2020_Kurita}
Kensuke Kurita and Takashi Koretsune.
\newblock Systematic first-principles study of the on-site spin-orbit coupling in crystals.
\newblock {\em Phys. Rev. B}, 102:045109, 2020.

\bibitem{PhysRevB_104_195104_2021_Cuadrado}
R.~Cuadrado, R.~Robles, A.~García, M.~Pruneda, P.~Ordejón, J.~Ferrer, and Jorge~I. Cerdá.
\newblock Validity of the on-site spin-orbit coupling approximation.
\newblock {\em Phys. Rev. B}, 104:195104, 2021.

\bibitem{PRL2012Xiao}
Di~Xiao, Gui-Bin Liu, Wanxiang Feng, Xiaodong Xu, and Wang Yao.
\newblock Coupled spin and valley physics in monolayers of ${\mathrm{mos}}_{2}$ and other group-vi dichalcogenides.
\newblock {\em Phys. Rev. Lett.}, 108:196802, 2012.

\bibitem{PhysRevB_93_180501_Zhou}
Benjamin~T. Zhou, Noah F.~Q. Yuan, Hong-Liang Jiang, and K.~T. Law.
\newblock Ising superconductivity and majorana fermions in transition-metal dichalcogenides.
\newblock {\em Phys. Rev. B}, 93:180501, 2016.

\bibitem{PhysRevB_52_14531_1995_McKinnon}
B.~A. McKinnon and T.~C. Choy.
\newblock Significance of nonorthogonality in tight-binding models.
\newblock {\em Phys. Rev. B}, 52:14531, 1995.

\bibitem{PhysRevB_47_12995_1993_Dorantes_Davila}
J.~Dorantes-Dávila, A.~Vega, and G.~M. Pastor.
\newblock Self-consistent theory of overlap interactions in the tight-binding method.
\newblock {\em Phys. Rev. B}, 47:12995(R), 1993.

\bibitem{Electronic_structure_and_the_properties_of_solids_Harrison}
Walter~A. Harrison.
\newblock {\em Electronic structure and the properties of solids}.
\newblock Dover publications, Inc., New York, 1 edition, 1989.

\end{thebibliography}

\newpage

\title[Supplementary Materials]{Supplementary Materials on "Construction of optimized tight-binding models using \textit{ab initio} Hamiltonian: Application to monolayer $2H$-transition metal dichalcogenides"}

\author{Sejoong Kim}
\address{University of Science and Technology (UST), Gajeong-ro 217, Daejeon 34113, Republic of Korea}
\address{Korea Institute for Advanced Study, Hoegiro 85, Seoul 02455, Republic of Korea}
\ead{sejoong@alum.mit.edu}
\vspace{10pt}

\section*{\label{appendix:energy_integrals}Energy integrals}
Here we provide a full list of energy integrals in the tight-binding (TB) model consisting of five $d$-orbitals of transition metal atom and six $p$-orbitals of chalcogen atoms as well as on-site Hamiltonians at transition metal atoms and chalcogen ones.  

\subsection{On-site Hamiltonians}
The transition metal atom site $\textrm{M}$ possesses the following on-site Hamiltonian $\mathcal{H}_{\textrm{M}}$, 
\begin{equation}
\label{Eq:H_M}\mathcal{H}_{\textrm{M}}=\left[\begin{array}{ccccc}
E_{z^{2}} & 0 & 0 & 0 & 0\\
0 & E_{xy} & 0 & 0 & 0\\
0 & 0 & E_{xy} & 0 & 0\\
0 & 0 & 0 & E_{zx} & 0\\
0 & 0 & 0 & 0 & E_{zx}
\end{array}\right],    
\end{equation}
where is written with the basis of $|d_{z^2}\rangle$, $|d_{x^2-y^2}\rangle$, $|d_{xy}\rangle$, $|d_{zx}\rangle$, and $|d_{yz}\rangle$.

The Hamiltonian $\mathcal{H}_{\textrm{X}}$ at chalcogen atom sites reads 
\begin{equation}
\label{Eq:H_X}\mathcal{H}_{\textrm{X}}=\left[\begin{array}{cccccc}
E_{uz} & 0 & 0 & E_{uz,lz}^{(0)} & 0 & 0\\
0 & E_{ux} & 0 & 0 & E_{ux,lx}^{(0)} & 0\\
0 & 0 & E_{ux} & 0 & 0 & E_{ux,lx}^{(0)}\\
E_{uz,lz}^{(0)} & 0 & 0 & E_{lz} & 0 & 0\\
0 & E_{ux,lx}^{(0)} & 0 & 0 & E_{lx} & 0\\
0 & 0 & E_{ux,lx}^{(0)} & 0 & 0 & E_{lx}
\end{array}\right],
\end{equation}
where is written with the basis of $|p_{uz}\rangle$, $|p_{ux}\rangle$, $|p_{uy}\rangle$, $|p_{lz}\rangle$, $|p_{lx}\rangle$, and $|p_{ly}\rangle$. Here $u$ and $l$ stand for upper and lower chalcogen layers. Equation~(\ref{Eq:H_M}) includes $p$ orbitals of chalcogen atoms at upper and lower layers, which have the same $x$ and $y$ coordinates. 
\begin{figure*}[t]
\begin{center}
\includegraphics[width=1.0\textwidth, clip=true]{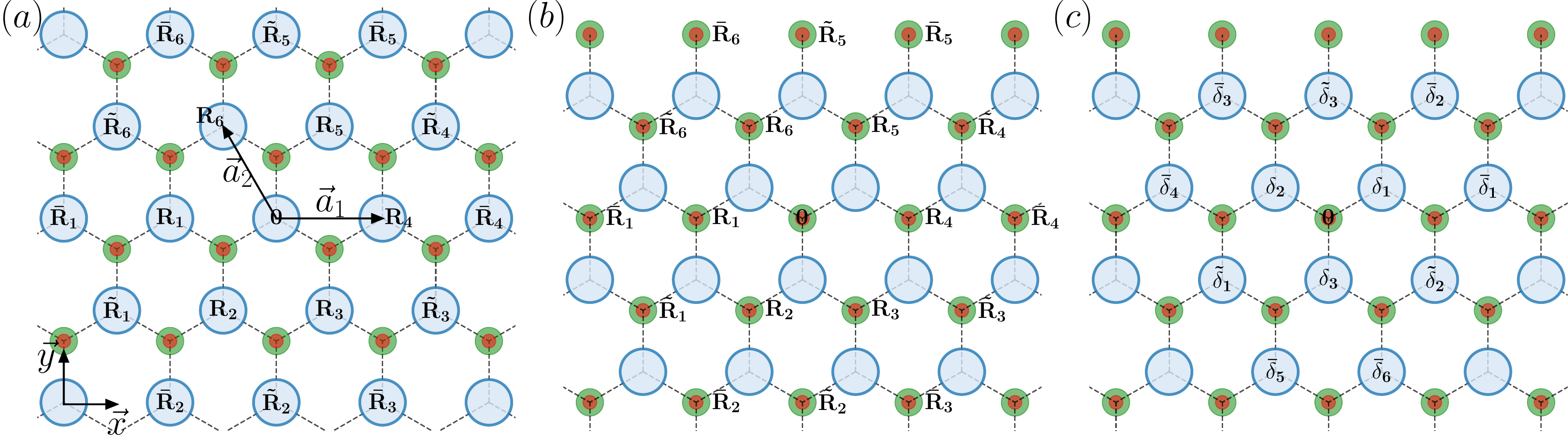} 
\end{center}
\caption{\label{Fig:neighbors} (Color online) (a) First, second and third nearest neighbors among $d$ orbitals at transition metal atom sites (b) First, second and third nearest neighbors among $p$ orbitals at chalcogen atoms (c) First, second and third nearest neighbors between transition metal $d$ orbitals and chalcogen $p$ orbitals}
\end{figure*}

\subsection{Energy integrals between $d$-orbitals}
\subsubsection{The nearest-neighbors}
A transition metal atom is surrounded by six nearest neighbors $\mathbf{R}_{i}$ for $i=1,\cdots,6$ as illustrated in Fig.~\ref{Fig:neighbors}(a). In terms of lattice vectors $\vec{a}_{1}$ and $\vec{a}_{2}$, the six nearest neighbors are written as $\mathbf{R}_{1}=-\vec{a}_{1}$, $\mathbf{R}_{2}=-\vec{a}_{1}-\vec{a}_{2}$, $\mathbf{R}_{3}=-\vec{a}_{2}$, $\mathbf{R}_{4}=\vec{a}_{1}$, $\mathbf{R}_{5}=\vec{a}_{1}+\vec{a}_{2}$, and $\mathbf{R}_{6}=\vec{a}_{2}$. 
Energy integrals among nearest-neighboring $d$ orbitals are written as follows: 
\begin{eqnarray}
\label{Eq:E_11_1}
E_{z^2,z^2}(\mathbf{R}_{i}) &=& E_{z^2,z^2}^{(1)}\textrm{ for }i=1,\cdots,6 \\
E_{z^2, x^2-y^2}(\mathbf{R}_{1}) &=& E_{z^2, x^2-y^2}^{(1)} \\
E_{z^2, x^2-y^2}(\mathbf{R}_{2}) &=& -\frac{1}{2}E_{z^2, x^2-y^2}^{(1)}+\frac{\sqrt{3}}{2}E_{z^2, xy}^{(1)} \\
E_{z^2, x^2-y^2}(\mathbf{R}_{3}) &=& -\frac{1}{2}E_{z^2, x^2-y^2}^{(1)}+\frac{\sqrt{3}}{2}E_{z^2, xy}^{(1)} \\
E_{z^2, x^2-y^2}(\mathbf{R}_{4}) &=& E_{z^2, x^2-y^2}^{(1)} \\
E_{z^2, x^2-y^2}(\mathbf{R}_{5}) &=& -\frac{1}{2}E_{z^2, x^2-y^2}^{(1)}-\frac{\sqrt{3}}{2}E_{z^2, xy}^{(1)} \\
E_{z^2, x^2-y^2}(\mathbf{R}_{6}) &=& -\frac{1}{2}E_{z^2, x^2-y^2}^{(1)}-\frac{\sqrt{3}}{2}E_{z^2, xy}^{(1)} \\
E_{z^2, xy}(\mathbf{R}_{1}) &=& E_{z^2, xy}^{(1)} \\
E_{z^2, xy}(\mathbf{R}_{2})&=& \frac{\sqrt{3}}{2}E_{z^2, x^2-y^2}^{(1)}+\frac{1}{2}E_{z^2, xy}^{(1)} \\ 
E_{z^2, xy}(\mathbf{R}_{3}) &=& -\frac{\sqrt{3}}{2}E_{z^2, x^2-y^2}^{(1)}-\frac{1}{2}E_{z^2, xy}^{(1)} \\ 
E_{z^2, xy}(\mathbf{R}_{4})&=& -E_{z^2, xy}^{(1)} \\
E_{z^2, xy}(\mathbf{R}_{5})&=& \frac{\sqrt{3}}{2}E_{z^2, x^2-y^2}^{(1)}-\frac{1}{2}E_{z^2, xy}^{(1)} \\
E_{z^2, xy}(\mathbf{R}_{6}) &=& -\frac{\sqrt{3}}{2}E_{z^2, x^2-y^2}^{(1)}+\frac{1}{2}E_{z^2, xy}^{(1)} \\
E_{z^2, zx}(\mathbf{R}_{1}) &=& E_{z^2, zx}^{(1)} \\
E_{z^2, zx}(\mathbf{R}_{2}) &=& \frac{1}{2}E_{z^2, zx}^{(1)}+\frac{\sqrt{3}}{2}E_{z^2, yz}^{(1)} \\
E_{z^2, zx}(\mathbf{R}_{3}) &=& -\frac{1}{2}E_{z^2, zx}^{(1)}-\frac{\sqrt{3}}{2}E_{z^2, yz}^{(1)} \\
E_{z^2, zx}(\mathbf{R}_{4}) &=& - E_{z^2, zx}^{(1)} \\
E_{z^2, zx}(\mathbf{R}_{5}) &=& -\frac{1}{2}E_{z^2, zx}^{(1)}+\frac{\sqrt{3}}{2}E_{z^2, yz}^{(1)} \\
E_{z^2, zx}(\mathbf{R}_{6}) &=& \frac{1}{2}E_{z^2, zx}^{(1)}-\frac{\sqrt{3}}{2}E_{z^2, yz}^{(1)} \\
E_{z^2, yz}(\mathbf{R}_{1}) &=& E_{z^2, yz}^{(1)} \\
E_{z^2, yz}(\mathbf{R}_{2}) &=& \frac{\sqrt{3}}{2}E_{z^2, zx}^{(1)}-\frac{1}{2}E_{z^2, yz}^{(1)} \\
E_{z^2, yz}(\mathbf{R}_{3}) &=& \frac{\sqrt{3}}{2}E_{z^2, zx}^{(1)}-\frac{1}{2}E_{z^2, yz}^{(1)} \\
E_{z^2, yz}(\mathbf{R}_{4}) &=& E_{z^2, yz}^{(1)} \\
E_{z^2, yz}(\mathbf{R}_{5}) &=& -\frac{\sqrt{3}}{2}E_{z^2, zx}^{(1)}-\frac{1}{2}E_{z^2, yz}^{(1)} \\
E_{z^2, yz}(\mathbf{R}_{6}) &=& -\frac{\sqrt{3}}{2}E_{z^2, zx}^{(1)}-\frac{1}{2}E_{z^2, yz}^{(1)} \\
E_{x^2-y^2, x^2-y^2}(\mathbf{R}_{2}) &=& \frac{1}{4}E_{x^2-y^2, x^2-y^2}^{(1)}+\frac{3}{4}E_{xy,xy}^{(1)} \nonumber \\\\
E_{x^2-y^2, x^2-y^2}(\mathbf{R}_{4}) &=& E_{x^2-y^2, x^2-y^2}^{(1)} \\
E_{x^2-y^2, x^2-y^2}(\mathbf{R}_{5}) &=& \frac{1}{4}E_{x^2-y^2, x^2-y^2}^{(1)}+\frac{3}{4}E_{xy,xy}^{(1)} \nonumber \\\\
E_{x^2-y^2, x^2-y^2}(\mathbf{R}_{3}) &=& \frac{1}{4}E_{x^2-y^2, x^2-y^2}^{(1)}+\frac{3}{4}E_{xy,xy}^{(1)} \nonumber \\\\
E_{x^2-y^2, x^2-y^2}(\mathbf{R}_{1}) &=& E_{x^2-y^2, x^2-y^2}^{(1)} \\
E_{x^2-y^2, x^2-y^2}(\mathbf{R}_{6}) &=& \frac{1}{4}E_{x^2-y^2, x^2-y^2}^{(1)}+\frac{3}{4}E_{xy,xy}^{(1)} \nonumber \\\\
E_{x^2-y^2, xy}(\mathbf{R}_{1}) &=& E_{x^2-y^2, xy}^{(1)} \\
E_{x^2-y^2, xy}(\mathbf{R}_{2}) &=& -\frac{\sqrt{3}}{4}E_{x^2-y^2, x^2-y^2}^{(1)}-E_{x^2-y^2, xy}^{(1)}\nonumber \\
&&+\frac{\sqrt{3}}{4}E_{xy,xy}^{(1)}  \\
E_{x^2-y^2, xy}(\mathbf{R}_{3}) &=& \frac{\sqrt{3}}{4}E_{x^2-y^2, x^2-y^2}^{(1)}+E_{x^2-y^2, xy}^{(1)}\nonumber \\
&&-\frac{\sqrt{3}}{4}E_{xy,xy}^{(1)} \\
E_{x^2-y^2, xy}(\mathbf{R}_{4}) &=& -E_{x^2-y^2, xy}^{(1)} \\
E_{x^2-y^2, xy}(\mathbf{R}_{5}) &=& -\frac{\sqrt{3}}{4}E_{x^2-y^2, x^2-y^2}^{(1)}+E_{x^2-y^2, xy}^{(1)}\nonumber \\ 
&&+\frac{\sqrt{3}}{4}E_{xy,xy}^{(1)} \\
E_{x^2-y^2, xy}(\mathbf{R}_{6}) &=& \frac{\sqrt{3}}{4}E_{x^2-y^2, x^2-y^2}^{(1)}-E_{x^2-y^2, xy}^{(1)}\nonumber \\
&&-\frac{\sqrt{3}}{4}E_{xy,xy}^{(1)} \\
E_{x^2-y^2, zx}(\mathbf{R}_{1}) &=& E_{x^2-y^2, zx}^{(1)} \\
E_{x^2-y^2, zx}(\mathbf{R}_{2}) &=& -\frac{1}{4}E_{x^2-y^2, zx}^{(1)}+\frac{\sqrt{3}}{4}E_{xy,zx}^{(1)}\nonumber\\
&& -\frac{\sqrt{3}}{4}E_{x^2-y^2, yz}^{(1)}+\frac{3}{4}E_{xy,yz}^{(1)} \\
E_{x^2-y^2, zx}(\mathbf{R}_{3}) &=& \frac{1}{4}E_{x^2-y^2, zx}^{(1)}-\frac{\sqrt{3}}{4}E_{xy,zx}^{(1)}\nonumber\\
&&+\frac{\sqrt{3}}{4}E_{x^2-y^2, yz}^{(1)}-\frac{3}{4}E_{xy,yz}^{(1)} \\
E_{x^2-y^2, zx}(\mathbf{R}_{4}) &=& -E_{x^2-y^2, zx}^{(1)} \\
E_{x^2-y^2, zx}(\mathbf{R}_{5}) &=& \frac{1}{4}E_{x^2-y^2, zx}^{(1)}+\frac{\sqrt{3}}{4}E_{xy,zx}^{(1)}\nonumber\\
&&-\frac{\sqrt{3}}{4}E_{x^2-y^2, yz}^{(1)}-\frac{3}{4}E_{xy,yz}^{(1)} \\
E_{x^2-y^2, zx}(\mathbf{R}_{6}) &=& -\frac{1}{4}E_{x^2-y^2, zx}^{(1)}-\frac{\sqrt{3}}{4}E_{xy,zx}^{(1)}\nonumber\\
&&+\frac{\sqrt{3}}{4}E_{x^2-y^2, yz}^{(1)}+\frac{3}{4}E_{xy,yz}^{(1)} \\
E_{x^2-y^2, yz}(\mathbf{R_{1}}) &=& E_{x^2-y^2, yz}^{(1)} \\
E_{x^2-y^2, yz}(\mathbf{R_{2}}) &=& -\frac{\sqrt{3}}{4}E_{x^2-y^2, zx}^{(1)}+\frac{3}{4}E_{xy, zx}^{(1)}\nonumber\\
&&+\frac{1}{4}E_{x^2-y^2, yz}^{(1)}-\frac{\sqrt{3}}{4}E_{xy, yz}^{(1)} \\
E_{x^2-y^2, yz}(\mathbf{R_{3}}) &=& -\frac{\sqrt{3}}{4}E_{x^2-y^2, zx}^{(1)}+\frac{3}{4}E_{xy, zx}^{(1)}\nonumber\\
&&+\frac{1}{4}E_{x^2-y^2, yz}^{(1)}-\frac{\sqrt{3}}{4}E_{xy, yz}^{(1)} \\
E_{x^2-y^2, yz}(\mathbf{R_{4}}) &=& E_{x^2-y^2, yz}^{(1)} \\
E_{x^2-y^2, yz}(\mathbf{R_{5}}) &=& \frac{\sqrt{3}}{4}E_{x^2-y^2, zx}^{(1)}+\frac{3}{4}E_{xy, zx}^{(1)}\nonumber\\
&&+\frac{1}{4}E_{x^2-y^2, yz}^{(1)}+\frac{\sqrt{3}}{4}E_{xy, yz}^{(1)} \\
E_{x^2-y^2, yz}(\mathbf{R_{6}}) &=& \frac{\sqrt{3}}{4}E_{x^2-y^2, zx}^{(1)}+\frac{3}{4}E_{xy, zx}^{(1)}\nonumber\\
&&+\frac{1}{4}E_{x^2-y^2, yz}^{(1)}+\frac{\sqrt{3}}{4}E_{xy, yz}^{(1)} \\
E_{xy, xy}(\mathbf{R}_{1}) &=& E_{xy, xy}^{(1)} \\
E_{xy, xy}(\mathbf{R}_{2}) &=& \frac{3}{4}E_{x^2-y^2,x^2-y^2}^{(1)}+\frac{1}{4}E_{xy, xy}^{(1)} \\
E_{xy, xy}(\mathbf{R}_{3}) &=& \frac{3}{4}E_{x^2-y^2,x^2-y^2}^{(1)}+\frac{1}{4}E_{xy, xy}^{(1)} \\
E_{xy, xy}(\mathbf{R}_{4}) &=& E_{xy, xy}^{(1)} \\
E_{xy, xy}(\mathbf{R}_{5}) &=& \frac{3}{4}E_{x^2-y^2,x^2-y^2}^{(1)}+\frac{1}{4}E_{xy, xy}^{(1)} \\
E_{xy, xy}(\mathbf{R}_{6}) &=& \frac{3}{4}E_{x^2-y^2,x^2-y^2}^{(1)}+\frac{1}{4}E_{xy, xy}^{(1)} \\
E_{xy, zx}(\mathbf{R}_{2}) &=& \frac{\sqrt{3}}{4}E_{x^2-y^2, zx}^{(1)}+\frac{1}{4}E_{xy, zx}^{(1)}\nonumber\\
&&+\frac{3}{4}E_{x^2-y^2, yz}^{(1)}+\frac{\sqrt{3}}{4}E_{xy, yz}^{(1)} \\
E_{xy, zx}(\mathbf{R}_{4}) &=& E_{xy, zx}^{(1)} \\
E_{xy, zx}(\mathbf{R}_{5}) &=& -\frac{\sqrt{3}}{4}E_{x^2-y^2, zx}^{(1)}+\frac{1}{4}E_{xy, zx}^{(1)}\nonumber\\
&&+\frac{3}{4}E_{x^2-y^2, yz}^{(1)}-\frac{\sqrt{3}}{4}E_{xy, yz}^{(1)} \\
E_{xy, zx}(\mathbf{R}_{3}) &=& \frac{\sqrt{3}}{4}E_{x^2-y^2, zx}^{(1)}+\frac{1}{4}E_{xy, zx}^{(1)}\nonumber\\
&&+\frac{3}{4}E_{x^2-y^2, yz}^{(1)}+\frac{\sqrt{3}}{4}E_{xy, yz}^{(1)} \\
E_{xy, zx}(\mathbf{R}_{4}) &=& E_{xy, zx}^{(1)} \\
E_{xy, zx}(\mathbf{R}_{6}) &=& -\frac{\sqrt{3}}{4}E_{x^2-y^2, zx}^{(1)}+\frac{1}{4}E_{xy, zx}^{(1)}\nonumber\\
&&+\frac{3}{4}E_{x^2-y^2, yz}^{(1)}-\frac{\sqrt{3}}{4}E_{xy, yz}^{(1)} \\
E_{xy, yz}(\mathbf{R}_{1}) &=& E_{xy, yz}^{(1)} \\
E_{xy, yz}(\mathbf{R}_{2}) &=& \frac{3}{4}E_{x^2-y^2, zx}^{(1)}+\frac{\sqrt{3}}{4}E_{xy, zx}^{(1)}\nonumber\\
&&-\frac{\sqrt{3}}{4}E_{x^2-y^2, yz}^{(1)}-\frac{1}{4}E_{xy, yz}^{(1)} \\
E_{xy, yz}(\mathbf{R}_{3}) &=& -\frac{3}{4}E_{x^2-y^2, zx}^{(1)}-\frac{\sqrt{3}}{4}E_{xy, zx}^{(1)}\nonumber\\
&&+\frac{\sqrt{3}}{4}E_{x^2-y^2, yz}^{(1)}+\frac{1}{4}E_{xy, yz}^{(1)} \\
E_{xy, yz}(\mathbf{R}_{4}) &=& -E_{xy, yz}^{(1)} \\
E_{xy, yz}(\mathbf{R}_{5}) &=& -\frac{3}{4}E_{x^2-y^2, zx}^{(1)}+\frac{\sqrt{3}}{4}E_{xy, zx}^{(1)}\nonumber\\
&&-\frac{\sqrt{3}}{4}E_{x^2-y^2, yz}^{(1)}+\frac{1}{4}E_{xy, yz}^{(1)} \\
E_{xy, yz}(\mathbf{R}_{6}) &=& \frac{3}{4}E_{x^2-y^2, zx}^{(1)}-\frac{\sqrt{3}}{4}E_{xy, zx}^{(1)}\nonumber\\
&&+\frac{\sqrt{3}}{4}E_{x^2-y^2, yz}^{(1)}-\frac{1}{4}E_{xy, yz}^{(1)} \\
E_{zx, zx}(\mathbf{R}_{1}) &=& E_{zx, zx}^{(1)} \\
E_{zx, zx}(\mathbf{R}_{2}) &=& \frac{1}{4}E_{zx, zx}^{(1)}+\frac{3}{4}E_{yz,yz}^{(1)} \\
E_{zx, zx}(\mathbf{R}_{3}) &=& \frac{1}{4}E_{zx, zx}^{(1)}+\frac{3}{4}E_{yz,yz}^{(1)} \\
E_{zx, zx}(\mathbf{R}_{4}) &=& E_{zx, zx}^{(1)} \\
E_{zx, zx}(\mathbf{R}_{5}) &=& \frac{1}{4}E_{zx, zx}^{(1)}+\frac{3}{4}E_{yz,yz}^{(1)} \\
E_{zx, zx}(\mathbf{R}_{6}) &=& \frac{1}{4}E_{zx, zx}^{(1)}+\frac{3}{4}E_{yz,yz}^{(1)} \\
E_{zx, yz}(\mathbf{R}_{1}) &=& E_{zx, yz}^{(1)} \\
E_{zx, yz}(\mathbf{R}_{2}) &=& \frac{\sqrt{3}}{4}E_{zx,zx}^{(1)}-E_{zx, yz}^{(1)}-\frac{\sqrt{3}}{4}E_{yz,yz}^{(1)} \nonumber \\\\
E_{zx, yz}(\mathbf{R}_{3}) &=& -\frac{\sqrt{3}}{4}E_{zx,zx}^{(1)}+E_{zx, yz}^{(1)}+\frac{\sqrt{3}}{4}E_{yz,yz}^{(1)} \nonumber \\\\
E_{zx, yz}(\mathbf{R}_{4}) &=& -E_{zx, yz}^{(1)} \\
E_{zx, yz}(\mathbf{R}_{5}) &=& \frac{\sqrt{3}}{4}E_{zx,zx}^{(1)}+E_{zx, yz}^{(1)}-\frac{\sqrt{3}}{4}E_{yz,yz}^{(1)} \nonumber \\\\
E_{zx, yz}(\mathbf{R}_{6}) &=& -\frac{\sqrt{3}}{4}E_{zx,zx}^{(1)}-E_{zx, yz}^{(1)}+\frac{\sqrt{3}}{4}E_{yz,yz}^{(1)} \nonumber \\\\
E_{yz, yz}(\mathbf{R}_{1}) &=& E_{yz, yz}^{(1)} \\
E_{yz, yz}(\mathbf{R}_{2}) &=& \frac{3}{4}E_{zx,zx}^{(1)}+\frac{1}{4}E_{yz, yz}^{(1)} \\
E_{yz, yz}(\mathbf{R}_{3}) &=& \frac{3}{4}E_{zx,zx}^{(1)}+\frac{1}{4}E_{yz, yz}^{(1)} \\
E_{yz, yz}(\mathbf{R}_{4}) &=& E_{yz, yz}^{(1)} \\
E_{yz, yz}(\mathbf{R}_{5}) &=& \frac{3}{4}E_{zx,zx}^{(1)}+\frac{1}{4}E_{yz, yz}^{(1)} \\
\label{Eq:E_55_1}E_{yz, yz}(\mathbf{R}_{6}) &=& \frac{3}{4}E_{zx,zx}^{(1)}+\frac{1}{4}E_{yz, yz}^{(1)} 
\end{eqnarray}
Note that there are fifteen independent energy integrals among nearest neighboring $d$ orbitals: $E^{(1)}_{z^2, z^2}$, $E^{(1)}_{z^2, x^2-y^2}$, $E^{(1)}_{z^2,xy}$, $E^{(1)}_{z^2, zx}$, $E^{(1)}_{z^2, yz}$, $E^{(1)}_{x^2-y^2, x^2-y^2}$, $E^{(1)}_{x^2-y^2, xy}$, $E^{(1)}_{x^2-y^2, zx}$, $E^{(1)}_{x^2-y^2, yz}$, $E^{(1)}_{xy, xy}$, $E^{(1)}_{xy, zx}$, $E^{(1)}_{xy, yz}$, $E^{(1)}_{zx,zx}$, $E^{(1)}_{zx, yz}$, and $E^{(1)}_{yz, yz}$, where
\begin{equation}
E_{ij}^{(1)} \equiv E_{ij}(\mathbf{R_{1}}) = \langle d_i(\mathbf{0})| \mathcal{H} | d_j(\mathbf{R}_{1}) \rangle. 
\end{equation}

\subsubsection{The second nearest-neighbors}
As shown in Fig.~\ref{Fig:neighbors}(a) there are six second nearest neighbors $\tilde{\mathbf{R}}_{i}$ for $i=1,\cdots,6$, which are $\tilde{\mathbf{R}}_{1}=\mathbf{R}_{1}+\mathbf{R}_{2}=-2\vec{a}_{1}-\vec{a}_{2}$, $\tilde{\mathbf{R}}_{2}=\mathbf{R}_{2}+\mathbf{R}_{3}=-\vec{a}_{1}-2\vec{a}_{2}$,   
$\tilde{\mathbf{R}}_{3}=\mathbf{R}_{3}+\mathbf{R}_{4}=\vec{a}_{1}-\vec{a}_{2}$, 
$\tilde{\mathbf{R}}_{4}=\mathbf{R}_{4}+\mathbf{R}_{5}=2\vec{a}_{1}+\vec{a}_{2}$, 
$\tilde{\mathbf{R}}_{5}=\mathbf{R}_{5}+\mathbf{R}_{6}=\vec{a}_{1}+2\vec{a}_{2}$, 
$\tilde{\mathbf{R}}_{6}=\mathbf{R}_{6}+\mathbf{R}_{1}=-\vec{a}_{1}+\vec{a}_{2}$ in terms of lattice vectors $\vec{a}_{1}$ and $\vec{a}_{2}$. Energy integrals for second nearest neighboring $d$-orbitals read
\begin{eqnarray}
E_{z^2, z^2}(\mathbf{\tilde{R}}_{i}) &=& E_{z^2, z^2}^{(2)}\textrm{ for }i=1,\cdots,6 \\
E_{z^2, x^2-y^2}(\mathbf{\tilde{R}}_{1}) &=&  E_{z^2, x^2-y^2}^{(2)} \\
E_{z^2, x^2-y^2}(\mathbf{\tilde{R}}_{2}) &=& -E_{x^2-y^2, z^2 }^{(2)} \\
E_{z^2, x^2-y^2}(\mathbf{\tilde{R}}_{3}) &=&  E_{z^2, x^2-y^2}^{(2)}  \\
E_{z^2, x^2-y^2}(\mathbf{\tilde{R}}_{4}) &=&  E_{x^2-y^2, z^2 }^{(2)} \\
E_{z^2, x^2-y^2}(\mathbf{\tilde{R}}_{5}) &=& -E_{z^2, x^2-y^2}^{(2)} \\
E_{z^2, x^2-y^2}(\mathbf{\tilde{R}}_{6}) &=&  E_{x^2-y^2, z^2 }^{(2)} \\
E_{z^2, xy}(\mathbf{\tilde{R}}_{1}) &=& \sqrt{3}E_{z^2, x^2-y^2}^{(2)}  \\
E_{z^2, xy}(\mathbf{\tilde{R}}_{2}) &=& 0 \\
E_{z^2, xy}(\mathbf{\tilde{R}}_{4}) &=& \sqrt{3}E_{x^2-y^2, z^2}^{(2)}  \\
E_{z^2, xy}(\mathbf{\tilde{R}}_{3}) &=& -\sqrt{3}E_{z^2, x^2-y^2}^{(2)} \\
E_{z^2, xy}(\mathbf{\tilde{R}}_{5}) &=& 0 \\
E_{z^2, xy}(\mathbf{\tilde{R}}_{6}) &=& -\sqrt{3}E_{x^2-y^2, z^2}^{(2)} \\
E_{z^2, zx}(\mathbf{\tilde{R}}_{1}) &=& \sqrt{3}E_{z^2, yz}^{(2)} \\
E_{z^2, zx}(\mathbf{\tilde{R}}_{2}) &=& 0 \\
E_{z^2, zx}(\mathbf{\tilde{R}}_{4}) &=& \sqrt{3}E_{yz, z^2}^{(2)} \\
E_{z^2, zx}(\mathbf{\tilde{R}}_{3}) &=& -\sqrt{3}E_{z^2, yz}^{(2)} \\
E_{z^2, zx}(\mathbf{\tilde{R}}_{5}) &=& 0 \\
E_{z^2, zx}(\mathbf{\tilde{R}}_{6}) &=& -\sqrt{3}E_{yz, z^2}^{(2)} \\
\mathcal{H}_{z^2, yz}(\mathbf{\tilde{R}}_{1}) &=& E_{z^2, yz}^{(2)} \\
\mathcal{H}_{z^2, yz}(\mathbf{\tilde{R}}_{2}) &=& -2E_{yz, z^2}^{(2)} \\
\mathcal{H}_{z^2, yz}(\mathbf{\tilde{R}}_{3}) &=& E_{z^2, yz}^{(2)} \\
\mathcal{H}_{z^2, yz}(\mathbf{\tilde{R}}_{4}) &=& E_{yz, z^2}^{(2)} \\
\mathcal{H}_{z^2, yz}(\mathbf{\tilde{R}}_{5}) &=& -2E_{z^2, yz}^{(2)} \\
\mathcal{H}_{z^2, yz}(\mathbf{\tilde{R}}_{6}) &=& E_{yz, z^2}^{(2)} \\
E_{x^2-y^2, x^2-y^2}^{(2)}(\mathbf{\tilde{R}}_{1}) &=& E_{x^2-y^2, x^2-y^2}^{(2)} \\
E_{x^2-y^2, x^2-y^2}^{(2)}(\mathbf{\tilde{R}}_{2}) &=& E_{x^2-y^2, x^2-y^2}^{(2)} +\sqrt{3}E_{x^2-y^2,xy}^{(2)} \nonumber \\\\
E_{x^2-y^2, x^2-y^2}^{(2)}(\mathbf{\tilde{R}}_{3}) &=& E_{x^2-y^2, x^2-y^2}^{(2)} \\
E_{x^2-y^2, x^2-y^2}^{(2)}(\mathbf{\tilde{R}}_{4}) &=& E_{x^2-y^2, x^2-y^2}^{(2)} \\
E_{x^2-y^2, x^2-y^2}^{(2)}(\mathbf{\tilde{R}}_{5}) &=& E_{x^2-y^2, x^2-y^2}^{(2)} +\sqrt{3}E_{x^2-y^2,xy}^{(2)} \nonumber \\\\
E_{x^2-y^2, x^2-y^2}^{(2)}(\mathbf{\tilde{R}}_{6}) &=& E_{x^2-y^2, x^2-y^2}^{(2)} \\
E_{x^2-y^2, xy}^{(2)}(\mathbf{\tilde{R}}_{1}) &=& E_{x^2-y^2, xy}^{(2)}\\
E_{x^2-y^2, xy}^{(2)}(\mathbf{\tilde{R}}_{3}) &=& -E_{x^2-y^2, xy}^{(2)}\\
E_{x^2-y^2, xy}^{(2)}(\mathbf{\tilde{R}}_{4}) &=& E_{x^2-y^2, xy}^{(2)}\\
E_{x^2-y^2, xy}^{(2)}(\mathbf{\tilde{R}}_{6}) &=& -E_{x^2-y^2, xy}^{(2)}\\
E_{x^2-y^2, xy}^{(2)}(\mathbf{\tilde{R}}_{2}) &=& 0 \\
E_{x^2-y^2, xy}^{(2)}(\mathbf{\tilde{R}}_{5}) &=& 0 \\
E_{x^2-y^2, zx}^{(2)}(\mathbf{\tilde{R}}_{1}) &=& E_{x^2-y^2, zx}^{(2)} \\
E_{x^2-y^2, zx}^{(2)}(\mathbf{\tilde{R}}_{2}) &=& 0 \\
E_{x^2-y^2, zx}^{(2)}(\mathbf{\tilde{R}}_{4}) &=& E_{zx, x^2-y^2}^{(2)} \\
E_{x^2-y^2, zx}^{(2)}(\mathbf{\tilde{R}}_{3}) &=& -E_{x^2-y^2, zx}^{(2)} \\
E_{x^2-y^2, zx}^{(2)}(\mathbf{\tilde{R}}_{5}) &=& 0 \\
E_{x^2-y^2, zx}^{(2)}(\mathbf{\tilde{R}}_{6}) &=& -E_{zx, x^2-y^2}^{(2)} \\
E_{x^2-y^2, yz}^{(2)}(\mathbf{\tilde{R}}_{1}) &=&  E_{x^2-y^2, yz}^{(2)} \\
E_{x^2-y^2, yz}^{(2)}(\mathbf{\tilde{R}}_{2}) &=&  E_{zx,xy}^{(2)} + \frac{1}{\sqrt{3}}E_{zx, x^2-y^2}^{(2)} \\
E_{x^2-y^2, yz}^{(2)}(\mathbf{\tilde{R}}_{3}) &=&  E_{x^2-y^2, yz}^{(2)} \\
E_{x^2-y^2, yz}^{(2)}(\mathbf{\tilde{R}}_{4}) &=&  E_{zx,xy}^{(2)} - \frac{2}{\sqrt{3}}E_{zx, x^2-y^2}^{(2)} \\
E_{x^2-y^2, yz}^{(2)}(\mathbf{\tilde{R}}_{5}) &=&  E_{x^2-y^2, yz}^{(2)} + \sqrt{3}E_{x^2-y^2, zx}^{(2)} \\
E_{x^2-y^2, yz}^{(2)}(\mathbf{\tilde{R}}_{6}) &=&  E_{zx,xy}^{(2)} - \frac{2}{\sqrt{3}}E_{zx, x^2-y^2}^{(2)} \\
E_{xy, xy}^{(2)}(\mathbf{\tilde{R}}_{1}) &=& E_{x^2-y^2, x^2-y^2}^{(2)}+\frac{2}{\sqrt{3}}E_{x^2-y^2, xy}^{(2)} \nonumber\\ \\
E_{xy, xy}^{(2)}(\mathbf{\tilde{R}}_{2}) &=& E_{x^2-y^2, x^2-y^2}^{(2)}-\frac{1}{\sqrt{3}}E_{x^2-y^2, xy}^{(2)} \nonumber\\ \\
E_{xy, xy}^{(2)}(\mathbf{\tilde{R}}_{3}) &=& E_{x^2-y^2, x^2-y^2}^{(2)}+\frac{2}{\sqrt{3}}E_{x^2-y^2, xy}^{(2)} \nonumber\\ \\
E_{xy, xy}^{(2)}(\mathbf{\tilde{R}}_{4}) &=& E_{x^2-y^2, x^2-y^2}^{(2)}+\frac{2}{\sqrt{3}}E_{x^2-y^2, xy}^{(2)} \nonumber\\ \\
E_{xy, xy}^{(2)}(\mathbf{\tilde{R}}_{5}) &=& E_{x^2-y^2, x^2-y^2}^{(2)}-\frac{1}{\sqrt{3}}E_{x^2-y^2, xy}^{(2)} \nonumber\\ \\
E_{xy, xy}^{(2)}(\mathbf{\tilde{R}}_{6}) &=& E_{x^2-y^2, x^2-y^2}^{(2)}+\frac{2}{\sqrt{3}}E_{x^2-y^2, xy}^{(2)} \nonumber\\ \\
E_{xy, zx}^{(2)}(\mathbf{\tilde{R}}_{1}) &=& E_{x^2-y^2, yz}^{(2)} + \frac{2}{\sqrt{3}}E_{x^2-y^2, zx}^{(2)} \\
E_{xy, zx}^{(2)}(\mathbf{\tilde{R}}_{2}) &=& E_{zx, xy}^{(2)} -\sqrt{3}E_{zx, x^2-y^2}^{(2)} \\
E_{xy, zx}^{(2)}(\mathbf{\tilde{R}}_{3}) &=& E_{x^2-y^2, yz}^{(2)} + \frac{2}{\sqrt{3}}E_{x^2-y^2, zx}^{(2)} \\
E_{xy, zx}^{(2)}(\mathbf{\tilde{R}}_{4}) &=& E_{zx, xy}^{(2)} \\
E_{xy, zx}^{(2)}(\mathbf{\tilde{R}}_{5}) &=& E_{x^2-y^2, yz}^{(2)} - \frac{1}{\sqrt{3}}E_{x^2-y^2, zx}^{(2)} \\
E_{xy, zx}^{(2)}(\mathbf{\tilde{R}}_{6}) &=& E_{zx, xy}^{(2)} \\
E_{xy, yz}^{(2)}(\mathbf{\tilde{R}}_{1}) &=& E_{x^2-y^2, zx}^{(2)} \\
E_{xy, yz}^{(2)}(\mathbf{\tilde{R}}_{2}) &=& 0 \\
E_{xy, yz}^{(2)}(\mathbf{\tilde{R}}_{3}) &=& -E_{x^2-y^2, zx}^{(2)} \\
E_{xy, yz}^{(2)}(\mathbf{\tilde{R}}_{4}) &=& E_{zx, x^2-y^2}^{(2)} \\
E_{xy, yz}^{(2)}(\mathbf{\tilde{R}}_{5}) &=& 0 \\
E_{xy, yz}^{(2)}(\mathbf{\tilde{R}}_{6}) &=& -E_{zx, x^2-y^2}^{(2)} \\
E_{zx, zx}^{(2)}(\mathbf{\tilde{R}}_{1}) &=& E_{zx, zx}^{(2)} \\
E_{zx, zx}^{(2)}(\mathbf{\tilde{R}}_{2}) &=& E_{zx, zx}^{(2)} - \sqrt{3}E_{zx, yz}^{(2)} \\
E_{zx, zx}^{(2)}(\mathbf{\tilde{R}}_{3}) &=& E_{zx, zx}^{(2)} \\
E_{zx, zx}^{(2)}(\mathbf{\tilde{R}}_{4}) &=& E_{zx, zx}^{(2)} \\
E_{zx, zx}^{(2)}(\mathbf{\tilde{R}}_{5}) &=& E_{zx, zx}^{(2)} - \sqrt{3}E_{zx, yz}^{(2)} \\
E_{zx, zx}^{(2)}(\mathbf{\tilde{R}}_{6}) &=& E_{zx, zx}^{(2)} \\
E_{zx, yz}^{(3)}(\mathbf{\tilde{R}}_{1}) &=& E_{zx, yz}^{(2)} \\
E_{zx, yz}^{(3)}(\mathbf{\tilde{R}}_{2}) &=& 0 \\
E_{zx, yz}^{(3)}(\mathbf{\tilde{R}}_{3}) &=& -E_{zx, yz}^{(2)} \\
E_{zx, yz}^{(3)}(\mathbf{\tilde{R}}_{4}) &=& E_{zx, yz}^{(2)} \\
E_{zx, yz}^{(3)}(\mathbf{\tilde{R}}_{5}) &=& 0 \\
E_{zx, yz}^{(3)}(\mathbf{\tilde{R}}_{6}) &=& -E_{zx, yz}^{(2)} \\
E_{yz, yz}^{(2)}(\mathbf{\tilde{R}}_{1}) &=& E_{zx, zx}^{(2)}-\frac{2}{\sqrt{3}}E_{zx, yz}^{(2)} \\
E_{yz, yz}^{(2)}(\mathbf{\tilde{R}}_{2}) &=& E_{zx, zx}^{(2)}+\frac{1}{\sqrt{3}}E_{zx, yz}^{(2)} \\
E_{yz, yz}^{(2)}(\mathbf{\tilde{R}}_{3}) &=& E_{zx, zx}^{(2)}-\frac{2}{\sqrt{3}}E_{zx, yz}^{(2)} \\
E_{yz, yz}^{(2)}(\mathbf{\tilde{R}}_{4}) &=& E_{zx, zx}^{(2)}-\frac{2}{\sqrt{3}}E_{zx, yz}^{(2)} \\
E_{yz, yz}^{(2)}(\mathbf{\tilde{R}}_{5}) &=& E_{zx, zx}^{(2)}+\frac{1}{\sqrt{3}}E_{zx, yz}^{(2)} \\
E_{yz, yz}^{(2)}(\mathbf{\tilde{R}}_{6}) &=& E_{zx, zx}^{(2)}-\frac{2}{\sqrt{3}}E_{zx, yz}^{(2)}
\end{eqnarray}
Here parameters $E_{ij}^{(2)}$ denote energy integrals between the $d$-orbital $i$ at origin and the $d$-orbital $j$ located at the second nearest neighbor vector $\mathbf{\tilde{R}}_{1}$:
\begin{equation}
E_{ij}^{(2)} \equiv E_{ij}(\mathbf{\tilde{R}_{1}}) = \langle i(\mathbf{0})| \mathcal{H} | j(\mathbf{\tilde{R}}_{1}) \rangle.
\end{equation}
The TB model for the second nearest neighboring $d$-orbitals includes thirteen independent energy integrals, $E_{z^2,z^2}^{(2)}$, $E_{z^2,x^2-y^2}^{(2)}$, $E_{x^2-y^2,z^2}^{(2)}$, $E_{z^2,yz}^{(2)}$, $E_{yz, z^2}^{(2)}$, $E_{x^2-y^2,x^2-y^2}^{(2)}$, $E_{x^2-y^2, xy}^{(2)}$, $E_{x^2-y^2, zx}^{(2)}$, $E_{zx, x^2-y^2}^{(2)}$, $E_{x^2-y^2, yz}^{(2)}$, $E_{zx,xy}^{(2)}$, $E_{zx,zx}^{(2)}$, and $E_{zx,yz}^{(2)}$.  

\subsubsection{The third nearest-neighbors}
Third nearest neighbors are $\bar{\mathbf{R}}_{i}$ for $i=1,\cdots,6$, which are twice larger than nearest neighbors, i.e., $\bar{\mathbf{R}}_{i}=2\mathbf{R}_{i}$. 
Note that energy integrals for third nearest-neighboring $d$-orbitals have the same forms of 
energy integrals for nearest neighbors [Eqs.~(\ref{Eq:E_11_1})-(\ref{Eq:E_55_1})], but independent TB parameters $E_{ij}^{(1)}$ are replaced by $E_{ij}^{(3)}$, where
\begin{equation}
E_{ij}^{(3)} \equiv E_{ij}(\mathbf{\bar{R}}_{1}) = \langle i(\mathbf{0})| \mathcal{H} | j(2\mathbf{R_{1}}) \rangle. 
\end{equation}
The TB model relating third nearest neighboring $d$-orbitals contains fifteen independent parameters, $E^{(3)}_{z^2, z^2}$, $E^{(3)}_{z^2, x^2-y^2}$, $E^{(3)}_{z^2,xy}$, $E^{(3)}_{z^2, zx}$, $E^{(3)}_{z^2, yz}$, $E^{(3)}_{x^2-y^2, x^2-y^2}$, $E^{(3)}_{x^2-y^2, xy}$, $E^{(3)}_{x^2-y^2, zx}$, $E^{(3)}_{x^2-y^2, yz}$, $E^{(3)}_{xy, xy}$, $E^{(3)}_{xy, zx}$, $E^{(3)}_{xy, yz}$, $E^{(3)}_{zx,zx}$, $E^{(3)}_{zx, yz}$, and $E^{(3)}_{yz, yz}$. 

\subsection{\label{Appdenix_TBp}Energy integrals between $p$-orbitals}
The TB model between $p$-orbitals includes six $p$-orbitals, which are $p_{x}$, $p_{y}$, and $p_{z}$ orbitals at upper and lower sublayers of chalcogen atoms. 
Figure~\ref{Fig:neighbors}(b) demonstrates sets of first, second and third nearest neighbors among chalcogen atoms. Nearest neighboring vectors among chalcogen $p$-orbitals on the $xy$ plane are identical to those of transition metal $d$-orbitals, $\{\mathbf{R}_{i}\}$, $\{\tilde{\mathbf{R}}_{i}\}$, and $\{\bar{\mathbf{R}}_{i}\}$. When two $p$-orbitals involved in energy integrals are on the different chalcogen sublayers, nearest neighbors contain the additional vector along the $z$ axis relating upper and lower chalcogen sublayers $\vec{c}=\pm c\hat{z}$, where $c$ is the distance between the sublayers. This vector between chalcogen sublayers is not explicitly indicated in the following energy integrals. Instead, energy integrals have chalcogen sublayer indicies $u$ (upper) or $l$ (lower) that $p$-orbitals belong to. For example, $uz$ denotes the $p_{z}$ orbital on the upper chalcogen sublayer. 

\subsubsection{The nearest-neighbors}
\begin{eqnarray}
\label{Eq:Hzz1}E_{\alpha z,\beta z}^{(1)}(\mathbf{R}_{i}) &=& E_{\alpha z,\beta z}^{(1)}\textrm{ for }i=1,\cdots,6 \\
\mathcal{H}_{\alpha z,\beta y}^{(1)}(\mathbf{R}_{1}) &=& E_{\alpha z,\beta y}^{(1)} \\
\mathcal{H}_{\alpha z,\beta y}^{(1)}(\mathbf{R}_{2}) &=& -\frac{1}{2}E_{\alpha z,\beta y}^{(1)}+\frac{\sqrt{3}}{2}E_{\alpha z,\beta x}^{(1)} \\
\mathcal{H}_{\alpha z,\beta y}^{(1)}(\mathbf{R}_{3}) &=& -\frac{1}{2}E_{\alpha z,\beta y}^{(1)}+\frac{\sqrt{3}}{2}E_{\alpha z,\beta x}^{(1)} \\
\mathcal{H}_{\alpha z,\beta y}^{(1)}(\mathbf{R}_{4}) &=& E_{\alpha z,\beta y}^{(1)} \\
\mathcal{H}_{\alpha z,\beta y}^{(1)}(\mathbf{R}_{5}) &=& -\frac{1}{2}E_{\alpha z,\beta y}^{(1)}-\frac{\sqrt{3}}{2}E_{\alpha z,\beta x}^{(1)} \\
\mathcal{H}_{\alpha z,\beta y}^{(1)}(\mathbf{R}_{6}) &=& -\frac{1}{2}E_{\alpha z,\beta y}^{(1)}-\frac{\sqrt{3}}{2}E_{\alpha z,\beta x}^{(1)} \\
E_{\alpha z,\beta x}^{(1)}(\mathbf{R}_{1}) &=& E_{\alpha z,\beta x}^{(1)}  \\
E_{\alpha z,\beta x}^{(1)}(\mathbf{R}_{2}) &=&  \frac{\sqrt{3}}{2}E_{\alpha z,\beta y}^{(1)}+\frac{1}{2}E_{\alpha z,\beta x}^{(1)}\\
E_{\alpha z,\beta x}^{(1)}(\mathbf{R}_{3}) &=& -\frac{\sqrt{3}}{2}E_{\alpha z,\beta y}^{(1)}-\frac{1}{2}E_{\alpha z,\beta x}^{(1)}\\
E_{\alpha z,\beta x}^{(1)}(\mathbf{R}_{4}) &=& -E_{\alpha z,\beta x}^{(1)} \\
E_{\alpha z,\beta x}^{(1)}(\mathbf{R}_{5}) &=&  \frac{\sqrt{3}}{2}E_{\alpha z,\beta y}^{(1)}-\frac{1}{2}E_{\alpha z,\beta x}^{(1)}\\
E_{\alpha z,\beta x}^{(1)}(\mathbf{R}_{6}) &=&  -\frac{\sqrt{3}}{2}E_{\alpha z,\beta y}^{(1)}+\frac{1}{2}E_{\alpha z,\beta x}^{(1)}\\
E_{\alpha y,\alpha y}^{(1)}(\mathbf{R}_{1}) &=& E_{\alpha y,\alpha y}^{(1)} \\
E_{\alpha y,\alpha y}^{(1)}(\mathbf{R}_{4}) &=& E_{\alpha y,\alpha y}^{(1)} \\
E_{\alpha y,\alpha y}^{(1)}(\mathbf{R}_{2}) &=& \frac{1}{4}E_{\alpha y,\alpha y}^{(1)}+\frac{3}{4}E_{\alpha x,\alpha x}^{(1)} \\
E_{\alpha y,\alpha y}^{(1)}(\mathbf{R}_{3}) &=& \frac{1}{4}E_{\alpha y,\alpha y}^{(1)}+\frac{3}{4}E_{\alpha x,\alpha x}^{(1)} \\
E_{\alpha y,\alpha y}^{(1)}(\mathbf{R}_{5}) &=& \frac{1}{4}E_{\alpha y,\alpha y}^{(1)}+\frac{3}{4}E_{\alpha x,\alpha x}^{(1)} \\
E_{\alpha y,\alpha y}^{(1)}(\mathbf{R}_{6}) &=& \frac{1}{4}E_{\alpha y,\alpha y}^{(1)}+\frac{3}{4}E_{\alpha x,\alpha x}^{(1)} \\
E_{\alpha y,\alpha x}^{(1)}(\mathbf{R}_{1}) &=& E_{\alpha y,\alpha x}^{(1)} \\
E_{\alpha y,\alpha x}^{(1)}(\mathbf{R}_{2}) &=& -\frac{\sqrt{3}}{4}E_{\alpha y,\alpha y}^{(1)}-E_{\alpha y,\alpha x}^{(1)}+\frac{\sqrt{3}}{4}E_{\alpha x,\alpha x}^{(1)} \nonumber\\ \\
E_{\alpha y,\alpha x}^{(1)}(\mathbf{R}_{3}) &=& \frac{\sqrt{3}}{4}E_{\alpha y,\alpha y}^{(1)}+E_{\alpha y,\alpha x}^{(1)}-\frac{\sqrt{3}}{4}E_{\alpha x,\alpha x}^{(1)} \nonumber\\ \\
E_{\alpha y,\alpha x}^{(1)}(\mathbf{R}_{4}) &=& -E_{\alpha y,\alpha x}^{(1)} \\
E_{\alpha y,\alpha x}^{(1)}(\mathbf{R}_{5}) &=& -\frac{\sqrt{3}}{4}E_{\alpha y,\alpha y}^{(1)}+E_{\alpha y,\alpha x}^{(1)}+\frac{\sqrt{3}}{4}E_{\alpha x,\alpha x}^{(1)} \nonumber\\ \\
E_{\alpha y,\alpha x}^{(1)}(\mathbf{R}_{6}) &=& \frac{\sqrt{3}}{4}E_{\alpha y,\alpha y}^{(1)}-E_{\alpha y,\alpha x}^{(1)}-\frac{\sqrt{3}}{4}E_{\alpha x,\alpha x}^{(1)} \nonumber\\ \\
E_{\alpha x,\alpha x}^{(1)}(\mathbf{R}_{1}) &=& E_{\alpha x,\alpha x}^{(1)}\\
E_{\alpha x,\alpha x}^{(1)}(\mathbf{R}_{2}) &=& \frac{3}{4}E_{\alpha y,\alpha y}^{(1)}+\frac{1}{4}E_{\alpha x,\alpha x}^{(1)}\\
E_{\alpha x,\alpha x}^{(1)}(\mathbf{R}_{3}) &=& \frac{3}{4}E_{\alpha y,\alpha y}^{(1)}+\frac{1}{4}E_{\alpha x,\alpha x}^{(1)}\\
E_{\alpha x,\alpha x}^{(1)}(\mathbf{R}_{4}) &=& E_{\alpha x,\alpha x}^{(1)}\\
E_{\alpha x,\alpha x}^{(1)}(\mathbf{R}_{5}) &=& \frac{3}{4}E_{\alpha y,\alpha y}^{(1)}+\frac{1}{4}E_{\alpha x,\alpha x}^{(1)}\\
E_{\alpha x,\alpha x}^{(1)}(\mathbf{R}_{6}) &=& \frac{3}{4}E_{\alpha y,\alpha y}^{(1)}+\frac{1}{4}E_{\alpha x,\alpha x}^{(1)}\\
\mathcal{H}_{\alpha y,\bar{\alpha}y}^{(1)}(\mathbf{R}_{1}) &=& E_{\alpha y,\bar{\alpha}y}^{(1)}\\
\mathcal{H}_{\alpha y,\bar{\alpha}y}^{(1)}(\mathbf{R}_{2}) &=& \frac{1}{4}E_{\alpha y,\bar{\alpha}y}^{(1)}+\frac{\sqrt{3}}{4}E_{\bar{\alpha}y,\alpha x}^{(1)}\nonumber\\
&&-\frac{\sqrt{3}}{4}E_{\alpha y,\bar{\alpha}x}^{(1)}+\frac{3}{4}E_{\alpha x,\bar{\alpha}x}^{(1)}\\
\mathcal{H}_{\alpha y,\bar{\alpha}y}^{(1)}(\mathbf{R}_{3}) &=& \frac{1}{4}E_{\alpha y,\bar{\alpha}y}^{(1)}+\frac{\sqrt{3}}{4}E_{\bar{\alpha}y,\alpha x}^{(1)}\nonumber\\
&&-\frac{\sqrt{3}}{4}E_{\alpha y,\bar{\alpha}x}^{(1)}+\frac{3}{4}E_{\alpha x,\bar{\alpha}x}^{(1)}\\
\mathcal{H}_{\alpha y,\bar{\alpha}y}^{(1)}(\mathbf{R}_{4}) &=& E_{\alpha y,\bar{\alpha}y}^{(1)}\\
\mathcal{H}_{\alpha y,\bar{\alpha}y}^{(1)}(\mathbf{R}_{5}) &=& \frac{1}{4}E_{\alpha y,\bar{\alpha}y}^{(1)}-\frac{\sqrt{3}}{4}E_{\bar{\alpha}y,\alpha x}^{(1)}\nonumber\\
&&+\frac{\sqrt{3}}{4}E_{\alpha y,\bar{\alpha}x}^{(1)}+\frac{3}{4}E_{\alpha x,\bar{\alpha}x}^{(1)}\\
\mathcal{H}_{\alpha y,\bar{\alpha}y}^{(1)}(\mathbf{R}_{6}) &=& \frac{1}{4}E_{\alpha y,\bar{\alpha}y}^{(1)}-\frac{\sqrt{3}}{4}E_{\bar{\alpha}y,\alpha x}^{(1)}\nonumber\\
&&+\frac{\sqrt{3}}{4}E_{\alpha y,\bar{\alpha}x}^{(1)}+\frac{3}{4}E_{\alpha x,\bar{\alpha}x}^{(1)}\\
E_{\alpha y,\bar{\alpha}x}^{(1)}(\mathbf{R}_{1}) &=&  E_{\alpha y,\bar{\alpha}x}^{(1)} \\
E_{\alpha y,\bar{\alpha}x}^{(1)}(\mathbf{R}_{2}) &=& -\frac{\sqrt{3}}{4}E_{\alpha y,\bar{\alpha}y}^{(1)}-\frac{3}{4}E_{\bar{\alpha}y,\alpha x}^{(1)}\nonumber\\
&&-\frac{1}{4}E_{\alpha y,\bar{\alpha}x}^{(1)}+\frac{\sqrt{3}}{4}E_{\alpha x,\bar{\alpha}x}^{(1)}\\
E_{\alpha y,\bar{\alpha}x}^{(1)}(\mathbf{R}_{3}) &=& \frac{\sqrt{3}}{4}E_{\alpha y,\bar{\alpha}y}^{(1)}+\frac{3}{4}E_{\bar{\alpha}y,\alpha x}^{(1)}\nonumber\\
&&+\frac{1}{4}E_{\alpha y,\bar{\alpha}x}^{(1)}-\frac{\sqrt{3}}{4}E_{\alpha x,\bar{\alpha}x}^{(1)}\\
E_{\alpha y,\bar{\alpha}x}^{(1)}(\mathbf{R}_{4}) &=& -E_{\alpha y,\bar{\alpha}x}^{(1)} \\
E_{\alpha y,\bar{\alpha}x}^{(1)}(\mathbf{R}_{5}) &=& -\frac{\sqrt{3}}{4}E_{\alpha y,\bar{\alpha}y}^{(1)}+\frac{3}{4}E_{\bar{\alpha}y,\alpha x}^{(1)}\nonumber\\
&&+\frac{1}{4}E_{\alpha y,\bar{\alpha}x}^{(1)}+\frac{\sqrt{3}}{4}E_{\alpha x,\bar{\alpha}x}^{(1)}\\
E_{\alpha y,\bar{\alpha}x}^{(1)}(\mathbf{R}_{6}) &=& \frac{\sqrt{3}}{4}E_{\alpha y,\bar{\alpha}y}^{(1)}-\frac{3}{4}E_{\bar{\alpha}y,\alpha x}^{(1)}\nonumber\\
&&-\frac{1}{4}E_{\alpha y,\bar{\alpha}x}^{(1)}-\frac{\sqrt{3}}{4}E_{\alpha x,\bar{\alpha}x}^{(1)}\\
E_{\alpha x,\bar{\alpha}y}^{(1)}(\mathbf{R}_{1}) &=& E_{\alpha x,\bar{\alpha}y}^{(1)} \\
E_{\alpha x,\bar{\alpha}y}^{(1)}(\mathbf{R}_{2}) &=& -\frac{\sqrt{3}}{4}E_{\alpha y,\bar{\alpha}y}^{(1)}+\frac{1}{4}E_{\bar{\alpha}y,\alpha x}^{(1)}\nonumber\\
&&+\frac{3}{4}E_{\alpha y,\bar{\alpha}x}^{(1)}+\frac{\sqrt{3}}{4}E_{\alpha x,\bar{\alpha}x}^{(1)} \\
E_{\alpha x,\bar{\alpha}y}^{(1)}(\mathbf{R}_{3}) &=& \frac{\sqrt{3}}{4}E_{\alpha y,\bar{\alpha}y}^{(1)}-\frac{1}{4}E_{\bar{\alpha}y,\alpha x}^{(1)}\nonumber\\
&&-\frac{3}{4}E_{\alpha y,\bar{\alpha}x}^{(1)}-\frac{\sqrt{3}}{4}E_{\alpha x,\bar{\alpha}x}^{(1)} \\
E_{\alpha x,\bar{\alpha}y}^{(1)}(\mathbf{R}_{4}) &=& -E_{\alpha x,\bar{\alpha}y}^{(1)} \\
E_{\alpha x,\bar{\alpha}y}^{(1)}(\mathbf{R}_{5}) &=& -\frac{\sqrt{3}}{4}E_{\alpha y,\bar{\alpha}y}^{(1)}-\frac{1}{4}E_{\bar{\alpha}y,\alpha x}^{(1)}\nonumber\\
&&-\frac{3}{4}E_{\alpha y,\bar{\alpha}x}^{(1)}+\frac{\sqrt{3}}{4}E_{\alpha x,\bar{\alpha}x}^{(1)}\\
E_{\alpha x,\bar{\alpha}y}^{(1)}(\mathbf{R}_{6}) &=& \frac{\sqrt{3}}{4}E_{\alpha y,\bar{\alpha}y}^{(1)}+\frac{1}{4}E_{\bar{\alpha}y,\alpha x}^{(1)}\nonumber\\
&&+\frac{3}{4}E_{\alpha y,\bar{\alpha}x}^{(1)}-\frac{\sqrt{3}}{4}E_{\alpha x,\bar{\alpha}x}^{(1)}\\
E_{\alpha x,\bar{\alpha}x}^{(1)}(\mathbf{R}_{1}) &=& E_{\alpha x,\bar{\alpha}x}^{(1)} \\
E_{\alpha x,\bar{\alpha}x}^{(1)}(\mathbf{R}_{2}) &=& \frac{3}{4}E_{\alpha y,\bar{\alpha}y}^{(1)}-\frac{\sqrt{3}}{4}E_{\bar{\alpha}y,\alpha x}^{(1)}\nonumber\\
&&+\frac{\sqrt{3}}{4}E_{\alpha y,\bar{\alpha}x}^{(1)}+\frac{1}{4}E_{\alpha x,\bar{\alpha}x}^{(1)}\\
E_{\alpha x,\bar{\alpha}x}^{(1)}(\mathbf{R}_{3}) &=& \frac{3}{4}E_{\alpha y,\bar{\alpha}y}^{(1)}-\frac{\sqrt{3}}{4}E_{\bar{\alpha}y,\alpha x}^{(1)}\nonumber\\
&&+\frac{\sqrt{3}}{4}E_{\alpha y,\bar{\alpha}x}^{(1)}+\frac{1}{4}E_{\alpha x,\bar{\alpha}x}^{(1)}\\
E_{\alpha x,\bar{\alpha}x}^{(1)}(\mathbf{R}_{4}) &=& E_{\alpha x,\bar{\alpha}x}^{(1)} \\
E_{\alpha x,\bar{\alpha}x}^{(1)}(\mathbf{R}_{5}) &=& \frac{3}{4}E_{\alpha y,\bar{\alpha}y}^{(1)}+\frac{\sqrt{3}}{4}E_{\bar{\alpha}y,\alpha x}^{(1)}\nonumber\\
&&-\frac{\sqrt{3}}{4}E_{\alpha y,\bar{\alpha}x}^{(1)}+\frac{1}{4}E_{\alpha x,\bar{\alpha}x}^{(1)}\\
\label{Eq:Hxbx1} E_{\alpha x,\bar{\alpha}x}^{(1)}(\mathbf{R}_{6}) &=& \frac{3}{4}E_{\alpha y,\bar{\alpha}y}^{(1)}+\frac{\sqrt{3}}{4}E_{\bar{\alpha}y,\alpha x}^{(1)}\nonumber\\
&&-\frac{\sqrt{3}}{4}E_{\alpha y,\bar{\alpha}x}^{(1)}+\frac{1}{4}E_{\alpha x,\bar{\alpha}x}^{(1)}
\end{eqnarray}
Here $\alpha$, $\beta$ = $u$ or $l$ where $u$ and $l$ stands for upper and lower chalcogen sublayers, respectively.  
TB parameters $E_{\alpha i, \beta j}^{(1)}$ are energy integrals between $p$-orbital $i$ of the $\alpha$ chalcogen sublayer at origin and $p$-orbital $j$ of the $\beta$ chalcogen sublayer at the lattice vector $\mathbf{R}_{1}$:
\begin{equation}
E_{\alpha i, \beta j}^{(1)} = E_{\alpha i, \beta j}(\mathbf{R}_{1}) = \langle \alpha p_{i}(\mathbf{0}) | \mathcal{H} | \beta p_{j} (\mathbf{R}_{1})\rangle.    
\end{equation}
Here independent TB parameters $E_{\alpha i, \beta j}^{(1)}$ are $E_{uz,uz}^{(1)}$, $E_{lz,lz}^{(1)}$, $E_{uz,lz}^{(1)}$, $E_{uz,uy}^{(1)}$, $E_{uz,ux}^{(1)}$, $E_{uz,ly}^{(1)}$, $E_{uz,lx}^{(1)}$, $E_{lz,uy}^{(1)}$, $E_{lz,ux}^{(1)}$, $E_{lz,ly}^{(1)}$, $E_{lz,lx}^{(1)}$, $E_{uy,uy}^{(1)}$, $E_{ly,ly}^{(1)}$, $E_{ux,ux}^{(1)}$, $E_{lx,lx}^{(1)}$, $E_{uy,ux}^{(1)}$, $E_{ly,lx}^{(1)}$, $E_{uy,ly}^{(1)}$, $E_{ux,lx}^{(1)}$, $E_{uy,lx}^{(1)}$, and $E_{ly,lx}^{(1)}$, which are 21 in total.     

\subsubsection{The second nearest-neighbors}
Energy integrals of second nearest-neighboring $p$-orbitals are given as follows:
\begin{eqnarray}
\mathcal{H}_{\alpha z,\beta z}^{(2)}(\mathbf{\tilde{R}}_{1}) &=& E_{\alpha z,\beta z}^{(2)} \\
\mathcal{H}_{\alpha z,\beta z}^{(2)}(\mathbf{\tilde{R}}_{2}) &=& E_{\beta z,\alpha z}^{(2)} \\
\mathcal{H}_{\alpha z,\beta z}^{(2)}(\mathbf{\tilde{R}}_{3}) &=& E_{\alpha z,\beta z}^{(2)} \\
\mathcal{H}_{\alpha z,\beta z}^{(2)}(\mathbf{\tilde{R}}_{4}) &=& E_{\beta z,\alpha z}^{(2)} \\
\mathcal{H}_{\alpha z,\beta z}^{(2)}(\mathbf{\tilde{R}}_{5}) &=& E_{\alpha z,\beta z}^{(2)} \\
\mathcal{H}_{\alpha z,\beta z}^{(2)}(\mathbf{\tilde{R}}_{6}) &=& E_{\beta z,\alpha z}^{(2)} \\
E_{\alpha z,\beta y}^{(2)}(\mathbf{\tilde{R}}_{1}) &=&  E_{\alpha z,\beta y}^{(2)}  \\
E_{\alpha z,\beta y}^{(2)}(\mathbf{\tilde{R}}_{2}) &=& -2E_{\beta y,\alpha z}^{(2)} \\
E_{\alpha z,\beta y}^{(2)}(\mathbf{\tilde{R}}_{3}) &=&  E_{\alpha z,\beta y}^{(2)}  \\
E_{\alpha z,\beta y}^{(2)}(\mathbf{\tilde{R}}_{4}) &=&  E_{\beta y,\alpha z}^{(2)}  \\
E_{\alpha z,\beta y}^{(2)}(\mathbf{\tilde{R}}_{5}) &=& -2E_{\alpha z,\beta y}^{(2)} \\
E_{\alpha z,\beta y}^{(2)}(\mathbf{\tilde{R}}_{6}) &=&  E_{\beta y,\alpha z}^{(2)}  \\
E_{\alpha z,\beta x}^{(2)}(\mathbf{\tilde{R}}_{1}) &=&  \sqrt{3}E_{\alpha z,\beta y}^{(2)} \\
E_{\alpha z,\beta x}^{(2)}(\mathbf{\tilde{R}}_{2}) &=&  0 \\
E_{\alpha z,\beta x}^{(2)}(\mathbf{\tilde{R}}_{3}) &=& -\sqrt{3}E_{\alpha z,\beta y}^{(2)} \\
E_{\alpha z,\beta x}^{(2)}(\mathbf{\tilde{R}}_{4}) &=&  \sqrt{3}E_{\beta y,\alpha z}^{(2)}\\
E_{\alpha z,\beta x}^{(2)}(\mathbf{\tilde{R}}_{5}) &=&  0 \\
E_{\alpha z,\beta x}^{(2)}(\mathbf{\tilde{R}}_{6}) &=& -\sqrt{3}E_{\beta y,\alpha z}^{(2)}\\
E_{\alpha y,\beta y}^{(2)}(\mathbf{\tilde{R}}_{1}) &=& E_{\alpha y,\beta y}^{(2)} \\
E_{\alpha y,\beta y}^{(2)}(\mathbf{\tilde{R}}_{2}) &=& E_{\beta y,\alpha y}^{(2)}+\sqrt{3}E_{\beta y,\alpha x}^{(2)} \\
E_{\alpha y,\beta y}^{(2)}(\mathbf{\tilde{R}}_{3}) &=& E_{\alpha y,\beta y}^{(2)} \\
E_{\alpha y,\beta y}^{(2)}(\mathbf{\tilde{R}}_{4}) &=& E_{\alpha y,\beta y}^{(2)} \\
E_{\alpha y,\beta y}^{(2)}(\mathbf{\tilde{R}}_{5}) &=& E_{\beta y,\alpha y}^{(2)}+\sqrt{3}E_{\beta y,\alpha x}^{(2)} \\
E_{\alpha y,\beta y}^{(2)}(\mathbf{\tilde{R}}_{6}) &=& E_{\alpha y,\beta y}^{(2)} \\
E_{\alpha y,\beta x}^{(2)}(\mathbf{\tilde{R}}_{1}) &=& E_{\alpha y,\beta x}^{(2)}\\
E_{\alpha y,\beta x}^{(2)}(\mathbf{\tilde{R}}_{2}) &=& 0 \\
E_{\alpha y,\beta x}^{(2)}(\mathbf{\tilde{R}}_{3}) &=& -E_{\alpha y,\beta x}^{(2)}\\
E_{\alpha y,\beta x}^{(2)}(\mathbf{\tilde{R}}_{4}) &=& E_{\alpha y,\beta x}^{(2)}\\
E_{\alpha y,\beta x}^{(2)}(\mathbf{\tilde{R}}_{5}) &=& 0 \\
E_{\alpha y,\beta x}^{(2)}(\mathbf{\tilde{R}}_{6}) &=& -E_{\alpha y,\beta x}^{(2)}\\
E_{\alpha x,\beta y}^{(2)}(\mathbf{\tilde{R}}_{1}) &=& E_{\alpha y,\beta x}^{(2)}\\
E_{\alpha x,\beta y}^{(2)}(\mathbf{\tilde{R}}_{2}) &=& 0 \\
E_{\alpha x,\beta y}^{(2)}(\mathbf{\tilde{R}}_{3}) &=& -E_{\alpha y,\beta x}^{(2)}\\
E_{\alpha x,\beta y}^{(2)}(\mathbf{\tilde{R}}_{4}) &=& E_{\alpha y,\beta x}^{(2)}\\
E_{\alpha x,\beta y}^{(2)}(\mathbf{\tilde{R}}_{5}) &=& 0 \\
E_{\alpha x,\beta y}^{(2)}(\mathbf{\tilde{R}}_{6}) &=& -E_{\alpha y,\beta x}^{(2)}\\
\label{Hxx2}
E_{\alpha x,\beta x}^{(2)}(\mathbf{\tilde{R}}_{1}) &=& \frac{2}{\sqrt{3}}E_{\alpha y,\beta x}^{(2)}+E_{\alpha y,\beta y}^{(2)} \\
E_{\alpha x,\beta x}^{(2)}(\mathbf{\tilde{R}}_{2}) &=& E_{\alpha y,\beta y}^{(2)}-\frac{\sqrt{3}}{3}E_{\alpha y,\beta x}^{(2)} \\
E_{\alpha x,\beta x}^{(2)}(\mathbf{\tilde{R}}_{3}) &=& \frac{2}{\sqrt{3}}E_{\alpha y,\beta x}^{(2)}+E_{\alpha y,\beta y}^{(2)} \\
E_{\alpha x,\beta x}^{(2)}(\mathbf{\tilde{R}}_{4}) &=& \frac{2}{\sqrt{3}}E_{\alpha y,\beta x}^{(2)}+E_{\alpha y,\beta y}^{(2)} \\
E_{\alpha x,\beta x}^{(2)}(\mathbf{\tilde{R}}_{5}) &=& E_{\alpha y,\beta y}^{(2)}-\frac{\sqrt{3}}{3}E_{\alpha y,\beta x}^{(2)} \\
E_{\alpha x,\beta x}^{(2)}(\mathbf{\tilde{R}}_{6}) &=& \frac{2}{\sqrt{3}}E_{\alpha y,\beta x}^{(2)}+E_{\alpha y,\beta y}^{(2)},
\end{eqnarray}
where $\alpha$, $\beta$ = $u$ or $l$. 
Here the energy integrals are defined as
\begin{equation}
E_{\alpha i, \beta j}^{(2)} = E_{\alpha i, \beta j}(\mathbf{\tilde{R}}_{1}) = \langle \alpha p_{i}(\mathbf{0}) | \mathcal{H} | \beta p_{j} (\mathbf{\tilde{R}}_{1}) \rangle 
\end{equation}
The TB model $\mathcal{H}_{\alpha i, \beta j}^{(2)}$ for second nearest-neighbors of $p$-orbitals includes 19 independent parameters $E_{uz, uz}^{(2)}$, $E_{uz, lz}^{(2)}$, $E_{lz, lz}^{(2)}$, $E_{\alpha z,\beta y}^{(2)}$, $E_{\alpha y,\beta z}^{(2)}$, $E_{\alpha y,\beta y}^{(2)}$, and $E_{\alpha y,\beta x}^{(2)}$, where $\alpha$, $\beta$ = $u$ or $l$. 

\subsubsection{The third nearest-neighbors}
The TB model $\mathcal{H}_{\alpha i, \beta j}^{(3)}$ for third nearest-neighboring $p$-orbitals shares the same forms of Eqs.~(\ref{Eq:Hzz1})-(\ref{Eq:Hxbx1}), except for the fact that energy integral parameters $E_{\alpha i, \beta j}^{(1)}$ are replaced by $E_{\alpha i, \beta j}^{(3)}$, which 
\begin{equation}
E_{\alpha i, \beta j}^{(3)} = E_{\alpha i, \beta j}(2\mathbf{R}_{1}) = \langle \alpha p_{i}(\mathbf{0}) | \mathcal{H} | \beta p_{j} (2\mathbf{R}_{1})\rangle.    
\end{equation}
Here independent TB parameters $E_{\alpha i, \beta j}^{(3)}$ are $E_{uz,uz}^{(3)}$, $E_{lz,lz}^{(3)}$, $E_{uz,lz}^{(3)}$, $E_{uz,uy}^{(3)}$, $E_{uz,ux}^{(3)}$, $E_{uz,ly}^{(3)}$, $E_{uz,lx}^{(3)}$, $E_{lz,uy}^{(3)}$, $E_{lz,ux}^{(3)}$, $E_{lz,ly}^{(3)}$, $E_{lz,lx}^{(3)}$, $E_{uy,uy}^{(3)}$, $E_{ly,ly}^{(3)}$, $E_{ux,ux}^{(3)}$, $E_{lx,lx}^{(3)}$, $E_{uy,ux}^{(3)}$, $E_{ly,lx}^{(3)}$, $E_{uy,ly}^{(3)}$, $E_{ux,lx}^{(3)}$, $E_{uy,lx}^{(3)}$, and $E_{ly,ux}^{(3)}$, which are 21 in total.   

\subsection{\label{Appdenix_TBdp}Energy integrals between $d$ and $p$ orbitals}
As shown in Fig.~\ref{Fig:neighbors}(c), there are six nearest neighbors, six second nearest neighbors, and twelve third nearest neighbors between transition metal $d$ orbitals and chalcogen $p$ orbitals by using distances between the transition metal and chalcogen atoms on the $xy$ plane. In fact, since distance between the transiton metal plance and the chalcogen sublayer slightly differs depending on whether upper or lower chalcogen sublayer, aforementioned classification of nearest neighbors can be further categorized when the distance along the $z$ direction is included. Instead of doing that, our classification on nearest neighbors is based on distances on the $xy$ plane. As done in nearest neighbors among $p$-orbitals, nearest neighboring vector components along the $z$ axis is suppressed in the following energy integrals. The sublayer where $p$-orbital belongs is distinguished by chalcogen sublayer indices $\alpha=u$ (upper) or $l$ (lower) in front of $p$-orbital indices. 

\subsubsection{The nearest-neighbors}
Three nearest neighboring vectors between $d$ and $p$ orbitals on the $xy$ plane are written as $\boldsymbol{\delta}_{1}=\frac{2}{3}\vec{a}_{1}+\frac{1}{3}\vec{a}_{2}$, $\boldsymbol{\delta}_{2}=-\frac{1}{3}\vec{a}_{1}+\frac{1}{3}\vec{a}_{2}$, and $\boldsymbol{\delta}_{3}=-\frac{1}{3}\vec{a}_{1}-\frac{2}{3}\vec{a}_{2}$. Energy integrals for nearest-neighboring transition metal $d$-orbitals and chalcogen atom $p$-orbitals are summarized as follows:
\begin{eqnarray}
E_{\alpha z,z^{2}}^{(1)}(\boldsymbol{\delta}_{1}) &=& E_{\alpha z,z^{2}}^{(1)} \\
E_{\alpha z,z^{2}}^{(1)}(\boldsymbol{\delta}_{2}) &=& E_{\alpha z,z^{2}}^{(1)} \\
E_{\alpha z,z^{2}}^{(1)}(\boldsymbol{\delta}_{3}) &=& E_{\alpha z,z^{2}}^{(1)} \\
\label{Hz_x^2-y^2}
E_{\alpha z,x^{2}\textrm{-}y^{2}}^{(1)}(\boldsymbol{\delta}_{1}) &=& \frac{1}{\sqrt{3}}E_{\alpha z,xy}^{(1)}\\
E_{\alpha z,x^{2}\textrm{-}y^{2}}^{(1)}(\boldsymbol{\delta}_{2}) &=& \frac{1}{\sqrt{3}}E_{\alpha z,xy}^{(1)}\\
E_{\alpha z,x^{2}\textrm{-}y^{2}}^{(1)}(\boldsymbol{\delta}_{3}) &=& -2\frac{1}{\sqrt{3}}E_{\alpha z,xy}^{(1)}\\
E_{\alpha z, xy}^{(1)}(\boldsymbol{\delta}_{1}) &=&  E_{\alpha z,xy}^{(1)}\\
E_{\alpha z, xy}^{(1)}(\boldsymbol{\delta}_{2}) &=& -E_{\alpha z,xy}^{(1)}\\
E_{\alpha z, xy}^{(1)}(\boldsymbol{\delta}_{3}) &=&  0 \\
E_{\alpha z,yz}^{(1)}(\boldsymbol{\delta}_{1}) &=& \frac{1}{\sqrt{3}}E_{\alpha z,zx}^{(1)}\\ 
E_{\alpha z,yz}^{(1)}(\boldsymbol{\delta}_{2}) &=& \frac{1}{\sqrt{3}}E_{\alpha z,zx}^{(1)}\\ 
E_{\alpha z,yz}^{(1)}(\boldsymbol{\delta}_{3}) &=& -\frac{2}{\sqrt{3}}E_{\alpha z,zx}^{(1)}\\ 
E_{\alpha z, zx}^{(1)}(\boldsymbol{\delta}_{1}) &=&  E_{\alpha z,zx}^{(1)}\\
E_{\alpha z, zx}^{(1)}(\boldsymbol{\delta}_{2}) &=& -E_{\alpha z,zx}^{(1)}\\
E_{\alpha z, zx}^{(1)}(\boldsymbol{\delta}_{3}) &=&  0 \\
E_{\alpha y, z^2}^{(1)}(\boldsymbol{\delta}_{1}) &=& E_{\alpha y,z^{2}}^{(1)}\\
E_{\alpha y, z^2}^{(1)}(\boldsymbol{\delta}_{2}) &=& E_{\alpha y,z^{2}}^{(1)}\\
E_{\alpha y, z^2}^{(1)}(\boldsymbol{\delta}_{3}) &=& -2E_{\alpha y,z^{2}}^{(1)}\\
E_{\alpha x, z^2}^{(1)}(\boldsymbol{\delta}_{1}) &=&  \sqrt{3}E_{\alpha y,z^{2}}^{(1)}\\
E_{\alpha x, z^2}^{(1)}(\boldsymbol{\delta}_{2}) &=& -\sqrt{3}E_{\alpha y,z^{2}}^{(1)}\\
E_{\alpha x, z^2}^{(1)}(\boldsymbol{\delta}_{3}) &=&  0\\
E_{\alpha y, x^2-y^2}^{(1)}(\boldsymbol{\delta}_{1}) &=& E_{\alpha y,x^{2}-y^{2}}^{(1)}\\
E_{\alpha y, x^2-y^2}^{(1)}(\boldsymbol{\delta}_{2}) &=& E_{\alpha y,x^{2}-y^{2}}^{(1)}\\
E_{\alpha y, x^2-y^2}^{(1)}(\boldsymbol{\delta}_{3}) &=& -\frac{1}{2}E_{\alpha y,x^{2}-y^{2}}^{(1)}+\frac{3}{2}E_{\alpha x,xy}^{(1)}\\
E_{\alpha y, xy}^{(1)}(\boldsymbol{\delta}_{1}) &=& -\frac{\sqrt{3}}{2}E_{\alpha y,x^{2}-y^{2}}^{(1)}+\frac{\sqrt{3}}{2}E_{\alpha x,xy}^{(1)}\nonumber\\\\
E_{\alpha y, xy}^{(1)}(\boldsymbol{\delta}_{2}) &=& \frac{\sqrt{3}}{2}E_{\alpha y,x^{2}-y^{2}}^{(1)}-\frac{\sqrt{3}}{2}E_{\alpha x,xy}^{(1)}\nonumber\\\\
E_{\alpha y, xy}^{(1)}(\boldsymbol{\delta}_{3}) &=& 0 \\
E_{\alpha x, xy}^{(1)}(\boldsymbol{\delta}_{1}) &=& E_{\alpha x,xy}^{(1)}\\
E_{\alpha x, xy}^{(1)}(\boldsymbol{\delta}_{2}) &=& E_{\alpha x,xy}^{(1)}\\
E_{\alpha x, xy}^{(1)}(\boldsymbol{\delta}_{3}) &=& -\frac{1}{2}E_{\alpha x,xy}^{(1)}+\frac{3}{2}E_{\alpha y,x^{2}-y^{2}}^{(1)}\\
E_{\alpha x, x^2-y^2}^{(1)}(\boldsymbol{\delta}_{1}) &=& -\frac{\sqrt{3}}{2}E_{\alpha y,x^{2}-y^{2}}^{(1)}+\frac{\sqrt{3}}{2}E_{\alpha x,xy}^{(1)}\nonumber\\\\
E_{\alpha x, x^2-y^2}^{(1)}(\boldsymbol{\delta}_{2}) &=& \frac{\sqrt{3}}{2}E_{\alpha y,x^{2}-y^{2}}^{(1)}-\frac{\sqrt{3}}{2}E_{\alpha x,xy}^{(1)}\nonumber\\\\
E_{\alpha x, x^2-y^2}^{(1)}(\boldsymbol{\delta}_{3}) &=& 0 \\
E_{\alpha y, yz}^{(1)}(\boldsymbol{\delta}_{1}) &=& E_{\alpha y,yz}^{(1)}\\
E_{\alpha y, yz}^{(1)}(\boldsymbol{\delta}_{2}) &=& E_{\alpha y,yz}^{(1)}\\
E_{\alpha y, yz}^{(1)}(\boldsymbol{\delta}_{3}) &=& -\frac{1}{2}E_{\alpha y,yz}^{(1)}+\frac{3}{2}E_{\alpha x,zx}^{(1)}\\
E_{\alpha x, zx}^{(1)}(\boldsymbol{\delta}_{1}) &=& E_{\alpha x,zx}^{(1)}\\
E_{\alpha x, zx}^{(1)}(\boldsymbol{\delta}_{2}) &=& E_{\alpha x,zx}^{(1)}\\
E_{\alpha x, zx}^{(1)}(\boldsymbol{\delta}_{3}) &=& -\frac{1}{2}E_{\alpha x,zx}^{(1)}+\frac{3}{2}E_{\alpha y,yz}^{(1)}\\
E_{\alpha y, zx}^{(1)}(\boldsymbol{\delta}_{1}) &=& -\frac{\sqrt{3}}{2}E_{\alpha y,yz}^{(1)}+\frac{\sqrt{3}}{2}E_{\alpha x,zx}^{(1)}\\
E_{\alpha y, zx}^{(1)}(\boldsymbol{\delta}_{2}) &=& \frac{\sqrt{3}}{2}E_{\alpha y,yz}^{(1)}-\frac{\sqrt{3}}{2}E_{\alpha x,zx}^{(1)}\\
E_{\alpha y, zx}^{(1)}(\boldsymbol{\delta}_{3}) &=& 0\\
E_{\alpha x, yz}^{(1)}(\boldsymbol{\delta}_{1}) &=& -\frac{\sqrt{3}}{2}E_{\alpha y,yz}^{(1)}+\frac{\sqrt{3}}{2}E_{\alpha x,zx}^{(1)}\\
E_{\alpha x, yz}^{(1)}(\boldsymbol{\delta}_{2}) &=& \frac{\sqrt{3}}{2}E_{\alpha y,yz}^{(1)}-\frac{\sqrt{3}}{2}E_{\alpha x,zx}^{(1)}\\
E_{\alpha x, yz}^{(1)}(\boldsymbol{\delta}_{3}) &=& 0
\end{eqnarray}
Here energy integrals between $p$-orbitals $i$ at origin and $d$-orbitals of nearest neighbors are defined as 
\begin{equation}
E_{\alpha i, j}^{(1)} = E_{\alpha i, j}(\boldsymbol{\delta}_{1}) = \langle \alpha p_{i}(\mathbf{0}) | \mathcal{H} | d_{j} (\boldsymbol{\delta}_{1})\rangle,    
\end{equation}
where $\alpha=u$ or $l$. 
Here independent energy integrals are $E_{\alpha z,z^{2}}^{(1)}$, $E_{\alpha z,xy}^{(1)}$, $E_{\alpha z,zx}^{(1)}$, $E_{\alpha y,z^{2}}^{(1)}$, $E_{\alpha y,x^{2}-y^{2}}^{(1)}$, $E_{\alpha x,xy}^{(1)}$, $E_{\alpha y,yz}^{(1)}$, $E_{\alpha x,zx}^{(1)}$, for $\alpha$ = $u$ or $l$, which are sixteen in total. 

\subsubsection{The second nearest-neighbors}
The second nearest-neighboring vectors on the $xy$ plane are $\boldsymbol{\tilde{\delta}}_{1}=-\frac{4}{3}\vec{a}_{1}-\frac{2}{3}\vec{a}_{2}$, $\boldsymbol{\tilde{\delta}}_{2}=\frac{2}{3}\vec{a}_{1}-\frac{2}{3}\vec{a}_{2}$, and $\boldsymbol{\tilde{\delta}}_{3}=\frac{2}{3}\vec{a}_{1}+\frac{4}{3}\vec{a}_{2}$. 
Energy integrals for second nearest-neighbors between $p$-orbitals and $d$-orbitals $j$ are derived as follows:
\begin{eqnarray}
\label{Hz_z^2_2}
E_{\alpha z, z^2}^{(2)}(\boldsymbol{\tilde{\delta}}_{1}) &=& E_{\alpha z,z^{2}}^{(2)} \\
E_{\alpha z, z^2}^{(2)}(\boldsymbol{\tilde{\delta}}_{2}) &=& E_{\alpha z,z^{2}}^{(2)} \\
E_{\alpha z, z^2}^{(2)}(\boldsymbol{\tilde{\delta}}_{3}) &=& E_{\alpha z,z^{2}}^{(2)} \\
E_{\alpha z, x^2\textrm{-}y^2}^{(2)}(\boldsymbol{\tilde{\delta}}_{1}) &=& \frac{1}{\sqrt{3}}E_{\alpha z,xy}^{(2)}\\
E_{\alpha z, x^2\textrm{-}y^2}^{(2)}(\boldsymbol{\tilde{\delta}}_{2}) &=& \frac{1}{\sqrt{3}}E_{\alpha z,xy}^{(2)}\\
E_{\alpha z, x^2\textrm{-}y^2}^{(2)}(\boldsymbol{\tilde{\delta}}_{3}) &=& -\frac{2}{\sqrt{3}}E_{\alpha z,xy}^{(2)}\\
E_{\alpha z, xy}^{(2)}(\boldsymbol{\tilde{\delta}}_{1}) &=& E_{\alpha z,xy}^{(2)}\\
E_{\alpha z, xy}^{(2)}(\boldsymbol{\tilde{\delta}}_{2}) &=& -E_{\alpha z,xy}^{(2)}\\
E_{\alpha z, xy}^{(2)}(\boldsymbol{\tilde{\delta}}_{3}) &=& 0\\
E_{\alpha z, yz}^{(2)}(\boldsymbol{\tilde{\delta}}_{1}) &=& \frac{1}{\sqrt{3}}E_{\alpha z,yz}^{(2)}\\
E_{\alpha z, yz}^{(2)}(\boldsymbol{\tilde{\delta}}_{2}) &=& \frac{1}{\sqrt{3}}E_{\alpha z,yz}^{(2)}\\
E_{\alpha z, yz}^{(2)}(\boldsymbol{\tilde{\delta}}_{3}) &=& -\frac{2}{\sqrt{3}}E_{\alpha z,yz}^{(2)}\\
E_{\alpha z, zx}^{(2)}(\boldsymbol{\tilde{\delta}}_{1}) &=& E_{\alpha z,zx}^{(2)}\\
E_{\alpha z, zx}^{(2)}(\boldsymbol{\tilde{\delta}}_{2}) &=& -E_{\alpha z,zx}^{(2)}\\
E_{\alpha z, zx}^{(2)}(\boldsymbol{\tilde{\delta}}_{3}) &=& 0\\
E_{\alpha y, z^2}^{(2)}(\boldsymbol{\tilde{\delta}}_{1}) &=& E_{\alpha y,z^{2}}^{(2)}\\
E_{\alpha y, z^2}^{(2)}(\boldsymbol{\tilde{\delta}}_{2}) &=& E_{\alpha y,z^{2}}^{(2)}\\
E_{\alpha y, z^2}^{(2)}(\boldsymbol{\tilde{\delta}}_{3}) &=& -2E_{\alpha y,z^{2}}^{(2)}\\
E_{\alpha x, z^2}^{(2)}(\boldsymbol{\tilde{\delta}}_{1}) &=& \sqrt{3}E_{\alpha y,z^{2}}^{(2)}\\
E_{\alpha x, z^2}^{(2)}(\boldsymbol{\tilde{\delta}}_{2}) &=& -\sqrt{3}E_{\alpha y,z^{2}}^{(2)}\\
E_{\alpha x, z^2}^{(2)}(\boldsymbol{\tilde{\delta}}_{3}) &=& 0\\
E_{\alpha y, x^2\textrm{-}y^2}^{(2)}(\boldsymbol{\tilde{\delta}}_{1}) &=& E_{\alpha y,x^{2}-y^{2}}^{(2)}\\
E_{\alpha y, x^2\textrm{-}y^2}^{(2)}(\boldsymbol{\tilde{\delta}}_{2}) &=& E_{\alpha y,x^{2}-y^{2}}^{(2)}\\
E_{\alpha y, x^2\textrm{-}y^2}^{(2)}(\boldsymbol{\tilde{\delta}}_{3}) &=& -\frac{1}{2}E_{\alpha y,x^{2}-y^{2}}^{(2)}+\frac{3}{2}E_{\alpha x,xy}^{(2)}\\
E_{\alpha x, xy}^{(2)}(\boldsymbol{\tilde{\delta}}_{1}) &=& E_{\alpha y,x^{2}-y^{2}}^{(2)}\\
E_{\alpha x, xy}^{(2)}(\boldsymbol{\tilde{\delta}}_{2}) &=& E_{\alpha y,x^{2}-y^{2}}^{(2)}\\
E_{\alpha x, xy}^{(2)}(\boldsymbol{\tilde{\delta}}_{3}) &=& -\frac{1}{2}E_{\alpha y,x^{2}-y^{2}}^{(2)}+\frac{3}{2}E_{\alpha x,xy}^{(2)}\\
E_{\alpha y, xy}^{(2)}(\boldsymbol{\tilde{\delta}}_{1}) &=& -\frac{\sqrt{3}}{2}E_{\alpha y,x^{2}-y^{2}}^{(2)}+\frac{\sqrt{3}}{2}E_{\alpha x,xy}^{(2)}\nonumber\\\\
E_{\alpha y, xy}^{(2)}(\boldsymbol{\tilde{\delta}}_{2}) &=& \frac{\sqrt{3}}{2}E_{\alpha y,x^{2}-y^{2}}^{(2)}-\frac{\sqrt{3}}{2}E_{\alpha x,xy}^{(2)}\nonumber\\\\
E_{\alpha y, xy}^{(2)}(\boldsymbol{\tilde{\delta}}_{3}) &=& 0\\
E_{\alpha x, x^2-y^2}^{(2)}(\boldsymbol{\tilde{\delta}}_{1}) &=& -\frac{\sqrt{3}}{2}E_{\alpha y,x^{2}-y^{2}}^{(2)}+\frac{\sqrt{3}}{2}E_{\alpha x,xy}^{(2)}\nonumber\\\\
E_{\alpha x, x^2-y^2}^{(2)}(\boldsymbol{\tilde{\delta}}_{2}) &=& \frac{\sqrt{3}}{2}E_{\alpha y,x^{2}-y^{2}}^{(2)}-\frac{\sqrt{3}}{2}E_{\alpha x,xy}^{(2)}\nonumber\\\\
E_{\alpha x, x^2-y^2}^{(2)}(\boldsymbol{\tilde{\delta}}_{3}) &=& 0\\
E_{\alpha y, yz}^{(2)}(\boldsymbol{\tilde{\delta}}_{1}) &=& E_{\alpha y,yz}^{(2)}\\
E_{\alpha y, yz}^{(2)}(\boldsymbol{\tilde{\delta}}_{2}) &=& E_{\alpha y,yz}^{(2)}\\
E_{\alpha y, yz}^{(2)}(\boldsymbol{\tilde{\delta}}_{3}) &=& -\frac{1}{2}E_{\alpha y,yz}^{(2)}+\frac{3}{2}E_{\alpha x,zx}^{(2)}\\
E_{\alpha x, zx}^{(2)}(\boldsymbol{\tilde{\delta}}_{1}) &=& E_{\alpha x,zx}^{(2)}\\
E_{\alpha x, zx}^{(2)}(\boldsymbol{\tilde{\delta}}_{2}) &=& E_{\alpha x,zx}^{(2)}\\
E_{\alpha x, zx}^{(2)}(\boldsymbol{\tilde{\delta}}_{3}) &=& -\frac{1}{2}E_{\alpha x,zx}^{(2)}+\frac{3}{2}E_{\alpha y,yz}^{(2)}\\
E_{\alpha y, zx}^{(2)}(\boldsymbol{\tilde{\delta}}_{1}) &=& -\frac{\sqrt{3}}{2}E_{\alpha y,yz}^{(2)}+\frac{\sqrt{3}}{2}E_{\alpha x,zx}^{(2)}\\
E_{\alpha y, zx}^{(2)}(\boldsymbol{\tilde{\delta}}_{2}) &=& \frac{\sqrt{3}}{2}E_{\alpha y,yz}^{(2)}-\frac{\sqrt{3}}{2}E_{\alpha x,zx}^{(2)}\\
E_{\alpha y, zx}^{(2)}(\boldsymbol{\tilde{\delta}}_{3}) &=& 0\\
E_{\alpha x, yz}^{(2)}(\boldsymbol{\tilde{\delta}}_{1}) &=& -\frac{\sqrt{3}}{2}E_{\alpha y,yz}^{(2)}+\frac{\sqrt{3}}{2}E_{\alpha x,zx}^{(2)}\\
E_{\alpha x, yz}^{(2)}(\boldsymbol{\tilde{\delta}}_{2}) &=& \frac{\sqrt{3}}{2}E_{\alpha y,yz}^{(2)}-\frac{\sqrt{3}}{2}E_{\alpha x,zx}^{(2)}\\
E_{\alpha x, yz}^{(2)}(\boldsymbol{\tilde{\delta}}_{3}) &=& 0
\end{eqnarray}
Here TB parameters $E_{\alpha i, j}^{(2)}$ is defined as the following energy integrals,
\begin{equation}
E_{\alpha i, j}^{(2)} \equiv E_{\alpha i, j}(\boldsymbol{\tilde{\delta}}_{1}) = \langle \alpha p_{i} (\mathbf{0})|\mathcal{H}| d_{j}(\boldsymbol{\tilde{\delta}}_{1})\rangle.    
\end{equation}
The independent energy integrals are $E_{\alpha z,z^{2}}^{(2)}$, $E_{\alpha z, xy}^{(2)}$, $E_{\alpha z,zx}^{(2)}$, $E_{\alpha y,z^{2}}^{(2)}$, $E_{\alpha y, x^2-y^2}^{(2)}$, $E_{\alpha x, xy}^{(2)}$, $E_{\alpha y, yz}^{(2)}$, and $E_{\alpha x, zx}^{(2)}$ for $\alpha$ = $u$ or $l$, which are 16 in total. 

\subsubsection{The third nearest-neighbors}
The third nearest-neighboring vectors on the $xy$ plane are $\bar{\boldsymbol{\delta}}_{1}=\frac{5}{3}\vec{a}_{1}+\frac{1}{3}\vec{a}_{2}$, $\bar{\boldsymbol{\delta}}_{2}=\frac{5}{3}\vec{a}_{1}+\frac{4}{3}\vec{a}_{2}$, $\bar{\boldsymbol{\delta}}_{3}=-\frac{1}{3}\vec{a}_{1}+\frac{4}{3}\vec{a}_{2}$, $\bar{\boldsymbol{\delta}}_{4}=-\frac{4}{3}\vec{a}_{1}+\frac{1}{3}\vec{a}_{2}$, $\bar{\boldsymbol{\delta}}_{5}=-\frac{4}{3}\vec{a}_{1}-\frac{5}{3}\vec{a}_{2}$, and $\bar{\boldsymbol{\delta}}_{6}=-\frac{1}{3}\vec{a}_{1}-\frac{5}{3}\vec{a}_{2}$. 
Energy integrals of third nearest-neighbors between $p$-orbitals $i$ on the $\alpha$ chalcogen layer and $d$-orbitals $j$ reads
\begin{eqnarray}
\label{Hz_z^2_3}
E_{\alpha z,z^2}^{(3)}(\bar{\boldsymbol{\delta}}_{i}) &=& E_{\alpha z,z^{2}}^{(3)} \textrm{ for }i=1,\cdots,6\\
E_{\alpha z, x^2-y^2}^{(3)}(\bar{\boldsymbol{\delta}}_{1}) &=& E_{\alpha z,x^{2}-y^{2}}^{(3)}\\
E_{\alpha z, x^2-y^2}^{(3)}(\bar{\boldsymbol{\delta}}_{2}) &=& -\frac{1}{2}E_{\alpha z,x^{2}-y^{2}}^{(3)}+\frac{\sqrt{3}}{2}E_{\alpha z,xy}^{(3)}\\
E_{\alpha z, x^2-y^2}^{(3)}(\bar{\boldsymbol{\delta}}_{3}) &=& -\frac{1}{2}E_{\alpha z,x^{2}-y^{2}}^{(3)}+\frac{\sqrt{3}}{2}E_{\alpha z,xy}^{(3)}\\
E_{\alpha z, x^2-y^2}^{(3)}(\bar{\boldsymbol{\delta}}_{4}) &=& E_{\alpha z,x^{2}-y^{2}}^{(3)}\\
E_{\alpha z, x^2-y^2}^{(3)}(\bar{\boldsymbol{\delta}}_{5}) &=& -\frac{1}{2}E_{\alpha z,x^{2}-y^{2}}^{(3)}-\frac{\sqrt{3}}{2}E_{\alpha z,xy}^{(3)}\\
E_{\alpha z, x^2-y^2}^{(3)}(\bar{\boldsymbol{\delta}}_{6}) &=& -\frac{1}{2}E_{\alpha z,x^{2}-y^{2}}^{(3)}-\frac{\sqrt{3}}{2}E_{\alpha z,xy}^{(3)}\\
E_{\alpha z, xy}^{(3)}(\bar{\boldsymbol{\delta}}_{1}) &=&  E_{\alpha z,xy}^{(3)} \\
E_{\alpha z, xy}^{(3)}(\bar{\boldsymbol{\delta}}_{2}) &=& \frac{\sqrt{3}}{2}E_{\alpha z,x^{2}-y^{2}}^{(3)}+\frac{1}{2}E_{\alpha z,xy}^{(3)}\\
E_{\alpha z, xy}^{(3)}(\bar{\boldsymbol{\delta}}_{3}) &=& -\frac{\sqrt{3}}{2}E_{\alpha z,x^{2}-y^{2}}^{(3)}-\frac{1}{2}E_{\alpha z,xy}^{(3)}\\
E_{\alpha z, xy}^{(3)}(\bar{\boldsymbol{\delta}}_{4}) &=& -E_{\alpha z,xy}^{(3)} \\
E_{\alpha z, xy}^{(3)}(\bar{\boldsymbol{\delta}}_{5}) &=& \frac{\sqrt{3}}{2}E_{\alpha z,x^{2}-y^{2}}^{(3)}-\frac{1}{2}E_{\alpha z,xy}^{(3)}\\
E_{\alpha z, xy}^{(3)}(\bar{\boldsymbol{\delta}}_{6}) &=& -\frac{\sqrt{3}}{2}E_{\alpha z,x^{2}-y^{2}}^{(3)}+\frac{1}{2}E_{\alpha z,xy}^{(3)}\\
E_{\alpha z, yz}^{(3)}(\bar{\boldsymbol{\delta}}_{1}) &=& E_{\alpha z,yz}^{(3)}\\
E_{\alpha z, yz}^{(3)}(\bar{\boldsymbol{\delta}}_{2}) &=& -\frac{1}{2}E_{\alpha z,yz}^{(3)}+\frac{\sqrt{3}}{2}E_{\alpha z,zx}^{(3)}\\
E_{\alpha z, yz}^{(3)}(\bar{\boldsymbol{\delta}}_{3}) &=& -\frac{1}{2}E_{\alpha z,yz}^{(3)}+\frac{\sqrt{3}}{2}E_{\alpha z,zx}^{(3)}\\
E_{\alpha z, yz}^{(3)}(\bar{\boldsymbol{\delta}}_{4}) &=& E_{\alpha z,yz}^{(3)}\\
E_{\alpha z, yz}^{(3)}(\bar{\boldsymbol{\delta}}_{5}) &=& -\frac{1}{2}E_{\alpha z,yz}^{(3)}-\frac{\sqrt{3}}{2}E_{\alpha z,zx}^{(3)}\\
E_{\alpha z, yz}^{(3)}(\bar{\boldsymbol{\delta}}_{6}) &=& -\frac{1}{2}E_{\alpha z,yz}^{(3)}-\frac{\sqrt{3}}{2}E_{\alpha z,zx}^{(3)}\\
E_{\alpha z, zx}^{(3)}(\bar{\boldsymbol{\delta}}_{1}) &=& E_{\alpha z,zx}^{(3)}\\
E_{\alpha z, zx}^{(3)}(\bar{\boldsymbol{\delta}}_{2}) &=& \frac{\sqrt{3}}{2}E_{\alpha z,yz}^{(3)}+\frac{1}{2}E_{\alpha z,zx}^{(3)}\\
E_{\alpha z, zx}^{(3)}(\bar{\boldsymbol{\delta}}_{3}) &=& -\frac{\sqrt{3}}{2}E_{\alpha z,yz}^{(3)}-\frac{1}{2}E_{\alpha z,zx}^{(3)}\\
E_{\alpha z, zx}^{(3)}(\bar{\boldsymbol{\delta}}_{4}) &=& -E_{\alpha z,zx}^{(3)}\\
E_{\alpha z, zx}^{(3)}(\bar{\boldsymbol{\delta}}_{5}) &=& \frac{\sqrt{3}}{2}E_{\alpha z,yz}^{(3)}-\frac{1}{2}E_{\alpha z,zx}^{(3)}\\
E_{\alpha z, zx}^{(3)}(\bar{\boldsymbol{\delta}}_{6}) &=& -\frac{\sqrt{3}}{2}E_{\alpha z,yz}^{(3)}+\frac{1}{2}E_{\alpha z,zx}^{(3)}\\
E_{\alpha y, z^2}^{(3)}(\bar{\boldsymbol{\delta}}_{1}) &=& E_{\alpha y,z^{2}}^{(3)}\\
E_{\alpha y, z^2}^{(3)}(\bar{\boldsymbol{\delta}}_{2}) &=& -\frac{1}{2}E_{\alpha y,z^{2}}^{(3)}+\frac{\sqrt{3}}{2}E_{\alpha x,z^{2}}^{(3)}\\
E_{\alpha y, z^2}^{(3)}(\bar{\boldsymbol{\delta}}_{3}) &=& -\frac{1}{2}E_{\alpha y,z^{2}}^{(3)}+\frac{\sqrt{3}}{2}E_{\alpha x,z^{2}}^{(3)}\\
E_{\alpha y, z^2}^{(3)}(\bar{\boldsymbol{\delta}}_{4}) &=& E_{\alpha y,z^{2}}^{(3)}\\
E_{\alpha y, z^2}^{(3)}(\bar{\boldsymbol{\delta}}_{5}) &=& -\frac{1}{2}E_{\alpha y,z^{2}}^{(3)}-\frac{\sqrt{3}}{2}E_{\alpha x,z^{2}}^{(3)}\\
E_{\alpha y, z^2}^{(3)}(\bar{\boldsymbol{\delta}}_{6}) &=& -\frac{1}{2}E_{\alpha y,z^{2}}^{(3)}-\frac{\sqrt{3}}{2}E_{\alpha x,z^{2}}^{(3)}\\
E_{\alpha x, z^2}^{(3)}(\bar{\boldsymbol{\delta}}_{1}) &=& E_{\alpha x,z^{2}}^{(3)}\\
E_{\alpha x, z^2}^{(3)}(\bar{\boldsymbol{\delta}}_{2}) &=& \frac{\sqrt{3}}{2}E_{\alpha y,z^{2}}^{(3)}+\frac{1}{2}E_{\alpha x,z^{2}}^{(3)}\\
E_{\alpha x, z^2}^{(3)}(\bar{\boldsymbol{\delta}}_{3}) &=& -\frac{\sqrt{3}}{2}E_{\alpha y,z^{2}}^{(3)}-\frac{1}{2}E_{\alpha x,z^{2}}^{(3)}\\
E_{\alpha x, z^2}^{(3)}(\bar{\boldsymbol{\delta}}_{4}) &=& -E_{\alpha x,z^{2}}^{(3)}\\
E_{\alpha x, z^2}^{(3)}(\bar{\boldsymbol{\delta}}_{5}) &=& \frac{\sqrt{3}}{2}E_{\alpha y,z^{2}}^{(3)}-\frac{1}{2}E_{\alpha x,z^{2}}^{(3)}\\
E_{\alpha x, z^2}^{(3)}(\bar{\boldsymbol{\delta}}_{6}) &=& -\frac{\sqrt{3}}{2}E_{\alpha y,z^{2}}^{(3)}+\frac{1}{2}E_{\alpha x,z^{2}}^{(3)}\\
E_{\alpha y, x^2-y^2}^{(3)}(\bar{\boldsymbol{\delta}}_{1}) &=& E_{\alpha y,x^{2}-y^{2}}^{(3)}\\
E_{\alpha y, x^2-y^2}^{(3)}(\bar{\boldsymbol{\delta}}_{2}) &=& \frac{1}{4}E_{\alpha y,x^{2}-y^{2}}^{(3)}-\frac{\sqrt{3}}{4}E_{\alpha x,x^{2}-y^{2}}^{(3)}\nonumber\\
&&-\frac{\sqrt{3}}{4}E_{\alpha y,xy}^{(3)}+\frac{3}{4}E_{\alpha x,xy}^{(3)}\\
E_{\alpha y, x^2-y^2}^{(3)}(\bar{\boldsymbol{\delta}}_{3}) &=& \frac{1}{4}E_{\alpha y,x^{2}-y^{2}}^{(3)}-\frac{\sqrt{3}}{4}E_{\alpha x,x^{2}-y^{2}}^{(3)}\nonumber\\
&&-\frac{\sqrt{3}}{4}E_{\alpha y,xy}^{(3)}+\frac{3}{4}E_{\alpha x,xy}^{(3)}\\
E_{\alpha y, x^2-y^2}^{(3)}(\bar{\boldsymbol{\delta}}_{4}) &=& E_{\alpha y,x^{2}-y^{2}}^{(3)}\\
E_{\alpha y, x^2-y^2}^{(3)}(\bar{\boldsymbol{\delta}}_{5}) &=& \frac{1}{4}E_{\alpha y,x^{2}-y^{2}}^{(3)}+\frac{\sqrt{3}}{4}E_{\alpha x,x^{2}-y^{2}}^{(3)}\nonumber\\
&&+\frac{\sqrt{3}}{4}E_{\alpha y,xy}^{(3)}+\frac{3}{4}E_{\alpha x,xy}^{(3)}\\
E_{\alpha y, x^2-y^2}^{(3)}(\bar{\boldsymbol{\delta}}_{6}) &=& \frac{1}{4}E_{\alpha y,x^{2}-y^{2}}^{(3)}+\frac{\sqrt{3}}{4}E_{\alpha x,x^{2}-y^{2}}^{(3)}\nonumber\\
&&+\frac{\sqrt{3}}{4}E_{\alpha y,xy}^{(3)}+\frac{3}{4}E_{\alpha x,xy}^{(3)}\\
E_{\alpha y, xy}^{(3)}(\bar{\boldsymbol{\delta}}_{1}) &=& E_{\alpha y,xy}^{(3)}\\
E_{\alpha y, xy}^{(3)}(\bar{\boldsymbol{\delta}}_{2}) &=& -\frac{\sqrt{3}}{4}E_{\alpha y,x^{2}-y^{2}}^{(3)}+\frac{3}{4}E_{\alpha x,x^{2}-y^{2}}^{(3)}\nonumber\\
&&-\frac{1}{4}E_{\alpha y,xy}^{(3)}+\frac{\sqrt{3}}{4}E_{\alpha x,xy}^{(3)}\\
E_{\alpha y, xy}^{(3)}(\bar{\boldsymbol{\delta}}_{3}) &=& \frac{\sqrt{3}}{4}E_{\alpha y,x^{2}-y^{2}}^{(3)}-\frac{3}{4}E_{\alpha x,x^{2}-y^{2}}^{(3)}\nonumber\\
&&+\frac{1}{4}E_{\alpha y,xy}^{(3)}-\frac{\sqrt{3}}{4}E_{\alpha x,xy}^{(3)}\\
E_{\alpha y, xy}^{(3)}(\bar{\boldsymbol{\delta}}_{4}) &=& -E_{\alpha y,xy}^{(3)}\\
E_{\alpha y, xy}^{(3)}(\bar{\boldsymbol{\delta}}_{5}) &=& -\frac{\sqrt{3}}{4}E_{\alpha y,x^{2}-y^{2}}^{(3)}-\frac{3}{4}E_{\alpha x,x^{2}-y^{2}}^{(3)}\nonumber\\
&&+\frac{1}{4}E_{\alpha y,xy}^{(3)}+\frac{\sqrt{3}}{4}E_{\alpha x,xy}\\
E_{\alpha y, xy}^{(3)}(\bar{\boldsymbol{\delta}}_{6}) &=& \frac{\sqrt{3}}{4}E_{\alpha y,x^{2}-y^{2}}^{(3)}+\frac{3}{4}E_{\alpha x,x^{2}-y^{2}}^{(3)}\nonumber\\
&&-\frac{1}{4}E_{\alpha y,xy}^{(3)}-\frac{\sqrt{3}}{4}E_{\alpha x,xy}\\
E_{\alpha x, x^2-y^2}^{(3)}(\bar{\boldsymbol{\delta}}_{1}) &=& E_{\alpha x,x^{2}-y^{2}}^{(3)}\\
E_{\alpha x, x^2-y^2}^{(3)}(\bar{\boldsymbol{\delta}}_{2}) &=& -\frac{\sqrt{3}}{4}E_{\alpha y,x^{2}-y^{2}}^{(3)}-\frac{1}{4}E_{\alpha x,x^{2}-y^{2}}^{(3)}\nonumber\\
&&+\frac{3}{4}E_{\alpha y,xy}^{(3)}+\frac{\sqrt{3}}{4}E_{\alpha x,xy}^{(3)}\\
E_{\alpha x, x^2-y^2}^{(3)}(\bar{\boldsymbol{\delta}}_{3}) &=& \frac{\sqrt{3}}{4}E_{\alpha y,x^{2}-y^{2}}^{(3)}+\frac{1}{4}E_{\alpha x,x^{2}-y^{2}}^{(3)}\nonumber\\
&&-\frac{3}{4}E_{\alpha y,xy}^{(3)}-\frac{\sqrt{3}}{4}E_{\alpha x,xy}^{(3)}\\
E_{\alpha x, x^2-y^2}^{(3)}(\bar{\boldsymbol{\delta}}_{4}) &=& -E_{\alpha x,x^{2}-y^{2}}^{(3)}\\
E_{\alpha x, x^2-y^2}^{(3)}(\bar{\boldsymbol{\delta}}_{5}) &=& -\frac{\sqrt{3}}{4}E_{\alpha y,x^{2}-y^{2}}^{(3)}+\frac{1}{4}E_{\alpha x,x^{2}-y^{2}}^{(3)}\nonumber\\
&&-\frac{3}{4}E_{\alpha y,xy}^{(3)}+\frac{\sqrt{3}}{4}E_{\alpha x,xy}^{(3)}\\
E_{\alpha x, x^2-y^2}^{(3)}(\bar{\boldsymbol{\delta}}_{6}) &=& \frac{\sqrt{3}}{4}E_{\alpha y,x^{2}-y^{2}}^{(3)}-\frac{1}{4}E_{\alpha x,x^{2}-y^{2}}^{(3)}\nonumber\\
&&+\frac{3}{4}E_{\alpha y,xy}^{(3)}-\frac{\sqrt{3}}{4}E_{\alpha x,xy}^{(3)}\\
E_{\alpha x, xy}^{(3)}(\bar{\boldsymbol{\delta}}_{1}) &=& E_{\alpha x,xy}^{(3)}\\
E_{\alpha x, xy}^{(3)}(\bar{\boldsymbol{\delta}}_{2}) &=& \frac{3}{4}E_{\alpha y,x^{2}-y^{2}}^{(3)}+\frac{\sqrt{3}}{4}E_{\alpha x,x^{2}-y^{2}}^{(3)}\nonumber\\
&&+\frac{\sqrt{3}}{4}E_{\alpha y,xy}^{(3)}+\frac{1}{4}E_{\alpha x,xy}^{(3)}\\
E_{\alpha x, xy}^{(3)}(\bar{\boldsymbol{\delta}}_{3}) &=& \frac{3}{4}E_{\alpha y,x^{2}-y^{2}}^{(3)}+\frac{\sqrt{3}}{4}E_{\alpha x,x^{2}-y^{2}}^{(3)}\nonumber\\
&&+\frac{\sqrt{3}}{4}E_{\alpha y,xy}^{(3)}+\frac{1}{4}E_{\alpha x,xy}^{(3)}\\
E_{\alpha x, xy}^{(3)}(\bar{\boldsymbol{\delta}}_{4}) &=& E_{\alpha x,xy}^{(3)}\\
E_{\alpha x, xy}^{(3)}(\bar{\boldsymbol{\delta}}_{5}) &=& \frac{3}{4}E_{\alpha y,x^{2}-y^{2}}^{(3)}-\frac{\sqrt{3}}{4}E_{\alpha x,x^{2}-y^{2}}^{(3)}\nonumber\\
&&-\frac{\sqrt{3}}{4}E_{\alpha y,xy}^{(3)}+\frac{1}{4}E_{\alpha x,xy}^{(3)}\\
E_{\alpha x, xy}^{(3)}(\bar{\boldsymbol{\delta}}_{6}) &=& \frac{3}{4}E_{\alpha y,x^{2}-y^{2}}^{(3)}-\frac{\sqrt{3}}{4}E_{\alpha x,x^{2}-y^{2}}^{(3)}\nonumber\\
&&-\frac{\sqrt{3}}{4}E_{\alpha y,xy}^{(3)}+\frac{1}{4}E_{\alpha x,xy}^{(3)}\\
E_{\alpha y, yz}^{(3)}(\bar{\boldsymbol{\delta}}_{1}) &=& E_{\alpha y,yz}\\
E_{\alpha y, yz}^{(3)}(\bar{\boldsymbol{\delta}}_{2}) &=& \frac{1}{4}E_{\alpha y,yz}-\frac{\sqrt{3}}{4}E_{\alpha x,yz}\nonumber\\
&&-\frac{\sqrt{3}}{4}E_{\alpha y,zx}+\frac{3}{4}E_{\alpha x,zx}\\
E_{\alpha y, yz}^{(3)}(\bar{\boldsymbol{\delta}}_{3}) &=& \frac{1}{4}E_{\alpha y,yz}-\frac{\sqrt{3}}{4}E_{\alpha x,yz}\nonumber\\
&&-\frac{\sqrt{3}}{4}E_{\alpha y,zx}+\frac{3}{4}E_{\alpha x,zx}\\
E_{\alpha y, yz}^{(3)}(\bar{\boldsymbol{\delta}}_{4}) &=& E_{\alpha y,yz}\\
E_{\alpha y, yz}^{(3)}(\bar{\boldsymbol{\delta}}_{5}) &=& \frac{1}{4}E_{\alpha y,yz}+\frac{\sqrt{3}}{4}E_{\alpha x,yz}\nonumber\\
&&+\frac{\sqrt{3}}{4}E_{\alpha y,zx}+\frac{3}{4}E_{\alpha x,zx}\\
E_{\alpha y, yz}^{(3)}(\bar{\boldsymbol{\delta}}_{6}) &=& \frac{1}{4}E_{\alpha y,yz}+\frac{\sqrt{3}}{4}E_{\alpha x,yz}\nonumber\\
&&+\frac{\sqrt{3}}{4}E_{\alpha y,zx}+\frac{3}{4}E_{\alpha x,zx}\\
E_{\alpha y, zx}^{(3)}(\bar{\boldsymbol{\delta}}_{1}) &=& E_{\alpha y,zx}\\
E_{\alpha y, zx}^{(3)}(\bar{\boldsymbol{\delta}}_{2}) &=& -\frac{\sqrt{3}}{4}E_{\alpha y,yz}+\frac{3}{4}E_{\alpha x,yz}\nonumber\\
&&-\frac{1}{4}E_{\alpha y,zx}+\frac{\sqrt{3}}{4}E_{\alpha x,zx}\\
E_{\alpha y, zx}^{(3)}(\bar{\boldsymbol{\delta}}_{3}) &=& \frac{\sqrt{3}}{4}E_{\alpha y,yz}-\frac{3}{4}E_{\alpha x,yz}\nonumber\\
&&+\frac{1}{4}E_{\alpha y,zx}-\frac{\sqrt{3}}{4}E_{\alpha x,zx}\\
E_{\alpha y, zx}^{(3)}(\bar{\boldsymbol{\delta}}_{4}) &=& -E_{\alpha y,zx}\\
E_{\alpha y, zx}^{(3)}(\bar{\boldsymbol{\delta}}_{5}) &=& -\frac{\sqrt{3}}{4}E_{\alpha y,yz}-\frac{3}{4}E_{\alpha x,yz}\nonumber\\
&&+\frac{1}{4}E_{\alpha y,zx}+\frac{\sqrt{3}}{4}E_{\alpha x,zx}\\
E_{\alpha y, zx}^{(3)}(\bar{\boldsymbol{\delta}}_{6}) &=& \frac{\sqrt{3}}{4}E_{\alpha y,yz}+\frac{3}{4}E_{\alpha x,yz}\nonumber\\
&&-\frac{1}{4}E_{\alpha y,zx}-\frac{\sqrt{3}}{4}E_{\alpha x,zx}\\
E_{\alpha x, yz}^{(3)}(\bar{\boldsymbol{\delta}}_{1}) &=& E_{\alpha x,yz}\\
E_{\alpha x, yz}^{(3)}(\bar{\boldsymbol{\delta}}_{2}) &=& -\frac{\sqrt{3}}{4}E_{\alpha y,yz}-\frac{1}{4}E_{\alpha x,yz}\nonumber\\
&&+\frac{3}{4}E_{\alpha y,zx}+\frac{\sqrt{3}}{4}E_{\alpha x,zx}\\
E_{\alpha x, yz}^{(3)}(\bar{\boldsymbol{\delta}}_{3}) &=& \frac{\sqrt{3}}{4}E_{\alpha y,yz}+\frac{1}{4}E_{\alpha x,yz}\nonumber\\
&&-\frac{3}{4}E_{\alpha y,zx}-\frac{\sqrt{3}}{4}E_{\alpha x,zx}\\
E_{\alpha x, yz}^{(3)}(\bar{\boldsymbol{\delta}}_{4}) &=& -E_{\alpha x,yz}\\
E_{\alpha x, yz}^{(3)}(\bar{\boldsymbol{\delta}}_{5}) &=& -\frac{\sqrt{3}}{4}E_{\alpha y,yz}+\frac{1}{4}E_{\alpha x,yz}\nonumber\\
&&-\frac{3}{4}E_{\alpha y,zx}+\frac{\sqrt{3}}{4}E_{\alpha x,zx}\\
E_{\alpha x, yz}^{(3)}(\bar{\boldsymbol{\delta}}_{6}) &=& \frac{\sqrt{3}}{4}E_{\alpha y,yz}-\frac{1}{4}E_{\alpha x,yz}\nonumber\\
&&+\frac{3}{4}E_{\alpha y,zx}-\frac{\sqrt{3}}{4}E_{\alpha x,zx}\\
E_{\alpha x, zx}^{(3)}(\bar{\boldsymbol{\delta}}_{1}) &=& E_{\alpha x,zx}\\
E_{\alpha x, zx}^{(3)}(\bar{\boldsymbol{\delta}}_{2}) &=& \frac{3}{4}E_{\alpha y,yz}+\frac{\sqrt{3}}{4}E_{\alpha x,yz}\nonumber\\
&&+\frac{\sqrt{3}}{4}E_{\alpha y,zx}+\frac{1}{4}E_{\alpha x,zx}\\
E_{\alpha x, zx}^{(3)}(\bar{\boldsymbol{\delta}}_{3}) &=& \frac{3}{4}E_{\alpha y,yz}+\frac{\sqrt{3}}{4}E_{\alpha x,yz}\nonumber\\
&&+\frac{\sqrt{3}}{4}E_{\alpha y,zx}+\frac{1}{4}E_{\alpha x,zx}\\
E_{\alpha x, zx}^{(3)}(\bar{\boldsymbol{\delta}}_{4}) &=& E_{\alpha x,zx}\\
E_{\alpha x, zx}^{(3)}(\bar{\boldsymbol{\delta}}_{5}) &=& \frac{3}{4}E_{\alpha y,yz}-\frac{\sqrt{3}}{4}E_{\alpha x,yz}\nonumber\\
&&-\frac{\sqrt{3}}{4}E_{\alpha y,zx}+\frac{1}{4}E_{\alpha x,zx}\\
E_{\alpha x, zx}^{(3)}(\bar{\boldsymbol{\delta}}_{6}) &=& \frac{3}{4}E_{\alpha y,yz}-\frac{\sqrt{3}}{4}E_{\alpha x,yz}\nonumber\\
&&-\frac{\sqrt{3}}{4}E_{\alpha y,zx}+\frac{1}{4}E_{\alpha x,zx}
\end{eqnarray}
Here, energy integrals for $\mathcal{H}_{\alpha i, j}^{(3)}$ is 
\begin{equation}
E_{\alpha i, j}^{(3)} = E_{\alpha i, j}(2\boldsymbol{\delta}_{1}) = \langle \alpha p_{i}(\mathbf{0}) | \mathcal{H} | d_{j} (2\boldsymbol{\delta}_{1})\rangle.    
\end{equation}
The independent energy integrals are $E_{\alpha z,z^{2}}^{(3)}$, $E_{\alpha z,xy}^{(3)}$, $E_{\alpha z,x^2-y^2}^{(3)}$, $E_{\alpha z,zx}^{(3)}$, $E_{\alpha z,yz}^{(3)}$ $E_{\alpha y,z^{2}}^{(3)}$, $E_{\alpha x,z^{2}}^{(3)}$, $E_{\alpha y,x^{2}-y^{2}}^{(3)}$, $E_{\alpha y,xy}^{(3)}$, $E_{\alpha x,x^2-y^2}^{(3)}$, $E_{\alpha x,xy}^{(3)}$, $E_{\alpha y,yz}^{(3)}$, $E_{\alpha y,zx}^{(3)}$, $E_{\alpha x,zx}^{(3)}$, and $E_{\alpha x,yz}^{(3)}$ for $\alpha$ = $u$ or $l$, which are thirty in total.

\end{document}